\documentclass[acmsmall,screen,nonacm]{acmart}

\usepackage{url}
\usepackage{xspace}
\usepackage{wrapfig}
\usepackage{hyperxmp}
\usepackage{booktabs}
\usepackage{makecell}
\usepackage{hyperref}
\usepackage{graphicx}
\usepackage{colortbl}
\usepackage{multirow}
\usepackage{enumitem}
\usepackage{subfigure}
\usepackage{todonotes}
\usepackage{tcolorbox}
\usepackage{tablefootnote}
\usepackage{color, xcolor}
\usepackage{amsthm,amsmath,amsfonts}

\newcommand{\figref}[1]{Fig.~\ref{#1}\xspace}
\newcommand{\tabref}[1]{Table~\ref{#1}\xspace}

\setcopyright{acmlicensed}
\copyrightyear{2026}
\acmYear{2026}
\acmDOI{XXX.XXX}

\acmJournal{TOSEM}
\acmVolume{1}
\acmNumber{1}
\acmArticle{1}
\acmMonth{1}

\begin{document}

\title{Software Engineering for and with GUI Agent}

\author{Shengcheng Yu}
\affiliation{\institution{State Key Laboratory for Novel Software Technology, Nanjing University}\city{Nanjing}\country{China}\postcode{210093}}
\affiliation{\institution{Technical University of Munich}\city{Heilbronn}\country{Germany}\postcode{74076}}
\email{shengcheng.yu@tum.de}
\orcid{0000-0003-4640-8637}

\author{Yuchen Ling}
\affiliation{\institution{State Key Laboratory for Novel Software Technology, Nanjing University}\city{Nanjing}\country{China}\postcode{210093}}
\orcid{0009-0006-9227-3824}
\email{yuchenling@smail.nju.edu.cn}

\author{Junyang Xing}
\affiliation{\institution{State Key Laboratory for Novel Software Technology, Nanjing University}\city{Nanjing}\country{China}\postcode{210093}}
\email{xingjunyang@smail.nju.edu.cn}
\orcid{0009-0006-5180-6137}

\author{Quan Zhou}
\email{qzhou@smail.nju.edu.cn}
\orcid{0009-0008-1701-7366}
\affiliation{\institution{State Key Laboratory for Novel Software Technology, Nanjing University}\city{Nanjing}\country{China}}

\author{Chunrong Fang}
\authornote{Chunrong Fang is the corresponding author.}
\affiliation{\institution{State Key Laboratory for Novel Software Technology, Nanjing University}\city{Nanjing}\country{China}\postcode{210093}}
\email{fangchunrong@nju.edu.cn}
\orcid{0000-0002-9930-7111}

\author{Zhenyu Chen}
\affiliation{\institution{State Key Laboratory for Novel Software Technology, Nanjing University}\city{Nanjing}\country{China}\postcode{210093}}
\email{zychen@nju.edu.cn}
\orcid{0000-0002-9592-7022}


\begin{abstract}

Graphical user interface (GUI) agents have advanced rapidly in recent years, producing a growing body of frameworks, benchmarks, and applications. However, this growth has outpaced the maturity of the field. GUI agents remain technically brittle, incompletely engineered, and insufficiently validated for sustained real-world use. They are also evolving into closed-loop software systems. Within these systems, model reasoning is coupled with interface perception, execution feedback, recovery, and human oversight. This evolution calls for a software engineering perspective that remains largely absent from existing research. We address this gap by reviewing 336 GUI-agent papers from January 2018 to April 2026. Five research questions examine the research landscape, architectures, evaluation, software lifecycle concerns, and future opportunities. Our findings show that the field has expanded sharply since 2024, while mobile and web settings remain dominant. Architectures increasingly adopt modular perceive--reason--act loops, but recovery, human escalation, safety enforcement, and auditability remain underdeveloped. This architectural imbalance extends to evaluation. Evaluations are becoming more interactive, but they remain centered on task success and are difficult to compare across protocols. More broadly, existing studies provide limited support for testing beyond benchmarks and for maintaining agents after release. Observability, privacy engineering, and systematic human oversight are also underdeveloped. Together, these findings show that capability improvements alone cannot ensure deployment readiness. Future research should connect dependable execution with lifecycle-centered testing and reproducible evaluation. It should also integrate permission and privacy controls with cost-aware, human-centered governance. This integration is necessary to build dependable, maintainable, secure, and deployable GUI-agent systems.

\end{abstract}



\maketitle

\section{Introduction}
\label{sec:introduction}

Graphical user interface (GUI) agents have become a major research topic as foundation models have improved their ability to interpret instructions, perceive interfaces, and generate actions.
They now operate across web navigation, mobile-device control, desktop automation, and general computer-use settings.
GUI-specific model training and fine-tuning have been central to this progress.
CogAgent combines GUI-oriented pre-training with multi-task fine-tuning, SeeClick strengthens visual agents through GUI-grounding pre-training, and GUICourse uses staged GUI data to adapt general vision-language models for interface understanding, grounding, and action generation \cite{hong2024cogagent003, cheng2024seeclick012, chen2025guicourse011}.
At the system level, frameworks such as Agent~S, UFO, and AppAgent show how foundation models can be orchestrated for multi-step interaction across operating systems and applications \cite{agashe2024agent101, zhang2024ufo109, li2024appagent206}.
Benchmarks such as WebArena, AndroidWorld, and OSWorld then provide shared test environments for web, mobile, and desktop agents \cite{zhou2023webarena128, rawles2024androidworld148, xie2024osworld159}.
Together, these progress has made GUI interaction a shared challenge across several research communities.

This growth is accompanied by a change in what constitutes a GUI agent.
Early work often focused on individual capabilities such as interface understanding, grounding, or action prediction.
Recent systems increasingly combine these capabilities within a closed interaction loop.
One line of work strengthens perception and grounding through GUI-specialized models or parsers \cite{hong2024cogagent003, cheng2024seeclick012, lu2024omniparser004, wu2024os063, qin2025ui001, lin2025showui062, li2025screenspot051, luo2025gui066}.
Another embeds those capabilities in runtimes that maintain task state, execute actions, and respond to failure \cite{wen2024autodroid149, wang2024mobile208, wang2025mobile225, wu2024os213, tan2024cradle214, liu2025pc233, zhang2025ufo2249}.
GUI agents are therefore becoming integrated software systems.
Their behavior depends on the interaction between the model, its orchestration layer, the target application, and the user.
The resulting system boundary now extends well beyond the capabilities of the foundation model itself.

The maturity of the field has not kept pace with this increasing system complexity.
Technical performance remains brittle under long-horizon tasks, interface changes, and unexpected environment states.
Engineering support is less consistent.
Recovery and monitoring are often partial, while maintenance procedures, permission controls, and human takeover policies remain uncommon.
Evaluation has become more interactive, but it remains centered on task success and is often difficult to compare across protocols.
Evidence of sustained value in real-world applications is also limited.
An agent may complete a benchmark task while relying on excessive retries, exposing sensitive content, or failing after a minor interface update.
Recent benchmarks and safety studies have begun to expose these limitations across different tasks and platforms \cite{koh2024visualwebarena129, wang2024mobileagentbench154, xu2025androidlab152, xu2025crab167, kapoor2024omniact161, cao2024spider2163, sun2025scienceboard183, levy2024st139, tur2025safearena175, zharmagambetov2025agentdam179, zhang2025environmental017, liao2024eia322, evtimov2025wasp320, chen2025toward036}.
These gaps become more consequential as agents receive broader access and greater operational autonomy.

Existing surveys provide valuable maps of this fast-moving area.
Some surveys organize the field around agent capabilities and benchmarks.
Others focus on particular platforms, training strategies, or trustworthiness concerns \cite{zhang2024large005,nguyen2025gui006,wang2024gui008,tang2025a009,liu2025llm065,hu2024os285,shi2025towards007,li2025a075}.
These surveys clarify what GUI agents are and how their core capabilities are developed.
However, they provide limited synthesis of how GUI agents should be engineered and evaluated across the software lifecycle.
This perspective is increasingly important because deployment requires more than an accurate model.
The system also needs explicit interfaces and requirements, testable recovery behavior, operational logs, bounded permissions, and a policy for human handoff.
A software engineering synthesis can connect these concerns and reveal where capability gains do not yet translate into dependable deployment.
Without this view, recurring system risks remain separated across otherwise closely related research threads in the current literature.

This paper studies GUI agents from that software-engineering perspective.
We analyze 336 papers published or posted from January 2018 to April 2026.
The corpus covers research artifacts ranging from models and frameworks to benchmarks, evaluations, tools, and prior surveys.
The goal is to determine how the field has evolved and whether current progress supports dependable, maintainable, secure, and deployable systems.
Five research questions structure this analysis.
RQ1 maps the research landscape, RQ2 examines system architectures, and RQ3 evaluates the evidence behind capability claims.
RQ4 then assesses software-engineering coverage across the lifecycle, while RQ5 synthesizes the resulting gaps into a future research agenda.
This progression connects descriptive evidence about the field with prescriptive guidance for engineering practice.

The main contributions of this survey are as follows:
\begin{itemize}[leftmargin=8mm]
\item We map the development of GUI-agent research across publication trends, contribution types, platforms, applications, interface representations, and enabling models.
\item We synthesize GUI-agent architectures as closed-loop software systems and identify weaknesses in recovery, escalation, safety enforcement, and auditability.
\item We analyze how evaluation protocols support capability claims and why their results remain difficult to compare.
\item We assess the coverage of software-engineering concerns across the lifecycle, including testing, maintainability, observability, security, privacy, efficiency, and human oversight.
\item We derive a research agenda for connecting capability improvements with dependable execution, lifecycle management, reproducible evaluation, and risk-aware deployment.
\end{itemize}

The remainder of the paper follows this logic.
Section~\ref{sec:background} defines GUI agents, summarizes their technical evolution, and introduces the software-engineering concepts used in the survey.
Section~\ref{sec:methodology} describes the corpus, research questions, collection process, and coding procedure.
Sections~\ref{sec:rq1-landscape}--\ref{sec:rq5-opportunities} answer the five research questions.
Section~\ref{sec:discussion} draws cross-cutting insights from the findings, and the final sections position this survey relative to prior reviews, discuss threats to validity, and conclude with implications for future GUI-agent research.

\section{Background}
\label{sec:background}

This section establishes the concepts needed to analyze GUI agents.
It first defines their closed-loop interaction and technical evolution, then introduces the engineering concerns created by integrating perception, reasoning, execution, and human oversight within a runtime.

\subsection{GUI Agents: Definitions, Evolution, and Scope}
\label{sec:background-definitions}

A graphical user interface (GUI) agent is an autonomous or semi-autonomous system that completes tasks through the interface presented to human users.
It interprets an instruction, observes the GUI, and executes actions such as clicking, typing, scrolling, dragging, or using system shortcuts.
Tool-oriented agents usually invoke documented APIs, whereas GUI agents act through visual layouts and interaction conventions that may change without a stable interface contract.
The agent must translate a semantic goal into spatially and temporally grounded operations under partial observability.
GUI understanding, task planning, action grounding, and sequential decision making consequently form one interaction problem \cite{zhang2024large005, nguyen2025gui006, wang2024gui008, tang2025a009}.

\begin{figure}[t]
\centering
\includegraphics[width=0.90\textwidth]{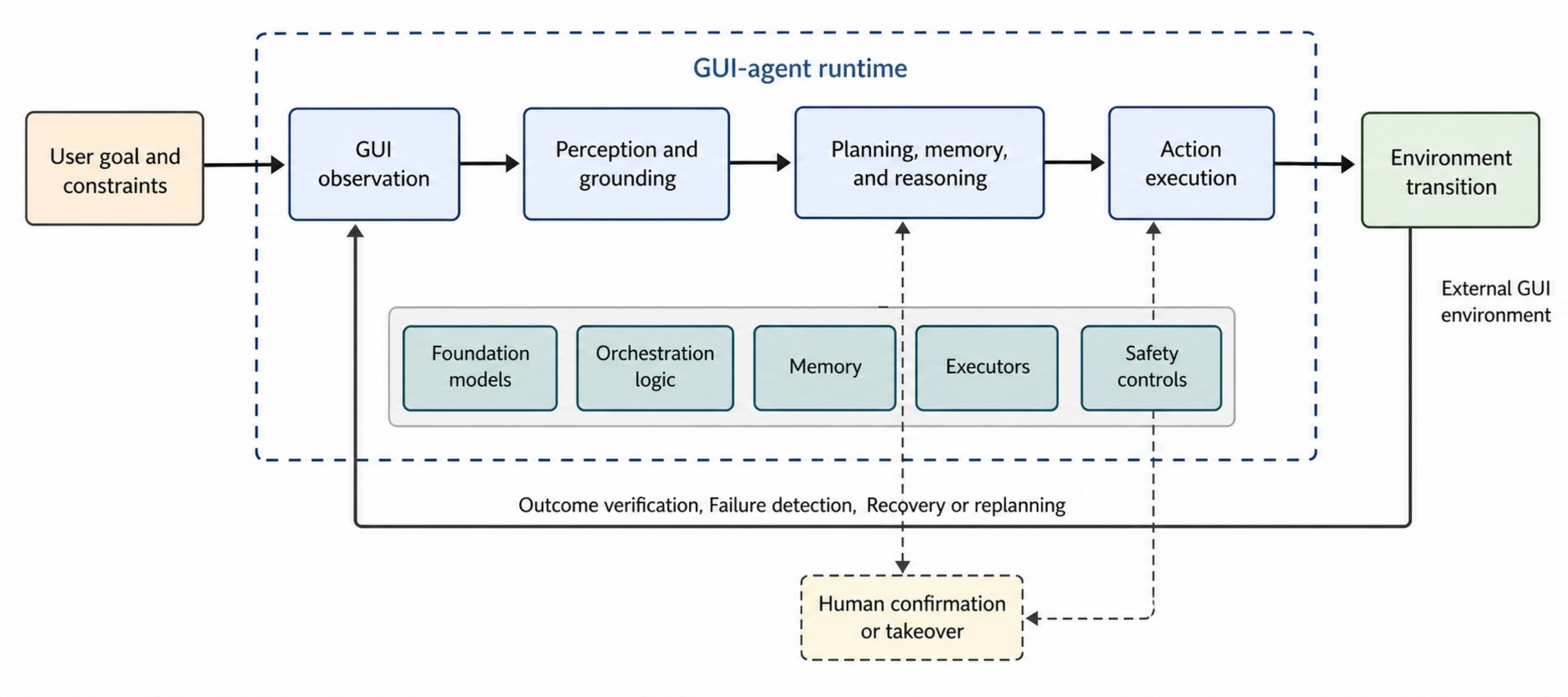}
\vspace{-3mm}
\caption{GUI-agent interaction loop.}
\label{fig:gui-agent-loop}
\end{figure}

Each action changes the interface and creates a closed feedback loop.
At step $t$, the agent combines the instruction with the current observation, updates its task state, executes an action, and observes the transition.
The loop ends when the goal is achieved, a stopping condition is reached, or control returns to the user.
Perception and grounding construct actionable interface state \cite{gou2024navigating002, cheng2024seeclick012, li2025screenspot051}.
Planning and memory maintain long-horizon progress \cite{zhang2024dynamic028, erdogan2025plan242, cheng2025mga025, wu2025auto032}, while verification and recovery respond to unexpected outcomes \cite{lee2025verisafe043, lu2025steve243, chen2025gui316}.
AppAgent and Mobile-Agent coordinate these capabilities through screenshots, view hierarchies, and interaction histories on mobile devices \cite{zhang2025appagent205, li2024appagent206, wang2024mobile208}.
Agent~S and UFO extend the loop to desktop and multi-application workflows \cite{agashe2024agent101, zhang2024ufo109, zhang2025ufo2249}.
We summarize the learned and deterministic components, verification paths, recovery, and human intervention in \figref{fig:gui-agent-loop}.

The closed-loop abstraction predates foundation models.
Earlier systems learned policies for constrained websites, grounded web-support instructions, mapped language to mobile actions, or used DOM and Android view hierarchies \cite{liu2018reinforcement124, xu2021grounding125, li2020mapping349, rawles2023android150}.
They established instruction-conditioned policies, interface state representations, finite action vocabularies, and environments that returned state transitions.
Their scope remained tied to specific platforms, tasks, or interface schemas and often required application-specific adaptation.
Moving to a new application commonly required additional demonstrations, handcrafted selectors, or a new representation of target interface.

Large language models and multimodal foundation models expanded this paradigm.
Language models improved instruction interpretation and task decomposition, while vision-language models made screenshots a practical observation channel \cite{baechler2024screenai309, lin2025showui062}.
SeeClick and CogAgent connect instructions to visual elements and coordinates \cite{cheng2024seeclick012, hong2024cogagent003}, and OmniParser converts screenshots into candidate interface elements \cite{lu2024omniparser004}.
Native agents such as UI-TARS, OS-ATLAS, ShowUI, and GUI-R1 integrate perception, reasoning, and action prediction within one model \cite{qin2025ui001, wu2024os063, lin2025showui062, luo2025gui066}.
These advances support general computer use across heterogeneous applications.
They also shift part of the interaction logic from platform-specific rules into learned visual and language representations.

Long-horizon benchmarks place agents in interactive websites, mobile applications, desktop software, and knowledge-work tasks \cite{zhou2023webarena128, koh2024visualwebarena129, rawles2024androidworld148, xie2024osworld159, bonatti2024windows160, drouin2024workarena134}.
They expose dependencies among perception, reasoning, execution, and recovery that a grounded click cannot capture.
Successful behavior in these settings requires persistent task state, appropriate termination, recovery from invalid actions, and control over irreversible effects.
GUI agents must therefore be analyzed as stateful systems whose behavior depends on learned components and surrounding infrastructure.

\subsection{Software Engineering of Agentic GUI Systems}
\label{sec:background-se-concepts}

Viewing GUI agents as stateful systems shifts attention to the complete runtime.
Observation adapters and memory construct state, executors and environment controllers apply decisions, and safety checks, logs, and confirmation mechanisms constrain behavior.
AutoDroid, AssistGUI, Agent~S, OS-Copilot, and MM-Pro combine foundation models with screenshots, structured metadata, tool wrappers, or deterministic controllers \cite{wen2024autodroid149, gao2024assistgui026, agashe2024agent101, wu2024os213, wu2025mmpro054}.
Failures can emerge from interactions among these components when interfaces move, pages load slowly, modals intercept input, or executors select the wrong window.
The model may produce a valid plan while the complete system still fails because of incorrect tool execution or environment feedback.

Requirements define intended capability and acceptable behavior.
Reliability and robustness concern repeated execution under interface variation, while operational quality covers latency, interaction cost, privacy, and resistance to malicious content.
Robustness, safety, privacy, and human-centered studies begin to isolate these attributes \cite{yang2025gui024, levy2024st139, zharmagambetov2025agentdam179, chen2025toward036}.
Their interactions require explicit operating boundaries.
Confirmation can reduce harmful actions while increasing user effort, and context compression can lower cost while weakening state tracking.
Privacy masking may also remove information needed for task completion.
Requirements must specify acceptable trade-offs among autonomy, reliability, cost, privacy, and human oversight.

Architecture assigns these responsibilities.
Framework-based agents expose perception, planning, execution, memory, and reflection as inspectable modules \cite{agashe2024agent101, zhang2024ufo109, wang2024mobile209, zhao2025cola114}.
Native models learn tighter mappings from observations and instructions to reasoning and actions \cite{qin2025ui001, wu2024os063, lin2025showui062}, while hybrid systems combine foundation models, specialized grounders, and deterministic controllers \cite{cheng2024seeclick012, lu2024omniparser004, lee2025verisafe043}.
These choices shape component replacement, failure isolation, observability, platform adaptation, and testing.
Modular systems expose intermediate state and policy boundaries more clearly, while native models reduce hand-built coordination and concentrate more responsibility inside learned behavior.

Testing must reflect the same boundaries.
Offline datasets isolate perception, grounding, and next-action prediction \cite{rawles2023android150, cheng2024seeclick012, li2025screenspot051}, while interactive benchmarks evaluate trajectories under environment feedback \cite{zhou2023webarena128, rawles2024androidworld148, xie2024osworld159}.
Offline tests omit timing, recovery, and side effects, and aggregate task success can hide retries, unsafe intermediate actions, or environment dependence.
Repeated runs, controlled perturbations, and stable regression oracles provide complementary evidence.
Reliability testing examines variation across repeated executions, while robustness testing changes interface content, timing, or environmental conditions.
Regression testing then checks whether model, prompt, wrapper, or application updates invalidate earlier behavior.
GUI-Robust, WABER, and environmental-injection benchmarks evaluate anomalies, web unreliability, and active perturbations \cite{yang2025gui024, kara2025waber321, chen2025evaluating318, chen2025ghostei315}.

Operation and evolution determine how long testing evidence remains valid.
Applications, models, prompts, and external tools change on separate schedules, so an agent can degrade without changes to its own code \cite{lu2025transbench123, zheng2024webolympus145, chezelles2024the168}.
Logs, state snapshots, trajectory replay, and failure classification distinguish model errors from environment or executor failures \cite{lu2025steve243, chen2025gui316, lu2025agentrewardbench261}.
Maintainability also requires explicit contracts for observations, actions, stopping criteria, fallback rules, and permissions.
Versioned interfaces and replayable traces help preserve system intent as individual components and target applications evolve.

Operation also creates a security and privacy boundary because agents may observe credentials or private documents and modify external state.
Environmental injection can redirect behavior or induce privacy leakage \cite{zhang2025environmental017, liao2024eia322, evtimov2025wasp320, lu2025eva323}.
GUIGuard, VeriOS, and VeriSafe address visual data protection, policy compliance, human involvement, and action verification \cite{wang2026guiguard033, wu2025verios327, lee2025verisafe043}.
Safety is therefore a runtime responsibility that must be specified, tested, monitored, and audited across the lifecycle.
Permission boundaries, confirmation policies, and logs connect that responsibility to concrete actions and later accountability.

\section{Survey Methodology}
\label{sec:methodology}

Our methodology links the research questions to corpus construction, structured coding, and evidence synthesis.
The complete corpus establishes the field-level landscape, while purpose-built analytical subsets provide the evidence required for the architectural, evaluation, and software-engineering analyses.
This layered design combines broad coverage with focused coding at the level appropriate to each research question.

\subsection{Research Questions}
\label{sec:method-rqs}

\begin{description}
\item[\textbf{RQ1}] How has GUI agent research evolved, and what are the major trends in research contributions, platforms, application domains, and enabling models?
\item[\textbf{RQ2}] How are GUI agents architected to perceive interfaces, reason about tasks, execute actions, and recover from failures?
\item[\textbf{RQ3}] How are GUI agents evaluated, and what protocol and system factors shape the validity, reproducibility, and comparability of evaluation evidence?
\item[\textbf{RQ4}] To what extent does existing GUI agent research address software engineering concerns across the system lifecycle?
\item[\textbf{RQ5}] What open challenges and research opportunities must be addressed to engineer dependable, maintainable, secure, and deployable GUI agents?
\end{description}

The research questions move from descriptive mapping to engineering interpretation.
RQ1 characterizes the field through publication trends, contribution types, platforms, applications, observation media, and model usage.
RQ2 moves inside the resulting systems to examine how they allocate perception, state representation, planning, memory, execution, verification, recovery, and human intervention.
RQ3 then assesses how benchmarks, protocols, and metrics support valid, reproducible, and comparable capability claims.
RQ4 broadens the unit of analysis from individual capabilities to lifecycle concerns such as requirements, testing, maintainability, observability, security, privacy, efficiency, and oversight.
Finally, RQ5 integrates the preceding results into a research agenda.
This sequence connects the scale and structure of the field to the evidence and engineering practices required for dependable deployment in practice.

\subsection{Literature Collection and Selection}
\label{sec:method-collection}

We defined the corpus around a functional criterion.
A paper was in scope when it treated a graphical interface as an observation or action environment for an autonomous or semi-autonomous agent.
The scope covers web pages, mobile applications, desktop operating systems, productivity software, and cross-platform computer-use environments.
Supporting work was also eligible when it directly contributed a dataset, benchmark, grounding model, UI parser, safety analysis, evaluation protocol, or engineering technique for GUI agents.
Together, these criteria center the corpus on studies that combine GUI interaction with autonomous or semi-autonomous operation.

Conventional GUI testing was incorporated through an agent-centered selection boundary.
That mature literature primarily studies test generation, event-sequence exploration, regression testing, crash detection, coverage, model-based testing, and record-and-replay validation.
Papers from this literature entered the corpus when they explicitly studied a GUI agent or directly supported agent perception, decision making, execution, benchmark design, safety, or lifecycle engineering.
This criterion concentrates the software-engineering evidence on techniques that inform GUI-agent construction and evaluation.

Corpus construction combined keyword search, backward and forward snowballing, and manual consolidation of existing paper lists.
The search vocabulary covered GUI agents, computer-use agents, web agents, mobile agents, Android agents, desktop agents, UI automation, interface grounding, multimodal agents, and GUI benchmarks.
We considered both peer-reviewed papers and preprints to represent rapidly emerging work alongside venue-reviewed evidence.
Publication source was recorded separately so that RQ1 could distinguish corpus growth from evidence maturity.
We manually consolidated duplicate records, title variants, and multiple versions of the same work.
When several versions were available, we selected the most complete and current bibliographic record and harmonized its metadata with the rest of the corpus.

The resulting corpus contains 336 unique papers published or posted between 2018 and April 2026.
Every retained paper contributes to at least one core category, namely framework, model, benchmark, evaluation study, assistive tool, survey, or software-engineering analysis.
This rule captures the multidisciplinary character of GUI-agent research while preserving a clear boundary around systems that perceive, reason over, act on, evaluate, or engineer graphical interfaces.
The corpus therefore supports both quantitative mapping and qualitative interpretation of the field.

\subsection{Data Extraction, Coding, and Synthesis}
\label{sec:method-coding}

The paper is the primary unit of analysis.
For each paper, we first recorded its bibliographic metadata and contribution type.
Technical coding covered the target platform, application domain, base model, observation medium, and agent modules.
Separate fields captured the reported techniques, evaluation design, metrics, and human-in-the-loop signals.
Contribution type is multi-label because one paper may introduce several artifacts, such as a framework and a benchmark.
Platform and observation medium are also multi-label because a system may span several platforms or combine screenshots with HTML, XML, accessibility trees, or generated descriptions.

We developed the coding scheme iteratively from coarse contribution categories to finer engineering dimensions.
The first layer distinguishes frameworks, benchmarks, models, evaluations, assistive tools, and surveys.
The second layer captures functions that may be implicit in a paper.
It follows the agent loop from perception and decision to execution, then records support for planning, memory, verification, reflection, and recovery.
Additional fields cover operational and governance concerns such as cost, observability, maintainability, safety, privacy, and human oversight.
When papers used different names for similar mechanisms, we coded the function performed in the agent loop.
For example, a verifier, critic, evaluator, reward model, or post-action checker may provide verification or reflection depending on how it affects execution.
This functional interpretation enables comparison across systems with different architectural vocabularies.

Each research question draws on an analytical population aligned with its evidence needs.
RQ1 uses the complete 336-paper corpus.
RQ2 draws module statistics from the 145 papers coded as frameworks because these papers expose explicit system structures.
RQ3 uses the 252 unique papers coded as frameworks, models, or evaluation studies, where capability evidence is most directly reported.
RQ4 uses the 327 records with complete software-engineering coding.
Each analytical subset has a distinct denominator, which accompanies all reported counts to support transparent interpretation.

We used quantitative synthesis for fields that supported consistent aggregation.
The field-level summaries describe publication patterns, research artifacts, target settings, and model use.
Architecture and evaluation are aggregated separately, as are the software-engineering signals used in RQ4.
Multi-label categories are reported as overlapping counts rather than mutually exclusive partitions.
We then used qualitative synthesis to explain the distributions.
This analysis examines how observation media shape architecture, how benchmark protocols condition performance claims, how systems evaluate recovery and verification, and how engineering concerns appear across the complete system lifecycle.

We checked related fields to strengthen internal coherence across the coding scheme.
A framework record was expected to expose perception, decision, and execution responsibilities.
A benchmark record required a task, environment, dataset, or evaluation-protocol contribution.
Safety and privacy signals required a concrete risk, attack, defense, policy, or sensitive-data concern.
For records requiring interpretation, we considered the contribution claim, method, and evaluation setup together with the terminology used by the authors.
These checks make interpretive judgments explicit and tie later claims to observable properties of each paper.

The remainder of the survey follows the same analytical sequence.
RQ1 establishes scale and distribution, RQ2 interprets architectural patterns, RQ3 examines evaluation evidence, and RQ4 assesses lifecycle coverage.
RQ5 then integrates the results into a research agenda.
Numerical summaries establish the prevalence of observed patterns, while representative papers explain their technical and engineering significance.

\section{RQ1: Research Landscape and Development Trends}
\label{sec:rq1-landscape}

We use RQ1 to establish the empirical landscape for the later analyses.
The corpus contains 336 papers from January 2018 to April 2026.
Only 19 appeared before 2024, while 317 were published or posted during 2024--2026.
This increase coincides with stronger multimodal models, interactive benchmarks across major platforms, and a shift from isolated GUI understanding to closed-loop GUI interaction \cite{hong2024cogagent003, cheng2024seeclick012, zhou2023webarena128, rawles2024androidworld148, xie2024osworld159}.
We examine this development through publication patterns, contribution types, platform and domain coverage, observation media, and model usage.

\begin{figure*}[t]
\centering
\includegraphics[width=0.8\textwidth]{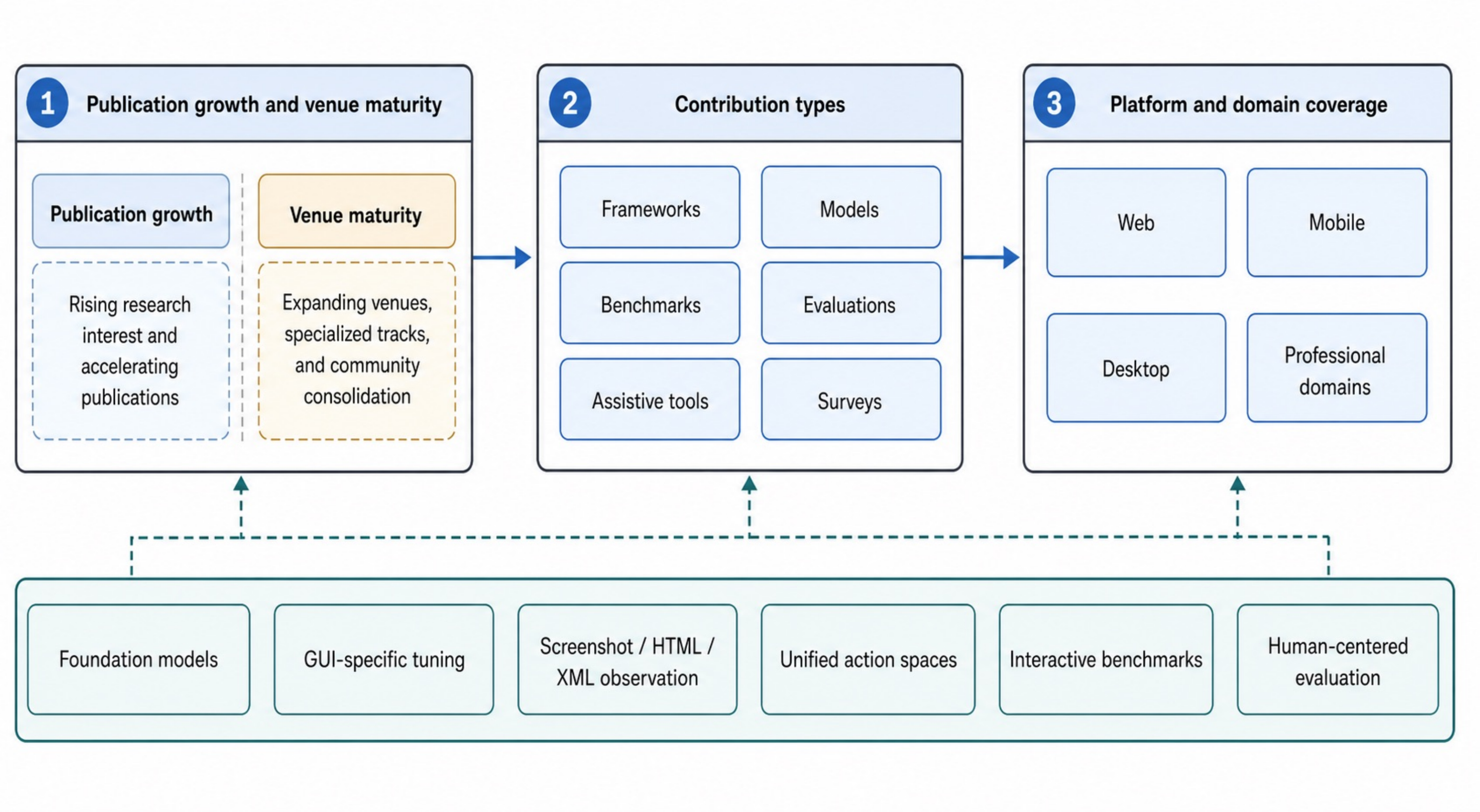}
\vspace{-5mm}
\caption{GUI-agent research landscape.}
\label{fig:rq1-landscape}
\end{figure*}

We use \figref{fig:rq1-landscape} to connect publication growth and source maturity with the technical dimensions that shape architectural and evaluation claims.
The recent expansion is distributed across systems, benchmarks, evaluation studies, assistive tools, and surveys \cite{agashe2024agent101, wang2025mmbench019, chen2025toward036, zhao2025appagent055, zhang2024large005}.
Many claims about GUI agents are produced inside tightly coupled research artifacts.
A new framework may introduce a benchmark, a benchmark may embed a particular observation interface, and a model paper may also define data generation and action-space assumptions.
We therefore interpret the literature as an evolving research ecosystem shaped by publication maturity, contribution type, platform boundary, task domain, observation medium, and model family.

\subsection{Rapid Growth and Uneven Publication Maturity}
\label{sec:rq1-publications}

\begin{figure*}[t]
\centering
\subfigure[Publication years]{%
\includegraphics[width=0.42\textwidth]{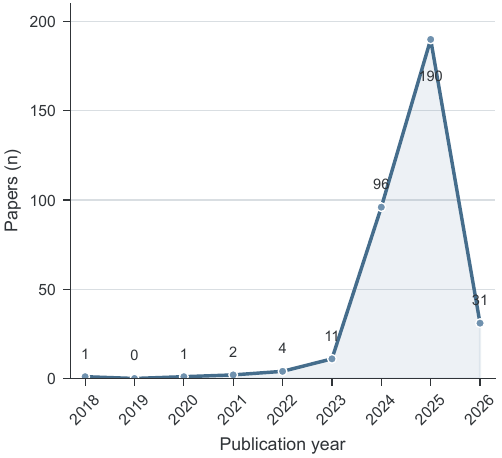}%
\label{fig:rq1-publication-years}%
}
\hspace{8mm}
\subfigure[Contribution types]{%
\includegraphics[width=0.42\textwidth]{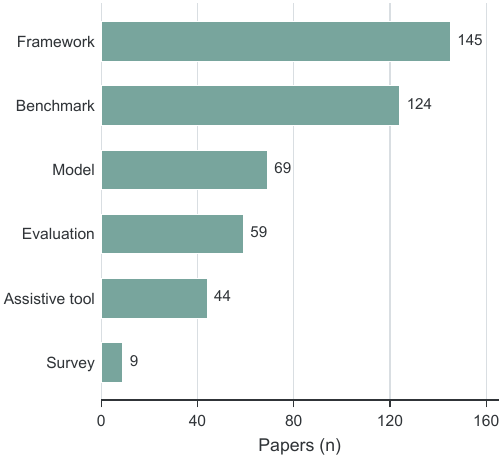}%
\label{fig:rq1-contribution-types}%
}
\vspace{-2mm}
\caption{Publication years and contribution types.}
\label{fig:rq1-year-contribution}
\end{figure*}

We observe in \figref{fig:rq1-year-contribution} that GUI-agent research moved from scattered precursors to a high-volume research topic within a short period.
Early work investigated web-interface reinforcement learning, web-support instruction grounding, and mobile GUI action prediction through narrower technical formulations \cite{liu2018reinforcement124, xu2021grounding125, li2020mapping349, rawles2023android150}.
WebShop and Mind2Web then helped frame GUI interaction as grounded language-to-action decision making in realistic web tasks \cite{yao2022webshop126, deng2023mind2web127}.
The post-2023 expansion is substantial.
We report 286 papers in 2024 and 2025 alone in the left panel of \figref{fig:rq1-year-contribution}, and the partial 2026 count already exceeds all pre-2024 work combined.
This acceleration coincides with WebArena, VisualWebArena, WebLINX, AndroidWorld, OSWorld, and later macOSWorld making interactive environments and cross-application tasks more visible as shared evaluation targets \cite{zhou2023webarena128, koh2024visualwebarena129, lu2024weblinx138, rawles2024androidworld148, xie2024osworld159, yang2025macosworld079}.
In parallel, GUI-specialized multimodal agent systems such as CogAgent, SeeClick, OmniParser, OS-ATLAS, UI-TARS, and ShowUI demonstrated that visual grounding, screenshot parsing, and action prediction could be trained or adapted at a scale that earlier GUI automation work did not support \cite{hong2024cogagent003, cheng2024seeclick012, lu2024omniparser004, wu2024os063, qin2025ui001, lin2025showui062}.
Contribution types are coded as multiple labels, so their counts do not sum to the corpus size.

\begin{table*}[t]
\centering
\scriptsize
\caption{Publication-source distribution.}
\label{tab:rq1-publication-source}
\resizebox{\textwidth}{!}{%
\begin{tabular}{lllrr}
\toprule
Source & Full name & Nature & Papers & Share \\
\midrule
arXiv & arXiv preprint repository & Informal & 204 & 60.7\% \\
ACL & Annual Meeting of the Association for Computational Linguistics & Conference & 20 & 6.0\% \\
Findings of ACL & Findings of the Association for Computational Linguistics: ACL & Conference & 16 & 4.8\% \\
AAAI & AAAI Conference on Artificial Intelligence & Conference & 12 & 3.6\% \\
NeurIPS & Conference on Neural Information Processing Systems & Conference & 12 & 3.6\% \\
EMNLP & Conference on Empirical Methods in Natural Language Processing & Conference & 9 & 2.7\% \\
Findings of EMNLP & Findings of the Association for Computational Linguistics: EMNLP & Conference & 8 & 2.4\% \\
CVPR & IEEE/CVF Conference on Computer Vision and Pattern Recognition & Conference & 7 & 2.1\% \\
ACM MM & ACM International Conference on Multimedia & Conference & 7 & 2.1\% \\
LNCS & Lecture Notes in Computer Science & Conference & 5 & 1.5\% \\
MobiCom & ACM International Conference on Mobile Computing and Networking & Conference & 4 & 1.2\% \\
KDD & ACM SIGKDD Conference on Knowledge Discovery and Data Mining & Conference & 4 & 1.2\% \\
ICLR & International Conference on Learning Representations & Conference & 3 & 0.9\% \\
CHI & ACM CHI Conference on Human Factors in Computing Systems & Conference & 3 & 0.9\% \\
CIKM & ACM International Conference on Information and Knowledge Management & Conference & 2 & 0.6\% \\
UIST & ACM Symposium on User Interface Software and Technology & Conference & 2 & 0.6\% \\
ACL Demo & ACL System Demonstrations & Conference & 2 & 0.6\% \\
EMFM Workshop & International Workshop on Edge and Mobile Foundation Models & Conference & 1 & 0.3\% \\
WWW & The Web Conference & Conference & 1 & 0.3\% \\
CVPR Workshop & IEEE/CVF Conference on Computer Vision and Pattern Recognition Workshops & Conference & 1 & 0.3\% \\
IC-NIDC & International Conference on Network Infrastructure and Digital Content & Conference & 1 & 0.3\% \\
MobiSys & ACM International Conference on Mobile Systems, Applications, and Services & Conference & 1 & 0.3\% \\
EMNLP Demo & EMNLP System Demonstrations & Conference & 1 & 0.3\% \\
Interspeech & Conference of the International Speech Communication Association & Conference & 1 & 0.3\% \\
COLM & Conference on Language Modeling & Conference & 1 & 0.3\% \\
VLDB & International Conference on Very Large Data Bases & Conference & 1 & 0.3\% \\
WSDM & ACM International Conference on Web Search and Data Mining & Conference & 1 & 0.3\% \\
ISSTA & ACM SIGSOFT International Symposium on Software Testing and Analysis & Conference & 1 & 0.3\% \\
SAC & ACM Symposium on Applied Computing & Conference & 1 & 0.3\% \\
ICLR Workshop & International Conference on Learning Representations Workshop & Conference & 1 & 0.3\% \\
IJCAI & International Joint Conference on Artificial Intelligence & Conference & 1 & 0.3\% \\
IEEE IoTJ & IEEE Internet of Things Journal & Journal & 1 & 0.3\% \\
Pattern Recognition & Pattern Recognition & Journal & 1 & 0.3\% \\
\bottomrule
\end{tabular}%
}
\end{table*}

We report the complete publication-source distribution of the 336-paper corpus in \tabref{tab:rq1-publication-source} to compare literature growth with formal publication cycles.
Preprints account for 204 papers, or 60.7\% of the corpus, so a large share of current evidence is still moving through review, revision, and venue consolidation.
The formally published papers are dispersed across several research communities.
NLP and vision contribute instruction following and visual grounding, while HCI studies the user-facing interaction.
Systems research supplies execution environments, and software engineering contributes methods for testing and lifecycle quality \cite{deng2023mind2web127, cheng2024seeclick012, zhang2025characterizing022, wen2024autodroid149, yang2025gui024}.
It also means that evidence maturity varies substantially.
A benchmark result reported in a preprint, an ACL paper on web instruction grounding, and a CVPR paper on screenshot understanding may all influence the field, yet they often carry different assumptions about task sampling, evaluation oracles, environment stability, and reproducibility.
Later sections interpret performance claims together with their evaluation settings and avoid direct score comparison when the underlying protocols differ in material ways.

Our contribution-type analysis shows a field that is still building its objects of study, as reported in the right panel of \figref{fig:rq1-year-contribution}.
Framework papers form the largest category, with 145 papers, followed by benchmarks with 124 papers.
Models and evaluation papers are smaller but substantial groups.
We observe an iterative loop between construction and measurement.
New agents expose missing task types, benchmarks make their weaknesses visible, and changes to perception or grounding create new evaluation needs.
Representative examples include framework-oriented systems for mobile and desktop automation \cite{zhang2025appagent205, agashe2024agent101, zhang2024ufo109}, benchmarks for web, mobile, and desktop interaction \cite{zhou2023webarena128, rawles2024androidworld148, xie2024osworld159}, and GUI-specific models for grounding or action prediction \cite{hong2024cogagent003, cheng2024seeclick012, qin2025ui001}.
Surveys have also appeared quickly \cite{zhang2024large005, nguyen2025gui006, wang2024gui008, tang2025a009}, which is typical of an area whose terminology and scope are expanding faster than its methodological conventions can stabilize across communities over time.

The contribution-type counts are multi-label, which is important for interpreting the right panel of \figref{fig:rq1-year-contribution}.
Many framework papers include benchmark construction or new evaluation protocols, while benchmark papers often encode implicit architectural assumptions through their observation format, action space, environment reset policy, and completion oracle \cite{wang2024mobileagentbench154, xu2025androidlab152, wang2025mmbench019}.
This coupling accelerates research while complicating cumulative evidence.
A reported improvement may come from the agent architecture or the foundation model.
It may instead reflect a more favorable action abstraction, tool wrapper, or task distribution.

\subsection{Mobile and Web Dominance in Current Research}
\label{sec:rq1-platforms}

\begin{figure*}[t]
\centering
\subfigure[Platform labels]{%
\includegraphics[width=0.42\textwidth]{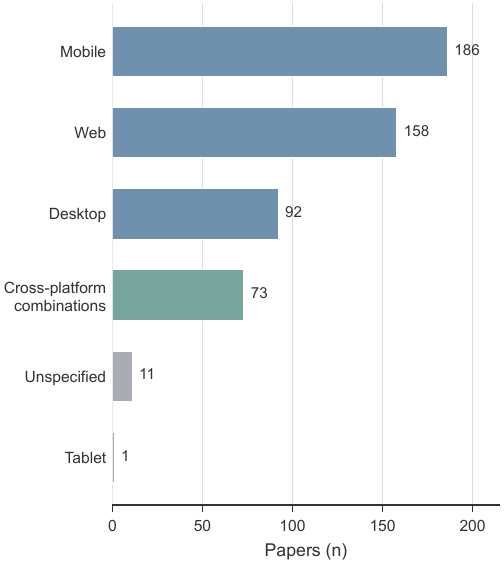}%
\label{fig:rq1-platform-labels}%
}
\hspace{8mm}
\subfigure[Application domains]{%
\includegraphics[width=0.42\textwidth]{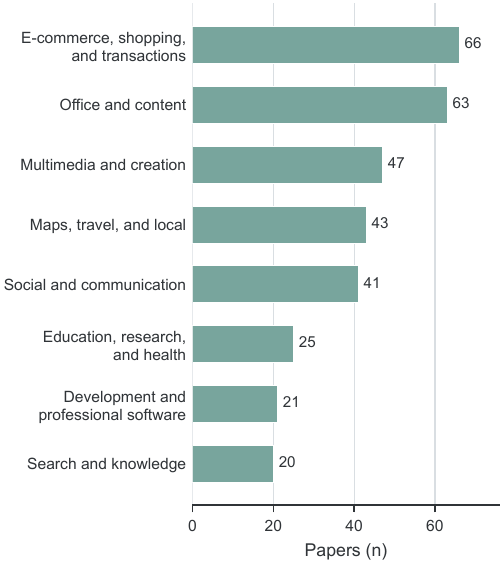}%
\label{fig:rq1-application-domains}%
}
\vspace{-2mm}
\caption{Platform and application-domain distributions.}
\label{fig:rq1-platform-domain}
\end{figure*}

We observe that mobile and web interfaces dominate the corpus, as reported in the platform panel of \figref{fig:rq1-platform-domain}.
Because platform labels are multi-label, the reported coverage overlaps and a paper can contribute to more than one platform category.
Under this coding, 186 papers cover mobile environments, 158 cover web environments, and 92 include desktop settings.
This pattern reflects both historical and practical factors.
Mobile interfaces combine a relatively constrained action space with app-scale task diversity.
Their structured view hierarchies can also complement screenshots \cite{rawles2023android150, zhang2023mobile146, wang2024mobileagentbench154, rawles2024androidworld148}.
Web environments are attractive for a different reason.
DOM structure and browser automation make public tasks easier to instrument at scale \cite{yao2022webshop126, deng2023mind2web127, zhou2023webarena128, chezelles2024the168, lu2024weblinx138}.
Desktop environments are harder to standardize because state is distributed across windows, files, applications, and operating-system services.
That complexity also makes them essential for evaluating general computer use \cite{xie2024osworld159, bonatti2024windows160, zhang2024ufo109, agashe2024agent101}.

Cross-platform work accounts for 73 papers, or 21.7\% of the corpus, as summarized in the platform panel of \figref{fig:rq1-platform-domain}.
These papers are important because they test whether a GUI agent can rely on abstractions that survive changes in device, layout convention, interaction primitive, and interface metadata.
Systems such as UI-TARS and OS-ATLAS explicitly pursue more general GUI action models across platforms \cite{qin2025ui001, wu2024os063}, while general-purpose computer-use frameworks and benchmarks such as Agent~S, OmniACT, CRAB, and MMBench-GUI aim to coordinate perception, planning, and tool use across broader computer-use settings \cite{agashe2024agent101, kapoor2024omniact161, xu2025crab167, wang2025mmbench019}.
Cross-platform claims carry a heavier evidentiary burden than single-platform claims.
A method that works on Android by using XML hierarchies may not transfer cleanly to arbitrary desktop applications.
A web agent that relies on HTML or accessibility trees may face different failure modes when the interface is image-heavy, dynamically generated, or partially hidden behind authentication and personalization.
Platform coverage therefore defines the assumptions under which an agent's perception, action, and recovery mechanisms are expected to operate.

Our application-domain analysis explains why these platforms became attractive testbeds, with the distribution reported in \figref{fig:rq1-platform-domain}.
The most common domains cover everyday activities such as commerce, office work, navigation, communication, and information access.
They share recognizable goals and visible intermediate states, which makes multi-step workflows easier to turn into repeatable tasks.
WebShop, WebArena, VisualWebArena, AndroidWorld, WorkArena, and OfficeBench make use of this structure by translating everyday activities into repeatable tasks \cite{yao2022webshop126, zhou2023webarena128, koh2024visualwebarena129, rawles2024androidworld148, drouin2024workarena134, wang2024officebench165}.
The same panel also shows a domain spread that extends beyond consumer tasks.
Scientific tools, engineering software, enterprise workflows, finance, health, and system settings point toward a harder deployment frontier.
Benchmarks such as WONDERBREAD, ScienceBoard, and Spider2-V illustrate this movement toward business processes, scientific workflows, and data-science or engineering software \cite{wornow2024wonderbread144, sun2025scienceboard183, cao2024spider2163}.
In these settings, task success may depend on business semantics, personal preferences, external documents, security policies, or irreversible side effects.
A click sequence can be syntactically correct while still producing a poor or unsafe real-world outcome.

This domain shift has a direct software engineering implication.
Early benchmarks can often define success through page state, selected items, form completion, or textual answer matching.
Real work settings require richer oracles.
The evaluator may need to inspect a generated artifact, determine whether a state change is reversible, or check whether the agent respected privacy and organizational constraints.
Safety, privacy, and human-centered benchmarks make this richer evaluation space explicit by measuring trustworthiness, leakage, and user-facing consequences in addition to task completion \cite{levy2024st139, tur2025safearena175, zharmagambetov2025agentdam179, zhang2025characterizing022}.
We accordingly distinguish task completion from reliability, reproducibility, safety, privacy, and human oversight in the later analysis.

\subsection{Screenshot-Centered Observation and Model Adaptation}
\label{sec:rq1-models}

\begin{figure*}[t]
\centering
\includegraphics[width=0.95\textwidth]{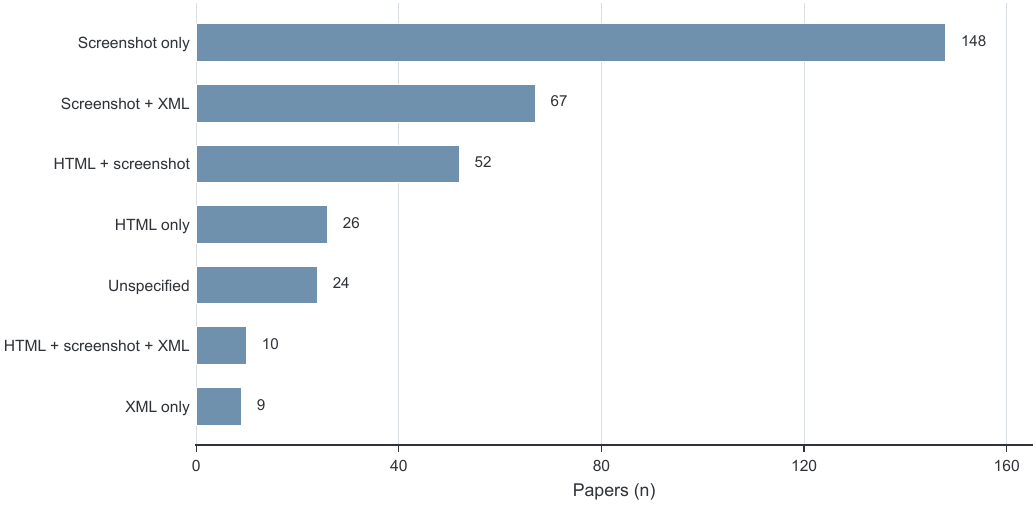}
\vspace{-2mm}
\caption{Observation-medium distribution.}
\label{fig:rq1-observation-media}
\end{figure*}

The enabling technology behind recent GUI agents is also changing.
We summarize the observation media used to interpret interfaces in \figref{fig:rq1-observation-media} and report model-family usage and model usage patterns in \tabref{tab:rq1-model-family} and \tabref{tab:rq1-model-usage-patterns}.
Screenshots appear in 277 papers, and 148 papers use screenshots as the only recorded GUI interpretation medium.
This dominance reflects the appeal of vision as a platform-independent observation channel.
A screenshot preserves layout, colors, icons, visual hierarchy, and spatial relations even when the underlying interface metadata is unavailable or inconsistent.
Pure-vision systems and tools such as CogAgent, SeeClick, OmniParser, ScreenAgent, and You Only Look at Screens therefore occupy a central position in the field \cite{hong2024cogagent003, cheng2024seeclick012, lu2024omniparser004, niu2024screenagent207, zhang2024you122}.
The categories in \figref{fig:rq1-observation-media} are exact observation-medium combinations and therefore sum to the full 336-paper corpus used in RQ1.
The two model tables also use paper-level counts from this corpus.
Model-family labels are multi-label because one paper can use or compare multiple families, whereas the usage-pattern categories summarize the \texttt{BASE\_MODEL} coding field.
In the latter table, ``uses'' means that at least one model satisfies the condition, ``only'' means that all specified models satisfy it, and ``combination'' means that both conditions occur in the same paper.

\begin{table*}[t]
\centering
\scriptsize
\caption{Model-family distribution.}
\label{tab:rq1-model-family}
\resizebox{\textwidth}{!}{%
\begin{tabular}{lrrl}
\toprule
Model family & Papers & Share & Representative variants or coding scope \\
\midrule
GPT-4/4o family & 96 & 28.6\% & GPT-4, GPT-4V, GPT-4o, GPT-4o-mini variants \\
Qwen/Qwen-VL family & 91 & 27.1\% & Qwen, Qwen-VL, Qwen2-VL, Qwen2.5-VL, Qwen3-VL variants \\
LLaMA/Llama family & 24 & 7.1\% & LLaMA/Llama variants excluding LLaVA-specific entries \\
Gemini family & 21 & 6.2\% & Gemini and Gemini Pro/Flash/Vision variants \\
Claude family & 17 & 5.1\% & Claude 3/3.5/3.7/4 and Sonnet/Opus variants \\
InternVL family & 12 & 3.6\% & InternVL and InternVL-derived variants \\
GPT-3.5/ChatGPT family & 12 & 3.6\% & GPT-3.5 and ChatGPT-labelled entries \\
UI-TARS family & 11 & 3.3\% & UI-TARS and UI-TARS-SFT/DPO/API variants \\
LLaVA family & 10 & 3.0\% & LLaVA and LLaVA-OneVision variants \\
GLM/ChatGLM family & 7 & 2.1\% & GLM, ChatGLM, and AutoWebGLM variants \\
OS-ATLAS family & 7 & 2.1\% & OS-ATLAS and OS-ATLAS-Pro variants \\
CogVLM/CogAgent family & 6 & 1.8\% & CogVLM, CogAgent, and CogAgent-Chat variants \\
Mistral/Codestral family & 6 & 1.8\% & Mistral and Codestral variants \\
PaLM/PaLI/PaliGemma family & 6 & 1.8\% & PaLM, PaLI, and PaliGemma variants \\
Gemma family & 6 & 1.8\% & Gemma, CodeGemma, and SLiME-Gemma variants \\
DeepSeek family & 5 & 1.5\% & DeepSeek-VL, DeepSeek-V3, and DeepSeek-R1-distilled variants \\
BLIP/InstructBLIP family & 5 & 1.5\% & BLIP-2 and InstructBLIP variants \\
OpenAI o-series & 4 & 1.2\% & o1/o3-labelled entries \\
MiniCPM family & 3 & 0.9\% & MiniCPM and MiniCPM-V variants \\
GPT-5 family & 3 & 0.9\% & GPT-5-labelled entries in corpus metadata \\
QwQ family & 2 & 0.6\% & QwQ-labelled entries \\
Ferret-UI family & 1 & 0.3\% & Ferret/Ferret-UI entry \\
ShowUI family & 1 & 0.3\% & ShowUI entry \\
MobileVLM family & 1 & 0.3\% & MobileVLM entry \\
Grok family & 1 & 0.3\% & Grok-labelled entry \\
\bottomrule
\end{tabular}%
}
\end{table*}

\begin{table*}[t]
\centering
\scriptsize
\caption{Model usage patterns.}
\label{tab:rq1-model-usage-patterns}
\resizebox{\textwidth}{!}{%
\begin{tabular}{llrrl}
\toprule
Dimension & Usage pattern & Papers & Share & Interpretation \\
\midrule
Availability & Uses at least one open model & 163 & 48.5\% & The paper includes one or more open-weight/open-source models \\
Availability & Uses only open models & 95 & 28.3\% & All specified models are coded as open \\
Availability & Uses at least one closed model & 124 & 36.9\% & The paper includes one or more proprietary or API-only models \\
Availability & Uses only closed models & 56 & 16.7\% & All specified models are coded as closed \\
Availability & Combines open and closed models & 68 & 20.2\% & The paper uses or compares both open and closed models \\
\midrule
Adaptation & Uses at least one fine-tuned model & 128 & 38.1\% & The paper includes a model adapted through GUI/task-specific training or tuning \\
Adaptation & Uses at least one off-the-shelf model & 146 & 43.5\% & The paper includes a base model used without GUI/task-specific tuning \\
Adaptation & Uses only off-the-shelf models & 92 & 27.4\% & All specified models are used without GUI/task-specific tuning \\
Adaptation & Combines fine-tuned and off-the-shelf models & 54 & 16.1\% & The paper uses or compares both adapted and non-adapted models \\
\bottomrule
\end{tabular}%
}
\end{table*}

We use the observation-medium distribution in \figref{fig:rq1-observation-media} to explain why screenshots have become the default interface representation.
Screenshot-only settings account for 148 papers, and another 129 papers combine screenshots with HTML, XML, or both.
This pattern indicates that GUI interpretation is an alignment problem among visual appearance, structural metadata, and executable action targets.
Screenshots help agents perceive what a user sees, while HTML, XML, accessibility trees, or view hierarchies can provide element text, hierarchy, clickability, and stable coordinates.
Web agents often exploit DOM-level information or browser instrumentation, whereas mobile agents can use Android view hierarchies and XML states \cite{deng2023mind2web127, lu2024weblinx138, wen2024autodroid149, xu2025androidlab152}.
The combined-representation categories in \figref{fig:rq1-observation-media} therefore carry architectural meaning.
They reveal where an agent receives extra structure, where it must infer structure from pixels, and where benchmark results may depend on platform-specific metadata that another environment cannot reliably provide.

We observe a similarly uneven model-family pattern in \tabref{tab:rq1-model-family}.
GPT-4/4o and Qwen/Qwen-VL are the two most frequently mentioned families in the extracted metadata.
The remaining papers form a long tail that mixes other general models with GUI-specific variants.
The concentration around a few major model families reflects the practical role of strong general multimodal models as reasoning and perception engines.
The long tail of GUI-specific or open model families shows a parallel effort to build models whose perception, grounding, action syntax, and reasoning traces are better aligned with interface interaction.
This split is visible in framework-oriented systems that embed a foundation model inside a larger agent loop \cite{agashe2024agent101, zhang2024ufo109, wen2024autodroid149}, as well as in fine-tuned or GUI-native systems such as CogAgent, SeeClick, OS-ATLAS, UI-TARS, ShowUI, and MobileGUI-RL \cite{hong2024cogagent003, cheng2024seeclick012, wu2024os063, qin2025ui001, lin2025showui062, shi2025mobilegui015}.

We further separate model availability from model adaptation in \tabref{tab:rq1-model-usage-patterns}.
Nearly half of the corpus uses at least one open model, while 36.9\% uses a closed model.
One fifth combines both within the same paper.
Open models support inspection and deployment control.
Closed models remain attractive when rapid prototyping or high-end multimodal capability is the immediate goal.
The adaptation rows show a second axis of variation.
Papers that use at least one fine-tuned model account for 128 cases, while 146 papers use at least one off-the-shelf model and 92 rely only on off-the-shelf models.
These numbers point to two complementary paths.
One orchestrates general-purpose models within an external runtime.
The other adapts the model itself to GUI perception, grounding, and executable action prediction.

The adaptation path appears across mobile, web, and cross-platform research.
Mobile work emphasizes visual state abstraction, action prediction, and online device control \cite{nong2024mobileflow016, qian2024visual303, papoudakis2025appvlm278, wu2024mobilevlm293, shaw2023from030, zhang2025does042, ma2024coco116, song2024visiontasker202, li2024uinav310, wang2023enabling336, ding2024enhancing334}.
Web work increasingly uses curricula, large-scale trajectories, and real-world exploration \cite{qi2024webrl290, thil2024navigating304, trabucco2025insta277, he2025openwebvoyager291, zhang2026tongui013, zheng2025vem070, zhang2025breaking091, zhang2025breaking260, liu2025wepo224, pahuja2025explorer276, furuta2023multimodal312}.
Reinforcement-learning studies treat interaction as sequential control and use experience, evaluators, or critics to improve action selection and error detection \cite{lu2025arpo096,xu2025mobilerl099, bai2024digirl302, chen2025enhancing317, wanyan2025look325, liu2025infigui077, yuan2025enhancing092, xiao2025ui094, zhou2025hiconagent100, yang2026gui103, lian2025ui105, lu2026ui273}.
Cross-platform models further incorporate action histories, speech, reward learning, and generalized task execution \cite{xu2024aguvis067, wu2025gui069, han2025uitron097, wang2025inreact110, bai2025digi282, yang2025magma281, lu2026ui082}.

Perception and assistive tools connect model adaptation to executable agent loops.
Grounding methods convert screenshots into actionable regions or candidate elements \cite{chen2025guicourse011, kwak2025mega045, ye2025gg086, tang2025lpo104, xu2025attention288, yang2025aria340, chen2025mpr039, ma2026beyond095, anand2025afragent108, tao2025understanding111, park2025r119, pawlowski2025tinyclick217, hoscilowicz2024clickagent228, wu2025smoothing269, rahman2024v305, zhang2023reinforced311}, while representation methods add screen schemas, region consistency, and layout-aware reading \cite{du2026test046, jin2025scp060, wang2025mp272, fan2024read121, zhang2025ui093, yang2025rwkv279, shen2024falcon287, li2024ferret292, ziyang2024vga296, you2024ferret306, burns2024tell341, bai2021uibert348}.
Synthetic data and large-scale grounding resources then connect these representations to training and benchmark construction \cite{liu2025ui263, li2025autogui283, xu2025deskvision038, hui2025winclick270, fan2025gui339, chai2025amex113, wang2024e299, chawla2024guide307, li2024on301, li2024on346}.
Together, this work explains the heterogeneity in \tabref{tab:rq1-model-family}.
The model category includes foundation models, perception modules, and policy-learning systems, all of which may occupy different positions in the same larger agent runtime used during execution.

Three technical trends follow from the combined evidence in \figref{fig:rq1-observation-media}, \tabref{tab:rq1-model-family}, and \tabref{tab:rq1-model-usage-patterns}.
First, GUI-agent research is moving toward native multimodal action models that learn from screenshots, element descriptions, action traces, and interactive feedback, reducing reliance on hand-designed prompts alone \cite{qin2025ui001, wu2024os063, lin2025showui062, liu2026infiguiagent048}.
Second, observation design is becoming a first-order architectural choice.
The decision to use pixels, HTML, XML, accessibility trees, or their combination determines what the agent can perceive, how it grounds actions, what failures can be diagnosed, and how portable the system is across platforms \cite{lu2024omniparser004, lu2024weblinx138, wen2024autodroid149}.
Third, benchmark design and model design are becoming increasingly intertwined.
A benchmark that exposes DOM nodes encourages different agent strategies from one that permits screenshots only.
A benchmark with live interaction and delayed consequences places greater demands on memory and recovery than a static grounding dataset \cite{zhou2023webarena128, koh2024visualwebarena129, rawles2024androidworld148, xie2024osworld159}.

\textbf{Answer to RQ1.} We find that GUI-agent research has changed from a small collection of platform-specific automation studies into a rapidly expanding, multi-community field.
The concentration of papers in 2024--2026 and the prevalence of preprints, frameworks, and benchmarks indicate uneven evidence maturity and evolving methodological conventions.
Mobile and web environments remain dominant, while desktop, cross-platform, and professional applications extend the field toward more consequential forms of computer use.
Screenshots are the primary observation channel, often supplemented by structural metadata.
Model development follows two paths. Some studies orchestrate general multimodal models within modular systems, while others adapt GUI-specific models for perception, grounding, and action.
We conclude that systems, benchmarks, representations, and models are evolving together.
The architectural analysis in RQ2 examines their coupling, and RQ3 assesses whether the resulting evaluations support comparable system-level claims.

\section{RQ2: Architectures and Engineering Techniques}
\label{sec:rq2-architecture}

We use RQ2 to translate the field-level trends in RQ1 into system responsibilities and interfaces.
Module statistics use the 145 framework papers, while broader architectural and recovery signals draw on the SE-parseable corpus where stated.
Across platforms, a stable closed loop appears \cite{deng2023mind2web127, wen2024autodroid149, zhang2024ufo109, agashe2024agent101, xie2024osworld159}.
An agent observes the GUI, constructs an internal state, selects and executes an action, and uses the resulting state to continue, verify, recover, or stop.
This loop is shared by web agents that operate through DOMs and browser automation, mobile agents that exploit screenshots and Android view hierarchies, and desktop agents that must coordinate across windows, applications, and files.

\begin{figure*}[t]
\centering
\includegraphics[width=0.9\textwidth]{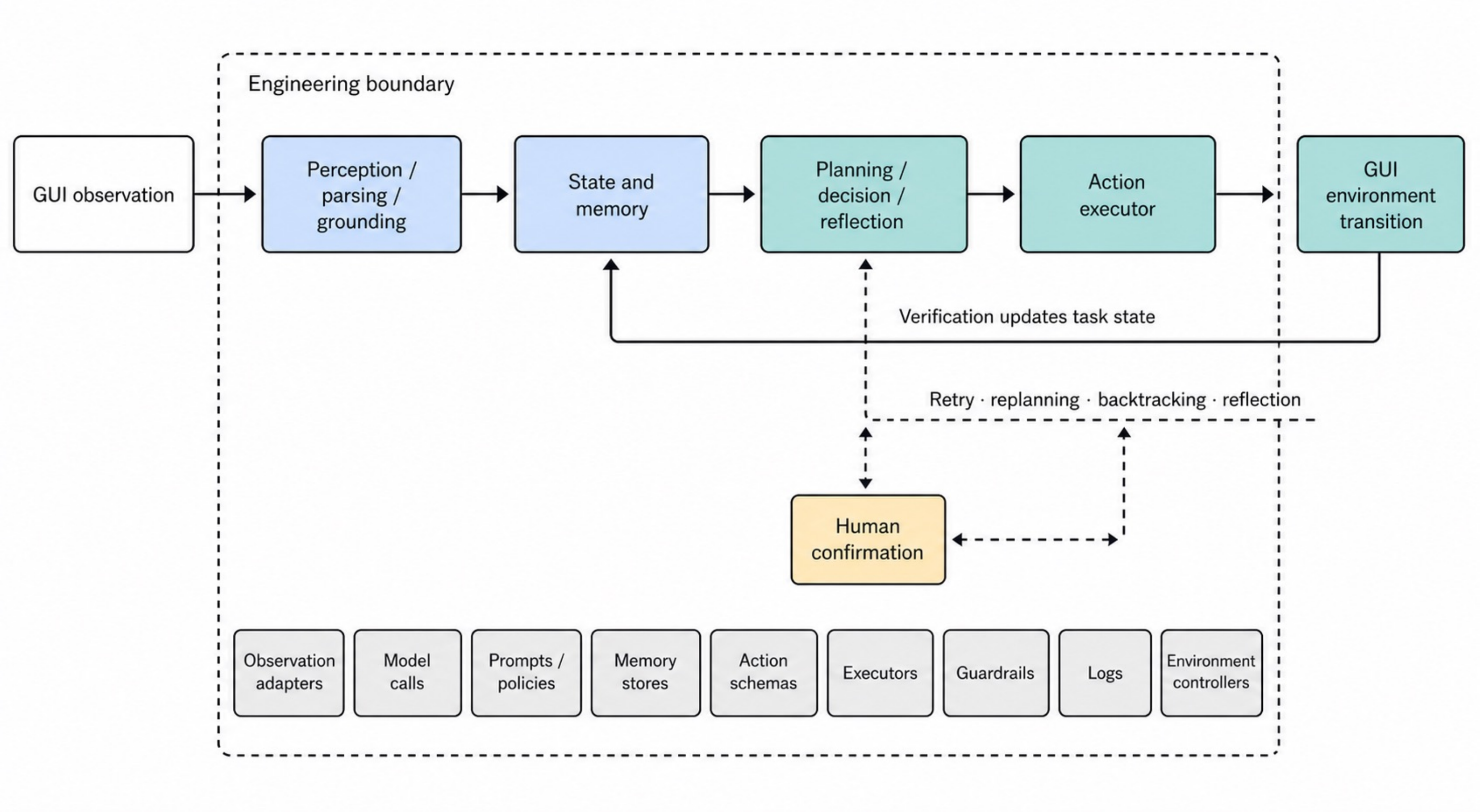}
\caption{Modular GUI-agent architecture.}
\label{fig:rq2-architecture-loop}
\end{figure*}

We use \figref{fig:rq2-architecture-loop} to separate the universal perceive--decide--execute loop from the modules that support longer tasks and safer operation.
The basic loop appears in all framework papers, while the placement of state, memory, verification, recovery, and human intervention varies.
Some systems invest in visual parsing and grounding to make screenshots actionable \cite{lu2024omniparser004, cheng2024seeclick012, li2025screenspot051}.
Long-horizon agents instead devote more architecture to planning, memory, and reflection \cite{zhang2024dynamic028, agashe2024agent101, tian2025agentprog106}.
Where runtime risk is central, verifiers and action guards are joined by safety filters or human queries \cite{dai2025advancing050, lee2025verisafe043, wu2025verios327}.
Our analysis follows the flow from representation through reasoning to execution and recovery.

\begin{wraptable}{r}{0.43\textwidth}
\centering
\scriptsize
\caption{Common architectural modules.}
\label{tab:rq2-common-modules}
\setlength{\tabcolsep}{4.5pt}
\begin{tabular}{lrr}
\toprule
Module & Papers & Share of frameworks \\
\midrule
Perception & 145 & 100.0\% \\
Decision & 145 & 100.0\% \\
Execution & 145 & 100.0\% \\
Verification & 93 & 64.1\% \\
Memory & 88 & 60.7\% \\
Planning & 79 & 54.5\% \\
Reflection & 61 & 42.1\% \\
\bottomrule
\end{tabular}
\end{wraptable}

Our module counts in \tabref{tab:rq2-common-modules} support interpreting GUI-agent frameworks as control systems with model components embedded in larger runtimes.
Perception, decision, and execution form the universal skeleton.
Perception constructs a usable state from the GUI observation.
Decision selects an operation from that state and the task history, after which execution applies it to the environment.
Verification, memory, planning, and reflection then indicate support for longer-horizon control.
Verification checks whether an action or trajectory satisfies an expected condition, memory stores task progress and reusable interaction knowledge, planning decomposes or revises goals, and reflection critiques failures or updates strategies.
We observe a maturation path from direct observation-to-action policies toward components that preserve state, reason over alternatives, check progress, and repair mistakes in long-horizon tasks.
The variation in module support also complicates architectural comparisons.
Two papers may both report a task-success rate, yet one may rely on a rich browser controller, structured memory, and a verifier, while another may ask a multimodal model to predict a coordinate from a screenshot.
We use module placement and responsibility as core evidence for architectural comparison.

\subsection{Observation-Driven State Construction}
\label{sec:rq2-perception}

The perception layer transforms raw GUI state into a representation that a planner or policy can use.
Screenshots dominate the corpus and provide a portable observation across platforms.
They do not directly expose which elements are clickable, editable, disabled, hidden, or semantically linked.
GUI agents therefore build perception pipelines that combine visual models, OCR, UI parsers, accessibility metadata, DOM/XML structures, coordinate normalization, and element-ranking rules.
The resulting representation becomes the contract between the external GUI and the agent's internal reasoning and subsequent action selection.

Pure-vision systems treat screenshots as the primary state.
CogAgent and SeeClick learn to ground instructions or elements in visual interface images, while OmniParser converts screenshots into structured visual elements that can be consumed by downstream agents \cite{hong2024cogagent003, cheng2024seeclick012, lu2024omniparser004}.
ScreenSpot and related grounding benchmarks isolate the accuracy of matching natural-language element references to screen targets \cite{li2025screenspot051}.
Screenshot-first agents such as ScreenAgent, You Only Look at Screens, ShowUI, OS-ATLAS, and UI-TARS further demonstrate how interface understanding can be framed as a multimodal perception-and-action problem \cite{niu2024screenagent207, zhang2024you122, lin2025showui062, wu2024os063, qin2025ui001}.
Screenshots remain usable when DOMs, accessibility trees, or app internals are unavailable.
The cost is that an agent must infer interactivity, hierarchy, and hidden state from pixels or from model priors alone during execution.

Structured-interface systems push part of this burden into platform metadata.
Web agents can use HTML, DOM nodes, accessibility trees, browser APIs, and page-change feedback to identify action targets and monitor transitions \cite{deng2023mind2web127, lu2024weblinx138, zhou2023webarena128}.
Mobile agents can exploit Android XML, view hierarchies, and device-control APIs, as seen in AutoDroid, AndroidWorld-oriented systems, AppAgent-style mobile automation, and MobileAgentBench-style evaluation \cite{wen2024autodroid149, rawles2024androidworld148, li2024appagent206, wang2024mobileagentbench154}.
Desktop agents such as UFO, UFO2, Agent~S, and PC-Agent must handle accessibility trees, OCR, window selection, application context, and mouse/keyboard actions at the operating-system level \cite{zhang2024ufo109, zhang2025ufo2249, agashe2024agent101, liu2025pc233}.
These systems show that structured metadata can improve grounding and action validity while introducing platform dependence.
A DOM-aware method and an Android-XML-aware method may solve different engineering problems under the same label of ``GUI agent''.

\begin{wraptable}{r}{0.64\textwidth}
\centering
\scriptsize
\caption{Specialized module themes.}
\label{tab:rq2-special-modules}
\begin{tabular}{lrr}
\toprule
Specialized theme & Papers & Share of frameworks \\
\midrule
Memory and retrieval & 51 & 35.2\% \\
Perception enhancement & 47 & 32.4\% \\
Planning and search & 44 & 30.3\% \\
Reflection and self-improvement & 33 & 22.8\% \\
Safety, privacy, and verification & 19 & 13.1\% \\
Multi-agent and role decomposition & 12 & 8.3\% \\
Other or uncategorized modules & 28 & 19.3\% \\
\bottomrule
\end{tabular}
\end{wraptable}

Mobile-control research illustrates how representation choices accumulate into system design.
Earlier systems combine screenshots, XML hierarchies, task descriptions, and learned policies in different proportions \cite{sun2022meta029, zhang2024android014, wen2023droidbot203, yan2023gpt204, christianos2024lightweight211}.
Newer frameworks add exploration, page graphs, modality fusion, and adaptive control \cite{xie2025gui018, chen2025pg053, dong2025mt081, cheng2025os072, nong2025craft049, xia2025g084}.
Retrieval, knowledge graphs, world models, and backtracking then compensate for information absent from the current screen \cite{guan2025kg112, xu2025retrieval102, sun2024os068, luo2025vimo264, wu2025backtrackagent089}.
The common architectural point is that perception defines what later modules can store, verify, generalize, and recover.

We group specialized modules beyond the basic loop in \tabref{tab:rq2-special-modules}.
The counts are multi-label semantic groupings of the specialized-module field across the 145 framework papers.
These specialized modules serve different stages of control.
Memory and retrieval preserve experience beyond the current screen.
Perception enhancement turns raw observations into more actionable states, while planning and search explore how the agent should proceed from them.
Reflection modules add critics, reward signals, bootstrapping, retry logic, or post-action critique, while safety/privacy modules add filters, guards, validators, or sensitive-operation checks.
Perception enhancement is the second most common theme, appearing in 47 framework papers.
These modules usually sit between raw observation and decision making.
Representative systems intervene at different points in this process.
Less is More simplifies the GUI context before reasoning, whereas MGA combines spatial-semantic grounding with cross-step memory.
AUTO-Explorer and Explorer use parsing or systematic element collection to support data acquisition, and PC-Agent combines accessibility information with OCR for active perception \cite{chen2025less023, cheng2025mga025, xiangwu2025auto021, chaimalas2025explorer258, liu2025pc233}.
The architectural issue extends beyond perception accuracy.
A perception module also determines the size of the action space, the cost of each step, the information exposed to the model, and the failure modes available for diagnosis.
When perception produces candidate elements and confidence signals, downstream modules can verify or recover.
When perception is hidden inside a model prediction, the system may achieve a compact interface while making errors harder to localize during debugging.

Specialized modules further shift perception from screen understanding toward control-ready state construction.
They add reusable exploration, region structure, temporal reuse, modality fusion, and grounding confidence before an action is selected \cite{sun2025gui020, chen2026gui078, singh2025trishul085, huang2025gui107, wang2025mp272, xiong2025gui088, tang2025think271}.
We interpret perception modules as state constructors.
Pixel-only observation maximizes platform portability and aligns with general multimodal models.
Structured metadata improves action grounding and state tracking when available, and hybrid representations are common because real interfaces rarely provide a complete view of the task state.
This choice shapes every later module.
A planner can reason only over the state it receives.
The same boundary constrains verification and safety because neither can inspect signals omitted by the observation layer during runtime diagnosis.

\subsection{Explicit State and Feedback for Long-Horizon Control}
\label{sec:rq2-reasoning}

The reasoning layer determines how an agent maintains the task goal across steps.
In short tasks, a model can often select the next action from the current screen and instruction.
In long-horizon GUI interaction, the current screen may omit earlier choices, hidden constraints, user preferences, or pending subtasks.
The agent must therefore preserve a compact task state, decompose objectives, track progress, decide when a plan is invalid, and incorporate feedback from executed actions.
Our coding identifies explicit planning in 79 framework papers, memory in 88, and reflection in 61, as reported in \tabref{tab:rq2-common-modules}.
These modules define how agents move from single-screen grounding to multi-step control and recovery.

Planning modules appear in several forms.
Dynamic planning systems revise action sequences as the GUI changes, while hierarchical planners separate high-level subgoals from low-level grounding.
Dynamic Planning for GUI Automation, ScaleTrack, HiconAgent, Octo-Planner, Agent~S, UFO, Agent-E, OS-Copilot, OSCAR, and PC-Agent illustrate variants of this separation \cite{zhang2024dynamic028, huang2025scaletrack076, zhou2025hiconagent100, chen2026octo298, agashe2024agent101, zhang2024ufo109, abuelsaad2024agent192, wu2024os213, wang2024oscar218, he2024pc223}.
Search-based methods expose alternative paths more directly.
Agent Alpha and Agent Q search over GUI states, with Agent Q adding MCTS-guided critique.
Mirage-1 organizes multimodal skills hierarchically, while WebPilot combines strategic exploration with multi-agent execution \cite{tang2026agent040, putta2024agent201, xie2025mirage098, zhang2025webpilot194}.
We find that planning becomes valuable when action consequences are uncertain, when backtracking is possible, or when a task requires information gathering before commitment.

Web agents make this design space especially visible because browser tasks combine explicit links and forms with long-horizon state changes.
Existing systems explore search, long-context planning, visual context, browser control, and lightweight agent baselines \cite{gur2023a185, he2024webvoyager187, lai2024autowebglm188, koh2024tree193, yang2024agentoccam196, murty2024nnetnav197, ma2023laser186, kil2024dual191, iong2024openwebagent198, tang2024steward199, zhang2025litewebagent237, srinivasan2025webnav257}.
Others learn environment dynamics, workflow traces, rollback rules, or reusable skills \cite{chae2024web184, gu2024is200, shen2024scribeagent221, xu2024agenttrek286, zheng2025skillweaver244, zhang2025webrollback246, zhou2024proposer222, zhang2025symbiotic232, wang2025inducing245}.
Web-agent planning spans navigation guidance, explicit search, learned environment models, specialized roles, and reusable skills.
Each design imposes different costs for debugging and reproducibility.

Memory modules address a different bottleneck.
GUI agents often operate under context-window limits, noisy histories, and repeated interface patterns.
Memory designs differ in what they preserve and how they reuse it.
AppAgentX and Agent~S retain experience for later tasks, whereas AgentProg manages the active context through program structure.
MOBA separates several forms of task and application state.
R2D2 and MobileGPT instead organize experience as graphs that support navigation or reuse \cite{jiang2025appagentx073, agashe2024agent101, tian2025agentprog106, zhu2025moba212, huang2025r2d2227, lee2024mobilegpt335}.
These systems suggest that memory is both a performance mechanism and an engineering boundary.
A memory store can reduce prompt length, improve reuse, and support recovery.
It also raises maintenance questions about validity after app updates, stale-trajectory detection, storage of private screen content, and the influence of retrieved memories on new tasks.

Reflection modules complete the control loop by interpreting failures and updating behavior.
Structured reflection for computer control, Mobile-Agent-E, AutoGLM, UItron, InfiGUIAgent, Guardian, and verifier-driven mobile agents all incorporate mechanisms for self-critique, outcome checking, curriculum improvement, or failure recovery \cite{li2023a215, wang2025mobile225, liu2024autoglm216, zeng2025uitron061, liu2026infiguiagent048, ran2024guardian295, dai2025advancing050}.
Reflection is architecturally distinct from verification.
Verification asks whether a condition holds.
Reflection explains what went wrong and suggests the next strategy.
The two are often coupled, and their distinction matters for testing.
A verifier can be unit-tested against expected states or action outcomes, while a reflection module is evaluated through its effect on subsequent trajectories.

Mobile and cross-platform systems extend reflection and memory from isolated failure handling to continual adaptation.
LearnAct, MobileSteward, Mobile-Agent-V, FedMobileAgent, ReachAgent, Learn-by-interact, CHOP, and PersonalAlign use demonstrations, self-evolution, decentralized user data, page reaching, interactive learning, optimized subtask planning, or long-term user records to make repeated GUI interaction more adaptive \cite{liu2025learnact044, liu2025mobilesteward235, wang2025mobile234, wang2025mobilea3gent230, wu2025reachagent229, su2025learn226, zhou2025chop238, lyu2026personalalign037}.
Process-reward and inference-time guidance methods provide a complementary route by shaping the agent's decisions during execution, as in GUI-PRA, process-reward VLM guidance, variational subgoal-conditioned RL, and language multi-agent learning with credit re-assignment \cite{xiong2025gui088, hu2025guiding248, wu2025advancing280, he2025advancing239}.
We observe adaptation at several time scales.
Grounding confidence acts within a step, reflection or backtracking acts within a trajectory, memory acts across tasks, and personalized or federated data acts across users.

\begin{wraptable}{r}{0.6\textwidth}
\centering
\scriptsize
\caption{Architectural patterns.}
\label{tab:rq2-architecture-patterns}
\begin{tabular}{lrr}
\toprule
Architectural pattern & Papers & Share \\
\midrule
Modular pipeline or multi-stage architecture & 154 & 47.1\% \\
External tool/browser/device/API execution layer & 139 & 42.5\% \\
Memory, reflection, or historical-context loop & 116 & 35.5\% \\
Planner-grounder or planner-executor separation & 63 & 19.3\% \\
End-to-end VLM or native-agent architecture & 18 & 5.5\% \\
\bottomrule
\end{tabular}
\end{wraptable}

We report multi-label architectural patterns from 327 SE-parseable papers in \tabref{tab:rq2-architecture-patterns} and compare them with the framework-module statistics in \tabref{tab:rq2-common-modules}.
The dominant architectural tendency is modular.
Modular or multi-stage pipelines expose responsibilities for inspection and replacement.
External execution layers move part of the system boundary into wrappers, browsers, device controllers, or operating-system tools.
State-history loops add continuity across screens and across runs.
Planner-grounder or planner-executor separation splits strategic task decomposition from concrete target selection and action application.
End-to-end native-agent designs compress several responsibilities into trained multimodal action models, making the interface compact while reducing component-level observability.
Native models remain technically central because systems such as UI-TARS, OS-ATLAS, ShowUI, and InfiGUIAgent improve the learned perception-action core \cite{qin2025ui001, wu2024os063, lin2025showui062, liu2026infiguiagent048}.
The corpus nevertheless indicates that most practical GUI-agent systems still externalize responsibilities into modules.
They do so because GUI control requires translating model output into a safe executable command, observing effects, updating state, and handling environment-specific constraints in addition to predicting an action token.

Multi-agent decomposition is a smaller but conceptually important version of modularity.
COLA, AssistEditor, WebPilot, MobileExperts, Agent~S2, PC-Agent, and UFO2 distribute tasks across managers, workers, planners, editors, grounders, verifiers, or specialized decision agents \cite{zhao2025cola114, gao2024assisteditor056, zhang2025webpilot194, zhang2024mobileexperts210, agashe2025agent247, liu2025pc233, zhang2025ufo2249}.
This design can align responsibilities with different knowledge sources or action spaces.
A manager can maintain global progress, a grounder can resolve coordinates, a verifier can assess completion, and a worker can execute local steps.
The same design increases coordination cost and introduces failure points involving stale shared state, inconsistent subtask boundaries, conflicting judgments, and accumulated latency.
We interpret multi-agent GUI frameworks as evidence of a shift from prompt engineering toward explicit software architecture.

Desktop and general-computer-use systems make this shift visible because they must coordinate multiple applications, files, windows, and specialized tools.
CCAgent, AgentStore, MMAC-Copilot, Programming with Pixels, COLA, and AgentStudio distribute responsibilities across collaborative agents, agent stores, copilot modules, software-engineering tasks, Windows UI automation, or virtual-agent toolkits \cite{chen2025ccagent052, jia2025agentstore219, song2024mmac220, aggarwal2025programming236, zhao2025cola274, zheng2024agentstudio166}.
Ponder \& Press, UIPro, SpiritSight Agent, Mobile-Agent-v3, Continual GUI Agents, and OmegaUse push the same architectural pressure toward cross-platform or general-purpose execution, where a single agent loop must handle diverse layouts, tasks, and operating contexts \cite{wang2025ponder064, li2025uipro034, huang2025spiritsight035, ye2025mobile010, liu2026continual031, zhang2026omegause083}.
We find that modularity remains common because the surrounding software environment still requires execution adapters, policy boundaries, logs, and recovery channels even when strong native action models are available.

\subsection{Gaps in Runtime Control and Action Governance}
\label{sec:rq2-execution}

The execution layer converts an agent's decision into an operation that changes the GUI environment.
It is the point where model uncertainty becomes external state change.
Across platforms, action execution ranges from simple coordinate clicks and text entry to browser commands, Android intents, accessibility-tree actions, Python scripts, API calls, long screenshots, window switching, and application-specific tools.
AutoDroid and AutoDroid-V2 illustrate how mobile execution can combine Android state abstraction with generated code \cite{wen2024autodroid149, wen2025autodroid087}.
On desktops, UFO and UFO2 divide control between host and application agents and add safeguards for sensitive actions \cite{zhang2024ufo109, zhang2025ufo2249}.
OS-Copilot, CRADLE, and OSCAR broaden the executor again by combining operating-system actions with tool generation and state-aware reasoning \cite{wu2024os213, tan2024cradle214, wang2024oscar218}.
These examples make the executor a major architectural component with its own state, constraints, and failure modes.

Action schemas are one source of variation.
Some systems use low-level primitives such as click, type, scroll, and wait.
Others introduce higher-level actions such as opening an app, calling an API, taking a long screenshot, selecting a window, querying an element, executing code, or handing control to a user.
MagicGUI includes an extended mobile action space with API calls, screenshots, takeover, waiting, and text entry \cite{tang2025magicgui071}.
PromptRPA and Prompt2Task translate user prompts into reusable mobile procedures and include assessment or intervention mechanisms \cite{huang2024promptrpa255,huang2025prompt2task231}.
Enterprise-oriented workflows use demonstrate-execute-validate patterns and standard operating procedures to constrain execution \cite{wornow2024automating240,ding2024mobileagent343}.
Higher-level actions can reduce trajectory length and improve reliability when correctly specified, but they move more semantics into the executor and require clearer preconditions, permissions, and failure handling.
The runtime-control signals in \tabref{tab:rq2-runtime-control} are not mutually exclusive.
Its exception-handling rows use the 327 SE-parseable papers, while the explicit human-in-the-loop row uses the 145-paper framework subset.

\begin{wraptable}{r}{0.56\textwidth}
\centering
\scriptsize
\caption{Runtime reliability and intervention signals.}
\label{tab:rq2-runtime-control}
\begin{tabular}{lrr}
\toprule
Runtime-control signal & Papers & Share \\
\midrule
Invalid/redundant/malformed action handling & 95 & 29.1\% \\
Reflection, correction, retry, or replanning & 86 & 26.3\% \\
Human request, confirmation, or takeover & 17 & 5.2\% \\
Explicit human-in-the-loop flag & 3 & 2.1\% \\
\bottomrule
\end{tabular}
\end{wraptable}

Our analysis of \tabref{tab:rq2-runtime-control} reveals a gap between autonomous execution and accountable runtime control.
Invalid-action handling appears in 95 SE-parseable papers, covering mechanisms that reject, filter, correct, or avoid actions that cannot be executed safely or usefully.
Reflection, correction, retry, or replanning appears in 86 papers, showing that many systems recognize execution failure as a normal runtime condition.
The lower counts for human request, confirmation, or takeover show that escalation remains less systematically engineered, and the explicit human-in-the-loop signal appears in only a very small subset of framework papers.
Verifier-driven agents such as V-Droid, VeriSafe, STEVE, and Guardian move the field toward explicit post-action checks and runtime validation \cite{dai2025advancing050, lee2025verisafe043, lu2025steve243, ran2024guardian295}.
Guardrail and privacy systems such as GUIGuard and GuardAgent further show that execution should be conditioned on policy, privacy, and safety reasoning alongside task progress \cite{wang2026guiguard033, xiang2025guardagent329}.

Failure recovery has three recurring forms.
Local retry or repair repeats an action, changes the target, waits for a page to load, or uses a backup grounding strategy.
Plan-level revision updates a subgoal, backtracks, or searches an alternative path, as in tree-search and MCTS-inspired agents \cite{tang2026agent040, putta2024agent201, zhang2025webpilot194}.
Knowledge update stores the failed trajectory, modifies memory, or evolves a shortcut or skill for future tasks \cite{jiang2025appagentx073, wang2025mobile225, huang2025r2d2227}.
Each recovery form requires different evidence.
Action-level outcomes can reveal local repair, while plan revision requires trajectory comparison.
Knowledge updates can be assessed only across repeated tasks or over time.
A single aggregate task-success metric can hide these differences.

Human intervention is the least developed but most consequential execution mechanism.
VeriOS explicitly studies proactive human-agent-GUI interaction for trustworthy OS agents, while MobileAgent and MobileGPT include human-machine interaction, SOP integration, or human repair mechanisms \cite{wu2025verios327, ding2024mobileagent343, lee2024mobilegpt335}.
Other systems use uncertainty-aware refinement, follow-up questions for ambiguous instructions, confirmation for sensitive actions, interactive modes, or user fallback without consistently treating the human as a first-class runtime component \cite{hao2025uncertainty090, cheng2025navi268, zhang2024ufo109, zhang2025ufo2249, abuelsaad2024agent192}.
Many GUI tasks contain decisions that are inappropriate to automate silently, including login, payment, deletion, permission granting, message sending, medical or financial operations, and enterprise workflow approval.
A robust architecture should define when to ask, what evidence to present, how user feedback changes the plan, and how the intervention is logged.

Voice, RPA, and toolkit-oriented systems provide additional execution interfaces through process automation, conversational control, reusable procedures, and constrained web policies \cite{xie2023openagents189, ye2023proagent250, guan2024intelligent251, vu2024gptvoicetasker253, pan2023autotask254, sodhi2023step337}.
Although some are not full autonomous agents, they show how users, procedures, and tools can restrict the executable action space.

\textbf{Answer to RQ2.} We find that GUI-agent architectures share a closed perceive--decide--execute loop whose responsibilities are increasingly distributed across modules.
All 145 framework papers expose perception, decision, and execution, while verification, memory, planning, and reflection provide additional support for long-horizon control.
Observation design determines the state available to the agent.
Reasoning modules preserve goals and revise strategies, while executors translate uncertain model outputs into external state changes.
The dominant architecture embeds foundation models within a larger runtime.
Adapters and memories construct state, planners and grounders select operations, and controllers apply them under partial verification or guardrails.
Recovery, human escalation, safety enforcement, and auditability remain less explicit than perception and planning.
This imbalance allows task success to conceal brittle or unsafe runtime behavior and establishes the evaluation problem examined in RQ3.
\section{RQ3: Evaluation and Benchmarking}
\label{sec:rq3-evaluation}

We use RQ3 to examine whether current evaluations provide credible evidence for the systems characterized in RQ2.
The analysis uses the 252 papers coded as frameworks, models, or evaluation studies.
GUI-agent evaluation connects offline capability tests, interactive trajectory execution, and oracles that judge outcomes and behavior.
The corpus contains extensive benchmark activity, while its evidence remains centered on task success.
We analyze benchmarks, modalities, metrics, and validity threats as parts of one evaluation pipeline.

\begin{figure*}[t]
\centering
\includegraphics[width=0.9\textwidth]{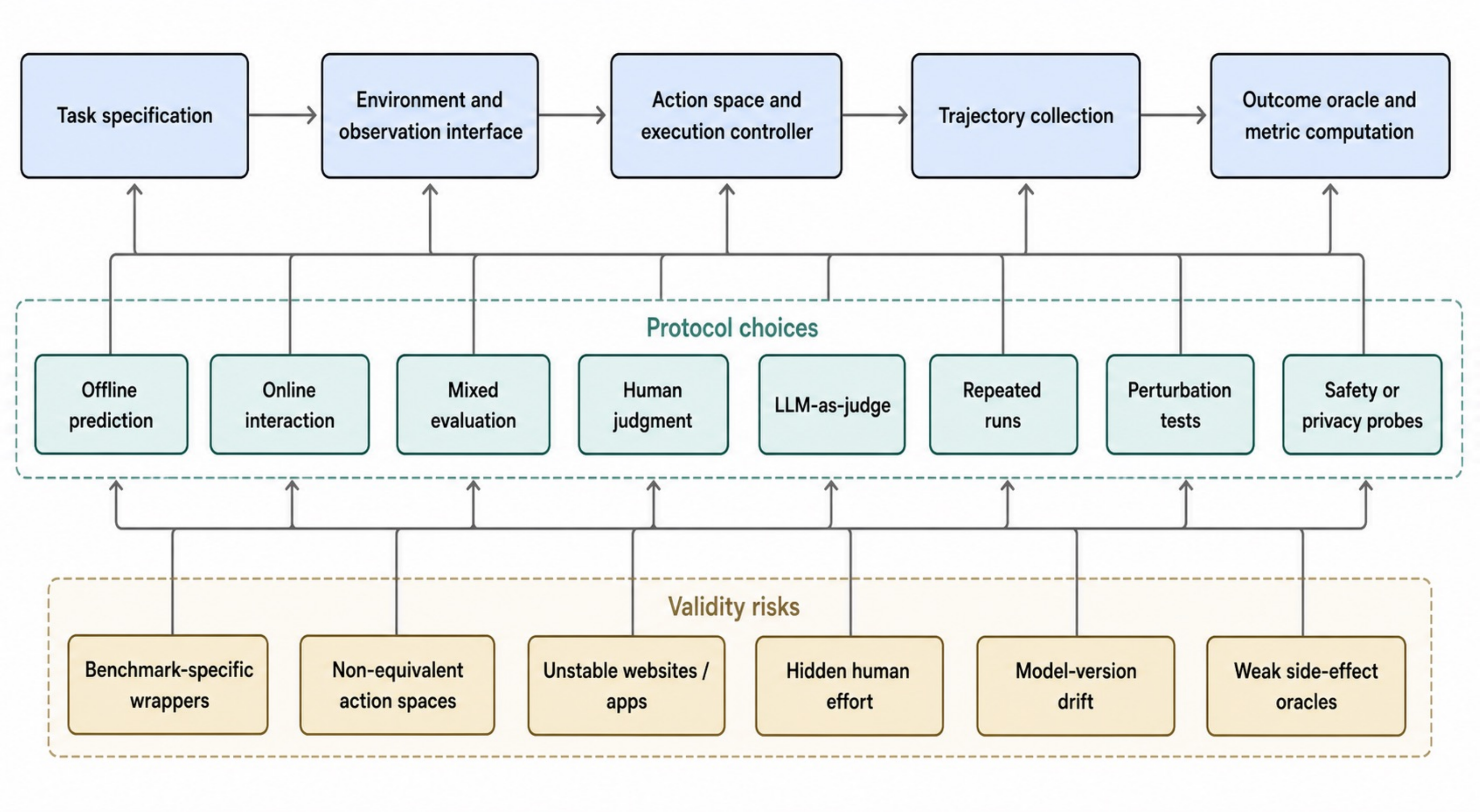}
\caption{GUI-agent evaluation pipeline.}
\label{fig:rq3-evaluation-pipeline}
\end{figure*}

We use \figref{fig:rq3-evaluation-pipeline} to show that a GUI-agent benchmark evaluates the agent and the surrounding interaction protocol together.
A reported score depends on task wording, allowed observations, action schema, environment reset policy, time or step budget, authentication state, external tools, and the oracle used to judge completion.
WebArena, VisualWebArena, AndroidWorld, OSWorld, and Windows Agent Arena made this dependency visible by moving evaluation into realistic web, mobile, and desktop environments \cite{zhou2023webarena128, koh2024visualwebarena129, rawles2024androidworld148, xie2024osworld159, bonatti2024windows160}.
Earlier and complementary datasets such as WebShop, Mind2Web, WebLINX, OmniACT, and ScreenSpot isolate grounded language-to-action prediction, website navigation, generalist action modeling, or screen grounding \cite{yao2022webshop126, deng2023mind2web127, lu2024weblinx138, kapoor2024omniact161, li2025screenspot051}.
The distinction matters because an offline action-prediction score and an online task-success rate answer different questions.

\subsection{Protocol Heterogeneity in Interactive Benchmarks}
\label{sec:rq3-benchmarks}

We use the 252-paper framework, model, and evaluation subset for \tabref{tab:rq3-evaluation-modality}.
Benchmark relation follows the coded paper type, while modality distinguishes offline data, live or interactive environments, and combinations of the two.

Our analysis shows that benchmark construction is a major contribution type without dominating all evaluation activity.
Among the 252 system-oriented papers, 77 contribute benchmarks, datasets, environments, or protocols, while 175 mainly evaluate systems or models through existing benchmarks, self-designed tasks, or experimental environments.
This balance is healthy for a young field because new benchmarks expose new capabilities and risks.
It also creates a comparability challenge.
When a framework introduces its own tasks, action abstractions, environment adapters, or success oracles, its reported improvement can reflect both agent design and evaluation design.
We interpret benchmark papers as technical artifacts that encode assumptions about what a GUI agent should observe, how it should act, and what counts as success.

The modality counts in \tabref{tab:rq3-evaluation-modality} reveal a strong preference for mixed evaluation.
Mixed protocols combine interactive execution with offline prediction, ablation, benchmark statistics, or trajectory analysis.
Offline protocols are more controlled and typically measure perception, grounding, action prediction, classification, or static outputs.
Online protocols place agents in live or simulated GUI environments and judge task completion or trajectory outcomes.
This pattern is visible in modern benchmarks that combine interactive execution with offline analysis.
The mixed protocols take several forms.
WebShop links interaction to grounded rewards, whereas Mind2Web uses offline prediction to study cross-site generalization \cite{yao2022webshop126, deng2023mind2web127}.
WebLINX and conversational web-agent studies add multi-turn data that supports turn-level prediction and broader trajectory analysis \cite{lu2024weblinx138, deng2024on130}.
Interactive benchmarks then expose the agent to environment feedback.
WebArena and VisualWebArena use realistic website state changes \cite{zhou2023webarena128, koh2024visualwebarena129}.
AndroidWorld and related infrastructure bring the same principle to device control, while OSWorld and Windows Agent Arena extend it to desktops \cite{rawles2024androidworld148, xu2025androidlab152, wang2024mobileagentbench154, xie2024osworld159, bonatti2024windows160}.

\begin{table*}[t]
\centering
\scriptsize
\caption{Evaluation relations and modalities.}
\label{tab:rq3-evaluation-modality}
\begin{tabular}{llrr}
\toprule
Dimension & Category & Papers & Share \\
\midrule
Benchmark relation & Existing/custom evaluation & 175 & 69.4\% \\
Benchmark relation & Self-built/contributed benchmark & 77 & 30.6\% \\
\midrule
Evaluation modality & Mixed online/offline & 164 & 65.1\% \\
Evaluation modality & Offline & 52 & 20.6\% \\
Evaluation modality & Online & 36 & 14.3\% \\
\bottomrule
\end{tabular}
\end{table*}

\begin{figure*}[t]
\centering
\includegraphics[width=0.9\textwidth]{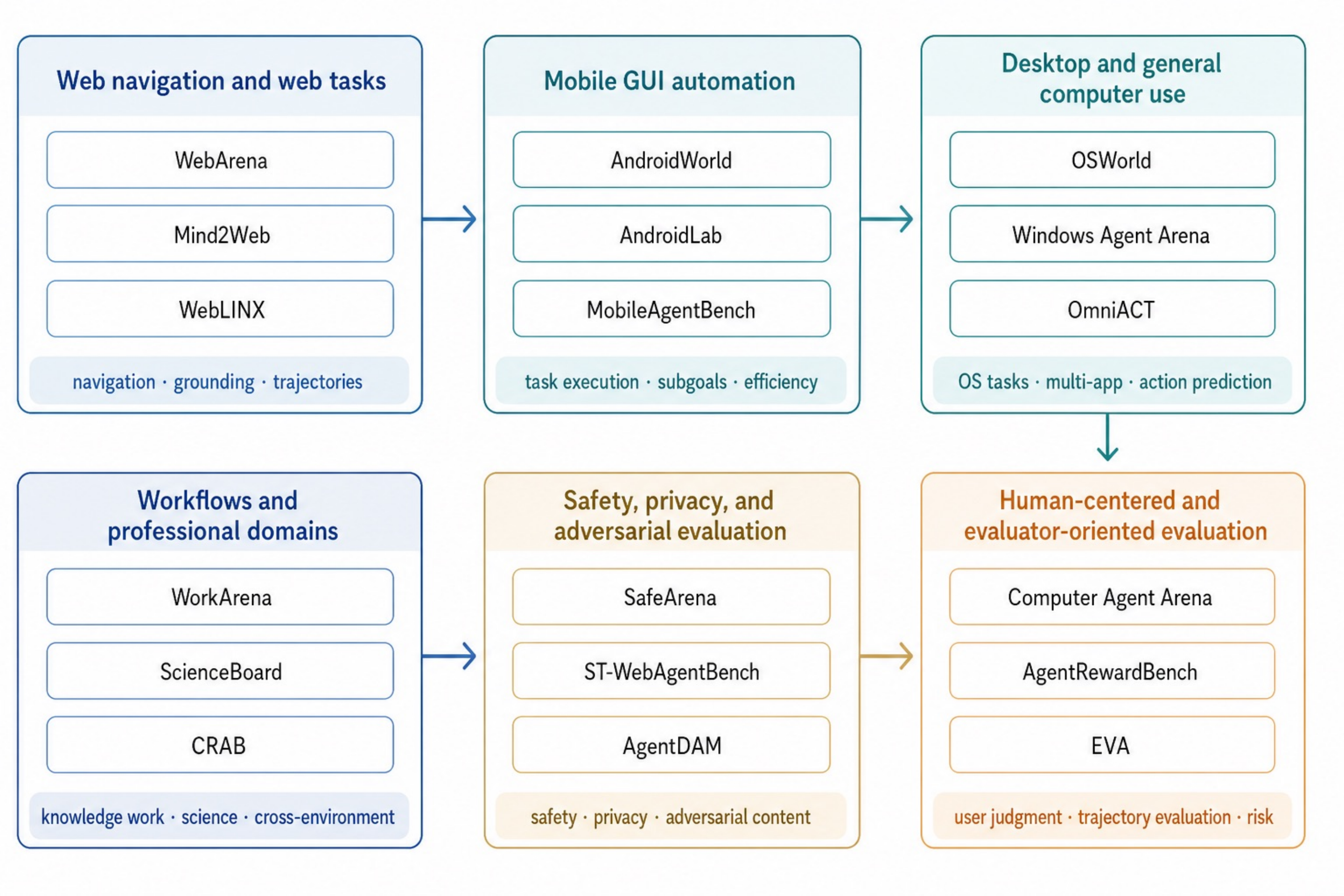}
\caption{GUI-agent benchmark and evaluation families.}
\label{fig:rq3-benchmark-families}
\end{figure*}

We use \figref{fig:rq3-benchmark-families} to trace the expansion of benchmark scope from web navigation and mobile action prediction to general computer use, professional workflows, safety, privacy, and human-centered evaluation.
Web resources such as WebShop, Mind2Web, WebArena, VisualWebArena, WebLINX, WebVLN, TurkingBench, NaviQAte, VisualWebBench, WebGames, RealWebAssist, and BrowserGym emphasize shopping, navigation, browser interaction, multimodal grounding, long-horizon assistance, and trajectory analysis \cite{yao2022webshop126, deng2023mind2web127, zhou2023webarena128, koh2024visualwebarena129, lu2024weblinx138, chen2024webvln137, xu2025turkingbench141, shahbandeh2024naviqate142, liu2024visualwebbench143, thomas2025webgames173, ye2026realwebassist262, chezelles2024the168}.
Mobile resources such as AndroidWorld, AndroidLab, A3, MobileAgentBench, FedMABench, GUI-Robust, MobileSafetyBench, and See--Think--Act focus on Android task execution, procedural state, decentralized data, efficiency, robustness, and safety \cite{rawles2024androidworld148, xu2025androidlab152, chai2025a3074, wang2024mobileagentbench154, wang2025fedmabench115, wang2025fedmabench267, yang2025gui024, lee2026mobilesafetybench156, wu2025see328}.
Desktop, game, and cross-environment benchmarks such as OSWorld, Windows Agent Arena, OmniACT, FlashAdventure, and CRAB test open-ended OS tasks, multi-application operation, desktop/web action prediction, long-horizon game interaction, and cross-environment task completion \cite{xie2024osworld159, bonatti2024windows160, kapoor2024omniact161, ahn2025flashadventure120, xu2025crab167}.
The workflow and risk-aware families extend the same logic to knowledge work, scientific workflows, safety, privacy, adversarial content, cultural awareness, and human judgment through WorkArena, WorkArena++, ScienceBoard, SafeArena, ST-WebAgentBench, AgentDAM, EIA, EVA, Computer Agent Arena, and AgentRewardBench \cite{drouin2024workarena134, boisvert2024workarena314, sun2025scienceboard183, tur2025safearena175, levy2024st139, zharmagambetov2025agentdam179, liao2024eia322, lu2025eva323, qiu2025evaluating324, wang2026computer180, lu2025agentrewardbench261}.
These resources move evaluation beyond a single success/failure endpoint.

The density of each benchmark family reflects different sources of difficulty.
Web benchmarks vary task horizon, visual demand, traversal structure, simulation control, sequential composition, and real-world assistance \cite{pan2024webcanvas132, tian2025mmina131, jang2024videowebarena135, wu2025webwalker170, song2025bearcubs182, garg2025real181, furuta2023exposing140, xue2025an178}.
Grounding resources isolate pixel-level understanding and the connection between visual evidence, user intent, and action targets \cite{yang2025pixelweb259, fereidouni2024grounded308}.
Together, they show that a single web score cannot distinguish semantic misunderstanding, grounding failure, lost context, or site-specific overfitting to individual sites.

Mobile benchmarks add variation in device configuration, app diversity, language, ambiguity, cross-app navigation, agent weaknesses, and reasoning--execution consistency \cite{deng2024mobile155, chen2024spa157, zhang2024llamatouch153, lee2024benchmarking147, gao2024mobileviews294, lu2024guiodyssey297, xing2024understanding151, dong2025say319}.
More diagnostic resources evaluate command feasibility, partial progress, failure categories, and evaluator reliability alongside final screen state \cite{burns2022a313, ran2025beyond171, sun2025autoeval174}.
Mobile evaluation covers action-target correctness, task feasibility, and execution faithfulness in the current device state.

Desktop, video, and general visual-agent benchmarks broaden the boundary to temporal procedures, multi-application work, and software-engineering-like tasks \cite{lin2024videogui162, wu2024gui164, chen2024gui300, nayak2025ui177, liu2024visualagentbench158, aggarwal2025programming236}.
We find across these families that a benchmark evaluates an agent together with a particular observation channel, action system, and environment controller.

Protocol design remains a central difficulty.
Offline evaluation is repeatable and helps isolate perception, grounding, action type, click target, or action-argument correctness.
It is especially useful for model development and ablation because the same examples can be reused across models.
Online evaluation is closer to deployment because it includes latency, environment state, side effects, recovery, and termination.
Mixed evaluation appears most often because researchers need both forms of evidence.
A complete benchmark should therefore specify observation permissions, action space, reset behavior, task timeout, repetition policy, judge type, and whether external tools or human interventions are allowed.
Without these details, a comparison between two agents may reflect hidden differences in scaffolding.

\subsection{Dominance of Task Success in Evaluation}
\label{sec:rq3-metrics}

We use the same 252-paper subset for the paper-level, multi-label counts in \tabref{tab:rq3-metric-families}.
The analysis covers high-frequency metrics and less frequent categories that capture system quality.
We find task success in 194 of the 252 system-oriented papers, making it the dominant metric.
Task-outcome metrics ask whether the trajectory reaches the intended final state.
Specialized settings may instead measure partial progress or the functional correctness of a produced artifact.
This dominance is understandable because GUI agents are ultimately judged by whether they complete user tasks, and task success can often be compared across agents with a single number.
However, the table also shows why a success-rate-centered view is incomplete.
A successful trajectory may still rely on excessive retries or hidden human repair.
It may also expose private information, violate a safety constraint, or depend on a benchmark-specific wrapper.
We separate outcome metrics from process, cost, reliability, safety, and human-centered metrics used in practice.

Grounding and action metrics remain essential because many GUI-agent failures arise before task-level judgment is possible.
Grounding metrics localize failure before the final outcome.
They test whether the agent selected the right target, action type, and arguments for an executable operation.
These metrics are especially important for model papers and offline datasets, including ScreenSpot, SeeClick, CogAgent-style grounding, OmniACT, DreamStruct, and mobile-interface understanding tasks \cite{li2025screenspot051, cheng2024seeclick012, hong2024cogagent003, kapoor2024omniact161, peng2024dreamstruct345, sunkara2022towards347}.
They also help diagnose online failures.
If a web task fails, the error may originate in high-level reasoning or element grounding.
The action type, its text argument, or delayed page state can produce the same final outcome.
Step-level metrics make these causes more visible than final task success alone.

\begin{wraptable}{r}{0.49\textwidth}
\centering
\scriptsize
\caption{Metric families.}
\label{tab:rq3-metric-families}
\resizebox{\linewidth}{!}{%
\begin{tabular}{llr}
\toprule
Family & Metric & Papers \\
\midrule
Task outcome & Task success rate & 194 \\
Task outcome & Task completion/progress & 44 \\
Task outcome & Exact match & 28 \\
Grounding/action & Benchmark accuracy & 73 \\
Grounding/action & Action prediction correctness & 51 \\
Grounding/action & UI grounding accuracy & 48 \\
Grounding/action & Action type accuracy & 44 \\
Efficiency/cost & Compute/token cost & 74 \\
Efficiency/cost & Interaction efficiency & 58 \\
Efficiency/cost & Runtime efficiency & 44 \\
Reliability & Error/failure rate & 22 \\
Safety/privacy & Attack success rate & 17 \\
Safety/privacy & Constraint/policy compliance & 14 \\
Human/judge & Human agreement & 14 \\
Human/judge & LLM-as-judge score & 14 \\
Human/judge & Usability score & 11 \\
\bottomrule
\end{tabular}%
}
\end{wraptable}

Efficiency metrics are becoming more visible because GUI agents often operate through expensive multimodal model calls and repeated environment interactions.
Compute or token cost appears in 74 papers, interaction efficiency in 58, and runtime efficiency in 44.
Systems such as MobileAgentBench, WABER, Windows Agent Arena, and on-device or distributed-control work make latency, interaction length, and resource use explicit \cite{wang2024mobileagentbench154, kara2025waber321, bonatti2024windows160, chen2024octopus342, wang2024distrl344}.
These metrics should be interpreted with platform context.
A five-step desktop task and a five-step mobile task may differ in action latency, rendering delay, state recovery, and cost per observation.
Efficiency also interacts with reliability.
Reducing the number of steps can improve user experience, while aggressive shortcuts can increase the risk of irreversible or poorly verified actions.

Safety, privacy, and human-centered metrics form a smaller but growing layer.
Safety and human-centered evaluations use a more varied set of measures.
Studies report attack or safeguard success, policy compliance, privacy leakage, human agreement, usability, and user effort, but each appears far less often than task success.
Their lower frequency should not be read as lower importance.
These metrics become central when an agent can observe sensitive content or change external state.
Submitting a form, sending a message, and altering a file carry risks that task success alone cannot represent.
SafeArena, ST-WebAgentBench, AgentDAM, EIA, EVA, and pop-up attack evaluations show how adversarial or policy-oriented benchmarks can reveal harms that ordinary task-completion benchmarks overlook \cite{tur2025safearena175, levy2024st139, zharmagambetov2025agentdam179, liao2024eia322, lu2025eva323, zhang2025attacking333}.
Human-centered evaluations and Computer Agent Arena add another missing dimension by asking which agent behavior users prefer and how they judge risk, usability, and step-wise behavior \cite{chen2025toward036,wang2026computer180}.

Risk-aware evaluation is now large enough to be treated as its own benchmark family.
Risk-aware benchmarks manipulate the GUI in several ways.
Some inject fine print or hidden instructions, while others introduce backdoors, unprivileged interference, or broader environmental distraction \cite{chen2025the047, cheng2025hidden118, liu2025hijacking058, wu2024wipi326, xu2024advweb330, yang2025mla331, yang2024systematic338, ma2025caution117, ma2025caution136}.
We use these studies to assess whether an agent follows malicious cues or leaks information under manipulation.
Other evaluation papers probe whether API-based web agents, grounded GPT-4V web agents, and preliminary Claude computer-use agents actually satisfy their claimed interaction capabilities under realistic protocols \cite{song2024beyond195, zheng2024gpt190,hu2024the027}.
Together with safety and privacy metrics in \tabref{tab:rq3-metric-families}, these studies shift evaluation from capability ranking toward behavioral auditing of complete agent trajectories.

The metric distribution also exposes a measurement gap around recovery and long-term dependability.
Error/failure rate is present in 22 papers, termination correctness in 4, and recovery success in only 2 according to the coded long-tail metrics.
This is small relative to the architectural emphasis on verification, reflection, memory, and replanning observed in RQ2.
We find that many systems implement recovery mechanisms while evaluating them indirectly through aggregate success.
A stronger evaluation design would measure how often agents enter loops, how quickly they detect failed actions, whether retry policies are bounded, whether recovery changes the failure mode, and whether human escalation occurs at appropriate times.
AgentRewardBench provides an important step by evaluating automatic assessments of web-agent trajectories, including side effects and repetition cycles \cite{lu2025agentrewardbench261}.

\subsection{System-Boundary Constraints on Cross-Benchmark Comparability}
\label{sec:rq3-validity}

Evaluation validity is the main pressure point for cumulative progress.
GUI-agent scores depend on the agent, the model version, the prompt, the observation adapter, the action wrapper, and the environment state.
A benchmark that exposes HTML or accessibility nodes may favor agents designed around structured metadata.
A benchmark that permits screenshots only may favor visually trained models.
A benchmark with deterministic resets and static tasks may understate failures that appear on live websites, dynamic mobile apps, or desktop workflows.
These dependencies define the conditions under which each score should be interpreted in later comparisons.

\begin{table*}[t]
\centering
\scriptsize
\caption{Evaluation validity and reproducibility concerns.}
\label{tab:rq3-validity-risks}
\begin{tabular}{p{0.27\textwidth}p{0.58\textwidth}}
\toprule
Concern & Primary affected protocol element \\
\midrule
Observation mismatch & Screenshots, DOM/XML, accessibility trees, OCR, tool descriptions \\
Action-space mismatch & Coordinates, element IDs, browser actions, device APIs, code/tool execution \\
Oracle ambiguity & Exact match, state match, reward, LLM-as-judge, human labels \\
Environment instability & Websites, apps, authentication, pop-ups, localization, model APIs \\
Cost and hidden assistance & Retries, prompt length, tool calls, human intervention, cached knowledge \\
Safety and privacy coverage & Malicious content, sensitive data, irreversible actions, policy constraints \\
\bottomrule
\end{tabular}
\end{table*}

We summarize the recurring threats to GUI-agent evaluation in \tabref{tab:rq3-validity-risks}.
Observation mismatch and action-space mismatch are especially important because they connect directly to RQ2.
Some agents receive screenshots only, while others receive DOM, XML, accessibility trees, OCR, element lists, or tool-generated descriptions.
Benchmarks also differ in whether they permit coordinate clicks, element IDs, browser actions, device APIs, code execution, high-level tools, or takeover options.
Two agents may both operate on the same task description while receiving different state representations and executing through different abstractions.
In such cases, the benchmark measures a system stack, not an isolated model.
This issue appears in comparisons between web agents using DOM-level information, mobile agents using XML hierarchies, and screenshot-only agents using visual grounding \cite{deng2023mind2web127, wen2024autodroid149, zhang2024you122, lu2024omniparser004}.
We identify observation channels, action schemas, tool permissions, and environment instrumentation as first-order protocol details for reproducible comparison.

Oracle ambiguity is another source of weak comparability.
Exact match works for tasks with a short textual answer or stable state label.
Functional correctness is more appropriate for generated files, data-analysis workflows, or scientific tasks.
Human judgment may be necessary for usability, risk, preference, and ambiguous outcomes.
LLM-as-judge can scale trajectory evaluation, but its agreement with human labels and sensitivity to prompt design must be established \cite{lu2025agentrewardbench261, wang2026computer180}.
WorkArena and ScienceBoard illustrate why richer oracles are needed for knowledge work and scientific workflows \cite{drouin2024workarena134, sun2025scienceboard183}.
A benchmark that judges only final page state may miss side effects, low-quality intermediate decisions, or dangerous actions that were reversed before completion.

Reproducibility is difficult because GUI environments are live software systems.
Websites update layouts, mobile apps change UI flows, operating-system dialogs appear unpredictably, and third-party services alter authentication or rate limits.
BrowserGym and WABER are useful because they focus attention on ecosystem support, reliability, and efficiency when evaluating web agents with existing benchmarks \cite{chezelles2024the168, kara2025waber321}.
AndroidLab and Windows Agent Arena also reflect a push toward systematic benchmarking infrastructure for mobile and OS agents \cite{xu2025androidlab152, bonatti2024windows160}.
Still, reproducible comparison requires more than publishing tasks.
It requires environment versioning, seed control, model-version reporting, action logs, trajectory replays, judge prompts, timeout rules, and failure categories for later diagnosis.

We treat cross-benchmark comparability cautiously.
A web-navigation score, an Android task-success score, a desktop OSWorld score, and an offline grounding score each measure different bundles of capability and infrastructure.
Leaderboards are useful for local progress within a benchmark family, while broader claims require triangulation across benchmarks with different observation channels, task horizons, domains, and risk profiles.
For example, strong results on WebArena and VisualWebArena support claims about web interaction under those protocols.
They do not establish mobile, desktop, cross-environment, or scientific-workflow generality without evidence from the corresponding benchmark families \cite{zhou2023webarena128, koh2024visualwebarena129, rawles2024androidworld148, xie2024osworld159, xu2025crab167, sun2025scienceboard183}.

\textbf{Answer to RQ3.} We find that GUI-agent evaluation is increasingly interactive and benchmark-rich while remaining dominated by task success.
Among 252 system-oriented papers, 77 contribute evaluation resources, 164 combine online and offline evidence, and 194 report task success.
The field has moved beyond static prediction without achieving uniform comparability.
Scores remain conditioned by what the agent can observe and do.
Wrappers, environment state, judge design, cost accounting, and hidden assistance further shape the result.
We interpret scores as evidence about a complete system and protocol.
Credible evaluation must connect outcomes to the process that produced them.
Benchmarks should disclose what the agent can observe and do, how the environment is controlled, and how outcomes are judged.
Reports should combine task success with the process, cost, risk, human burden, and reproducibility evidence required by the deployment claim.
RQ4 examines whether these qualities are engineered across the lifecycle and sustained beyond an isolated benchmark run.

\section{RQ4: GUI Agents through a Software Engineering Lens}
\label{sec:rq4-se-perspective}

We use RQ4 to examine whether the capabilities, architectures, and evaluations identified in RQ1--RQ3 are supported across the software lifecycle.
We analyze how requirements are specified, architectures accommodate change, quality is tested, runtime behavior is observed, and deployment risks are governed.
This perspective is necessary because GUI actions can change accounts, files, communications, permissions, or external services.
Engineering concerns that appear secondary in offline evaluation become central in real workflows.

\begin{figure*}[t]
\centering
\includegraphics[width=0.9\textwidth]{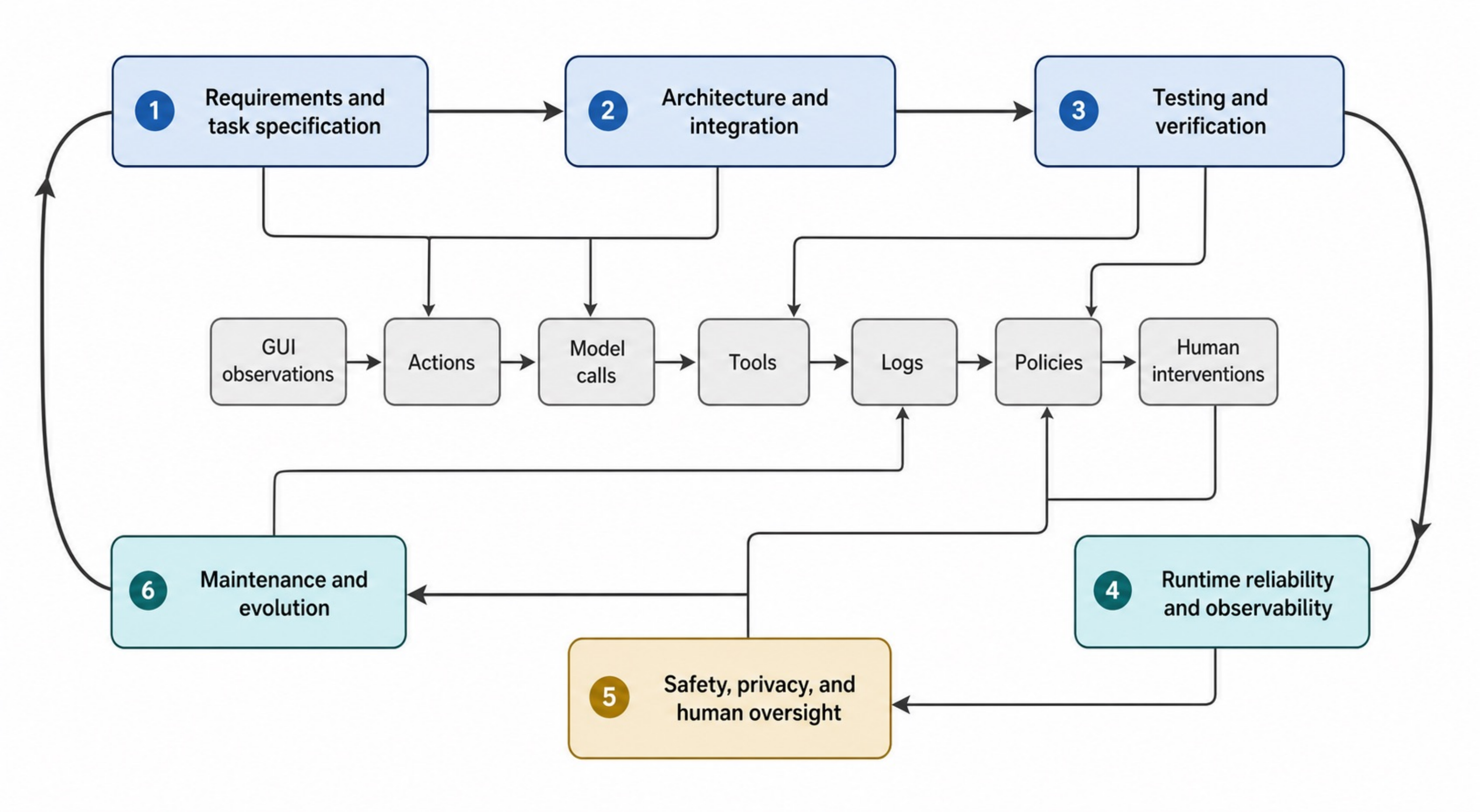}
\caption{Software-engineering lifecycle of GUI agents.}
\label{fig:rq4-se-lifecycle-map}
\end{figure*}

We organize the analysis from requirements and architecture to testing, operation, governance, and evolution in \figref{fig:rq4-se-lifecycle-map}.
Each stage reframes a capability question as a system question.
Perception and grounding become requirements on observation quality and action safety.
Planning becomes an architectural and verification concern.
Benchmarks become one part of a broader testing strategy.
Trajectories become potential runtime evidence for debugging, auditing, and regression analysis.
Human intervention becomes an engineering mechanism for boundary-setting and accountability.
We find that GUI-agent research has accumulated many relevant components, while several engineering practices remain underdeveloped as explicit system mechanisms across the lifecycle.

\subsection{Limited Testability of Capability Requirements}
\label{sec:rq4-requirements}

GUI-agent requirements span both task capability and operating quality.
In the 327-paper SE-parseable subset, most papers emphasize perception, localization, and action execution.
A smaller group addresses long-horizon control, memory, cross-platform generalization, and adaptation.
These requirements motivate modular frameworks across mobile, web, and desktop environments \cite{li2024appagent206, wen2024autodroid149, wu2024os213, zhang2024ufo109, agashe2024agent101, liu2025pc233}, as well as interface-grounding models such as SeeClick, ShowUI, and UI-TARS \cite{cheng2024seeclick012, lin2025showui062, qin2025ui001}.

\begin{table}[t]
\centering
\scriptsize
\caption{Requirement signals.}
\label{tab:rq4-requirement-signals}
\begin{tabular}{lrr}
\toprule
Requirement signal & Papers & Share \\
\midrule
GUI perception, localization, and action execution & 276 & 84.4\% \\
Cross-platform generalization and adaptation & 134 & 41.0\% \\
Planning, memory, and long-horizon task handling & 118 & 36.1\% \\
Low human intervention and autonomy & 24 & 7.3\% \\
\bottomrule
\end{tabular}
\end{table}

Our coding shows that GUI-agent requirements are dominated by the ability to observe and manipulate interfaces, followed by generalization and long-horizon task execution, as reported in \tabref{tab:rq4-requirement-signals}.
This distribution explains the architectural patterns identified in RQ2.
Modular pipelines, planner--executor separation, memory loops, tool layers, and reflection mechanisms are engineering responses to requirements that cannot be satisfied by a single prediction step.
The requirement profile changes with the platform.
Mind2Web and WebArena emphasize task decomposition and navigation on the web \cite{deng2023mind2web127, zhou2023webarena128}.
Mobile benchmarks add device state and application diversity, while desktop environments introduce files, windows, and operating-system control \cite{rawles2024androidworld148, xu2025androidlab152, wang2024mobileagentbench154, xie2024osworld159, bonatti2024windows160}.

From a software-engineering perspective, however, requirement specification remains less mature than capability description.
Many papers state that an agent should be autonomous or robust, while fewer define the contract that would make either claim testable.
Autonomy may refer to reduced human correction or to longer unassisted tasks, which are not equivalent.
Robustness may concern ordinary interface variation, hostile content, or transfer to a new domain.
Efficiency is similarly ambiguous unless the paper distinguishes machine cost from interaction time and human effort.
Without clearer requirement dimensions, systems with different assumptions can be compared under the same headline claim.
We identify explicit task boundaries, allowed observation channels, action authority, risk level, intervention policy, and environment assumptions as necessary elements of a testable requirement specification.

\subsection{Benchmark-Centered Testing and Underspecified Runtime Failures}
\label{sec:rq4-quality}

Current GUI-agent quality assurance is heavily benchmark-centered.
This is understandable because benchmarks provide a shared measurement substrate for a rapidly moving field, and RQ3 shows that interactive benchmarks have become one of the main drivers of progress.
Yet the SE statistics reveal a mismatch between evaluation practice and software testing practice.
Benchmark and ablation experiments are the main testing form in 160 papers, while only a small fraction explicitly discusses traditional testing gaps.
Exception handling is also limited.
Invalid-action handling and retry or replanning mechanisms appear in fewer than one third of the SE-parseable papers.

\begin{table}[t]
\centering
\scriptsize
\caption{Quality-assurance signals.}
\label{tab:rq4-quality-signals}
\begin{tabular}{lrr}
\toprule
Quality signal & Papers & Share \\
\midrule
Benchmark or ablation testing & 160 & 48.9\% \\
User study, human labels, or real-environment testing & 51 & 15.6\% \\
Traditional testing gap explicitly noted & 33 & 10.1\% \\
Invalid, redundant, or malformed action handling & 95 & 29.1\% \\
Reflection, retry, correction, or replanning & 86 & 26.3\% \\
Human help, confirmation, or takeover for failures & 17 & 5.2\% \\
\bottomrule
\end{tabular}
\end{table}

Our analysis of \tabref{tab:rq4-quality-signals} indicates that the field has invested far more in capability benchmarks than in systematic verification.
The difference matters because GUI-agent failures are often stateful.
A wrong click may change the screen and invalidate the next observation.
Repeated retries can create duplicate submissions, while an unsafe recovery step may hide side effects behind a correct final state.
Recent work on GUI testing, web-agent evaluation, and reward-based trajectory assessment begins to expose these problems by analyzing failure modes, action validity, and process-level behavior \cite{yang2025gui024, kara2025waber321, chen2025evaluating318, lu2025agentrewardbench261}.
STEVE, AgentRewardBench, VeriSafe, and related work further show that verification and recovery must reason over trajectories, intermediate states, and external effects, with final success serving as only one signal \cite{lu2025steve243, lee2025verisafe043, chen2025gui316}.

Robustness research has also expanded toward adversarial and environmental stressors.
Studies of malicious interfaces, GUI attacks, environmental injection, and web-agent security show that GUI agents can be misled through screen content, UI state, prompts embedded in pages, tool outputs, or environment manipulation \cite{zhang2025environmental017, evtimov2025wasp320, zhang2025attacking333, liao2024eia322, zharmagambetov2025agentdam179}.
These works are important because they move robustness beyond ordinary distribution shift.
A software system that controls GUIs must handle benign variation and hostile manipulation within the same execution loop.
This calls for testing at several levels.
Component checks can isolate perception and grounding, while integration tests exercise planner--executor coordination.
Saved trajectories support regression, and dynamic or adversarial environments provide system-level stress tests.
Existing benchmarks provide valuable starting points.
We find that deployed GUI agents need a broader testing pyramid that connects component checks, integration tests, trajectory regression, and system-level stress tests.

\subsection{Gaps in Maintainability and Observability}
\label{sec:rq4-operations}

Maintainability and observability are the clearest gaps when GUI agents are viewed as long-lived software systems.
Many architectures are modular at the research-prototype level, and RQ2 identified frequent use of separate perception, planning, grounding, memory, and execution modules.
However, maintainability is rarely articulated as an engineering goal.
The SE extraction shows that only 25 papers contain a modular or pluggable signal that can be interpreted as maintainability-related, while 279 papers lack discussion of maintenance processes, API evolution, or prompt maintenance.
This gap is consequential because several interfaces can change independently.
The model API and prompt may evolve without the target application, while permissions and browser or device wrappers follow their own versions.
Benchmark environments introduce another source of drift during long-term operation.

\begin{table}[t]
\centering
\scriptsize
\caption{Operational engineering signals.}
\label{tab:rq4-operational-signals}
\begin{tabular}{lrr}
\toprule
Operational signal & Papers & Share \\
\midrule
Modular or pluggable maintainability signal & 25 & 7.6\% \\
Missing maintenance process or API evolution & 279 & 85.3\% \\
Cross-platform or cross-task extension & 101 & 30.9\% \\
Data, training, or environment scale extension & 96 & 29.4\% \\
Module, tool, or model replacement extension & 100 & 30.6\% \\
Trajectory, history, log, or replay signal & 54 & 16.5\% \\
Missing monitoring, dashboard, or audit mechanism & 251 & 76.8\% \\
\bottomrule
\end{tabular}
\end{table}

We separate three related operational issues in \tabref{tab:rq4-operational-signals}.
First, extensibility is discussed more often than maintainability.
Cross-platform, data-scale, and module-replacement signals appear in roughly one third of the SE-parseable subset, often in systems that aim to generalize across apps, operating systems, or model backbones \cite{wu2024os063, zhang2024ufo109, zhang2025ufo2249, liu2025pc233}.
Second, maintainability remains implicit.
A modular diagram alone leaves open the need for versioned interfaces, compatibility tests, prompt migration rules, and a process for adapting to UI updates.
Third, observability is underdeveloped as a runtime capability.
Trajectories are often stored to train models or score benchmarks.
Far fewer systems turn the same evidence into operational logs, replay tools, alerts, or audit trails.

Efficiency has a stronger presence, although it is usually measured as a performance attribute more than as an operational budget.
Efficiency appears in several forms in the SE extraction.
Runtime or interaction efficiency is discussed in 249 papers, while 192 address token, compute, or training cost.
Another 98 discuss efficiency in parameters, data, or inference.
MobileAgentBench, Octopus, DistRL, and EcoAgent illustrate different efficiency directions, including mobile execution cost, on-device models, distributed reinforcement learning, and economical agent operation \cite{wang2024mobileagentbench154, chen2024octopus342, wang2024distrl344, yi2026ecoagent332}.
In deployment, these costs interact with reliability and oversight.
A system that retries aggressively may improve success rate while increasing latency, API cost, and duplicated side effects.
A system that asks for frequent human confirmation may reduce risk while increasing user burden.
A system that compresses observations may reduce cost while removing evidence needed for debugging or safety checks.
We frame efficiency as a lifecycle constraint that should be reported together with retries, action counts, model calls, context length, human-intervention frequency, and failure-recovery cost.

\subsection{Governance Gaps between Risk Recognition and Control}
\label{sec:rq4-governance}

Safety, security, privacy, and human oversight form the governance layer of GUI-agent engineering.
The corpus recognizes these risks more often than it implements controls for them.
Safety or security risks appear in 164 papers, while policy, permission, refusal, or defense mechanisms appear in 48.
Privacy shows the same asymmetry, with 212 risk signals but only 33 mitigation signals.
Human oversight is also limited.
Runtime confirmation, help, takeover, or correction appears in 47 papers, and only three framework papers are explicitly marked as human-in-the-loop in the framework subset used for RQ2.

Our analysis of \tabref{tab:rq4-governance-signals} shows that GUI-agent governance is widely recognized and remains thinly institutionalized.
This is a pressing issue because GUI agents combine broad observation with direct actuation.
The risk depends on both platform and task.
A web page can embed adversarial instructions, while a mobile app can expose private messages or credentials.
Desktop workflows often span files and communications and may reach terminals or payment dialogs.
GUI-specific safety work such as GUIGuard, SafeArena, VeriSafe, VeriOS, and EIA begins to address this problem by studying risk detection, unsafe actions, policy compliance, proactive verification, and environment-level attacks \cite{wang2026guiguard033, tur2025safearena175, lee2025verisafe043, wu2025verios327, liao2024eia322}.
Environmental injection, prompt injection through web pages, and visual adversarial manipulation further show that the GUI surface is both the agent's input channel and an attack channel \cite{zhang2025environmental017, levy2024st139, evtimov2025wasp320}.

\begin{wraptable}{r}{0.48\textwidth}
\centering
\scriptsize
\caption{Governance signals.}
\label{tab:rq4-governance-signals}
\resizebox{\linewidth}{!}{%
\begin{tabular}{lrr}
\toprule
Governance signal & Papers & Share \\
\midrule
Safety or security risk & 164 & 50.2\% \\
Policy, permission, refusal, or defense & 48 & 14.7\% \\
High-risk action gating & 45 & 13.8\% \\
Sensitive-data privacy risk & 212 & 64.8\% \\
Privacy mitigation & 33 & 10.1\% \\
Runtime human oversight & 47 & 14.4\% \\
Explicit HITL in framework subset & 3 & 2.1\% \\
\bottomrule
\end{tabular}%
}
\end{wraptable}

Privacy requires an equally concrete treatment.
Screenshots avoid exposing raw interface metadata, but they can still contain credentials, private communications, or proprietary content.
Conversely, DOM, XML, and accessibility representations may reveal hidden metadata or structured identifiers.
The choice of observation medium therefore changes the privacy boundary without eliminating privacy risk.
Systems such as VeriOS and human-centered studies of computer-use agents point toward finer-grained interaction policies, confirmation points, and user-aware control \cite{wu2025verios327, wang2026computer180}.
MobileAgent and MobileGPT also illustrate how human-machine interaction, SOP integration, and memory-like mechanisms can support more controllable mobile agents \cite{ding2024mobileagent343, lee2024mobilegpt335}.
Still, the low HITL count suggests that human oversight is rarely designed as a first-class lifecycle mechanism in deployed systems.

Defense and alignment papers show what a more concrete governance layer could contain.
LaSM studies pop-up attack defense through layer-wise scaling, while user-aligned web navigation frames task execution around ethical and personalized constraints \cite{yan2025lasm041, dammu2025towards241}.
These works complement the risk-oriented benchmarks by showing that mitigation can be introduced inside the perception, decision, or policy layer.
We argue that a GUI agent should expose where risk is detected, where policy is applied, and how a user or developer can inspect the resulting decision during later audits.

An SE-oriented governance model should begin with action authority and sensitive-data handling.
Policy enforcement and user confirmation constrain that authority, while audit logs preserve accountability.
These mechanisms should be tied to task risk.
Low-risk information lookup may tolerate higher autonomy.
External communication and irreversible changes require stricter gating, especially when credentials or payments are involved.
We treat governance as a lifecycle mechanism connected to requirements, architecture, evaluation, logging, and maintenance.
Otherwise, the system may appear safe in benchmark tasks while lacking the operational controls needed for deployment in real settings.

\textbf{Answer to RQ4.} We find that software-engineering concerns are present in GUI-agent research and unevenly covered across the lifecycle.
Capability requirements and modular architectures are common.
Testing beyond benchmarks and explicit verification are less developed, and the gap widens further for maintenance, observability, privacy controls, and systematic human oversight.
Among 327 SE-parseable papers, benchmark or ablation testing appears in 160, invalid-action handling in 95, retry or replanning in 86, trajectory or replay signals in 54, and runtime human oversight in 47.
By contrast, 279 lack a maintenance-process signal and 251 lack a monitoring or audit signal.
The field is moving from model capability toward system construction without yet establishing lifecycle-managed deployment in practice.
Closing this gap requires explicit requirements and testable execution contracts.
Observable behavior and maintainable interfaces must be supported by operational budgets and risk-aware governance throughout deployment and maintenance across the system lifecycle.

\section{RQ5: Open Challenges and Research Opportunities}
\label{sec:rq5-opportunities}

We use RQ5 to convert the preceding evidence into a research roadmap.
The field combines rapid and unevenly mature growth with weak runtime-control boundaries, success-centered evaluation, and limited lifecycle support.
These gaps require a connected system agenda spanning capability, evaluation, and deployment.

\begin{figure*}[b]
\centering
\includegraphics[width=0.9\textwidth]{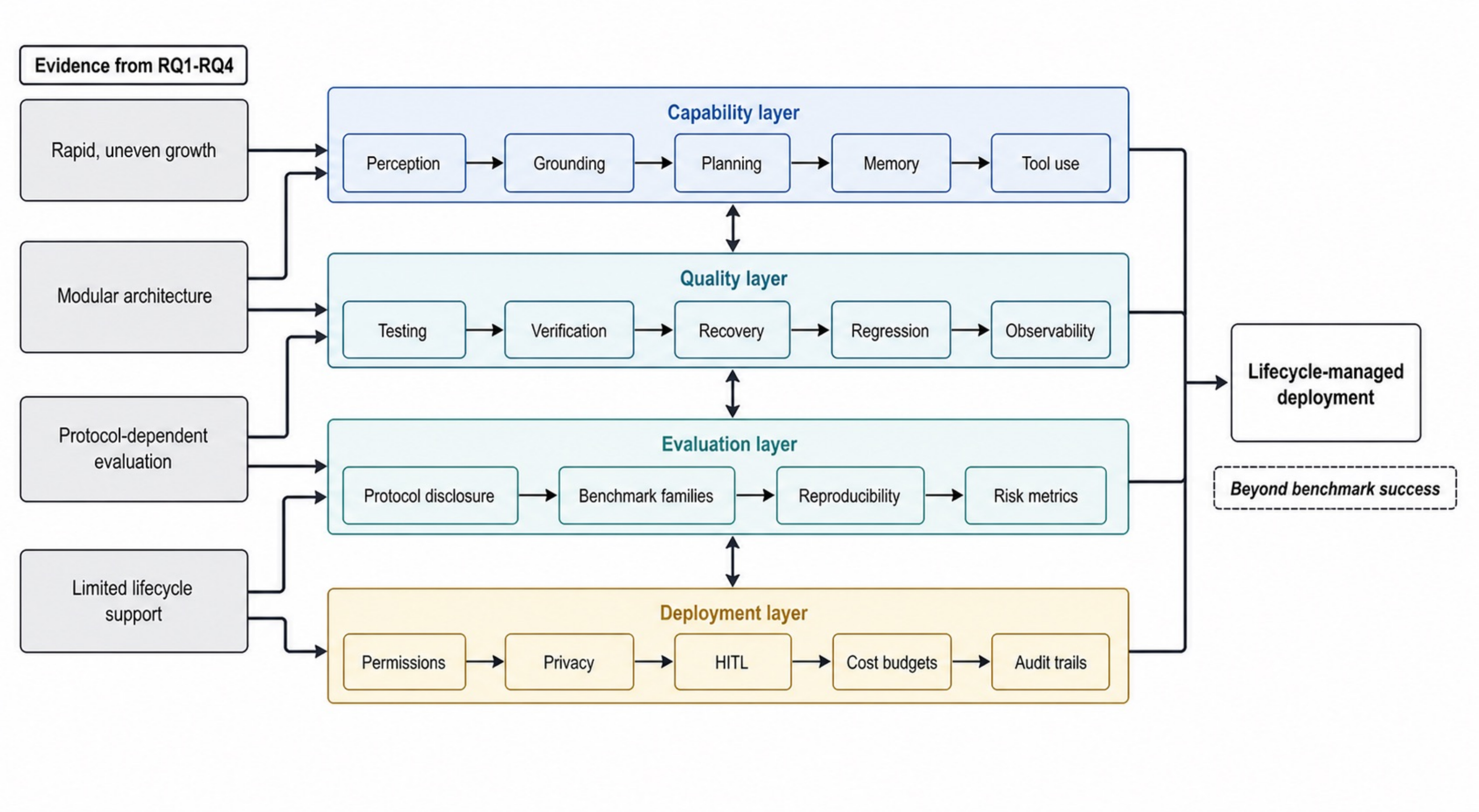}
\caption{Research roadmap for dependable GUI agents.}
\label{fig:rq5-research-roadmap}
\end{figure*}

We organize the agenda into capability, quality, evaluation, and deployment layers in \figref{fig:rq5-research-roadmap}.
The first concerns dependability in dynamic, heterogeneous, long-horizon environments.
The second concerns lifecycle-centered testing, debugging, and maintenance.
The third concerns continuous and risk-aware evaluation.
The fourth concerns safe, efficient, and human-centered deployment.
These opportunities are connected and involve trade-offs because GUI agents are closed-loop software systems.
A stronger planner may increase task success while making failures harder to diagnose.
Richer observations can improve grounding but expand the privacy boundary, and safer confirmation policies can increase user burden.
We identify a need for multi-objective engineering methods that assess capability improvements within their operational context.

\begin{table*}[t]
\centering
\scriptsize
\caption{Priority research gaps and opportunities.}
\label{tab:rq5-priority-gaps}
\begin{tabular}{>{\raggedright\arraybackslash}p{0.24\textwidth}>{\raggedright\arraybackslash}p{0.32\textwidth}>{\raggedright\arraybackslash}p{0.34\textwidth}}
\toprule
Gap area & Evidence signal & Research opportunity \\
\midrule
Dynamic dependability & 95 invalid-action handling, 86 retry/replanning, 17 human recovery & Recovery contracts, stop rules, rollback, and state-aware execution \\
Lifecycle quality & 160 benchmark/ablation tests, 33 testing-gap notes, 279 missing maintenance signals & GUI-agent test pyramid, trajectory regression, interface-version management \\
Evaluation comparability & Task success dominates, with protocols varying by observation, action space, oracle, and environment & Protocol disclosure, multi-objective leaderboards, reproducible benchmark infrastructure \\
Governance and deployment & 164 safety risks, 48 defenses, 212 privacy risks, 33 privacy mitigations, 3 HITL frameworks & Permission systems, privacy controls, audit trails, risk-adaptive human oversight \\
Efficiency and sustainability & 249 runtime-efficiency signals, 192 token/compute/training-cost signals & Cost budgets spanning model calls, actions, retries, latency, and supervision \\
\bottomrule
\end{tabular}
\end{table*}

We link each proposed direction to evidence reported earlier in the survey in \tabref{tab:rq5-priority-gaps}.
The dependability row follows from the limited recovery and human-escalation signals in RQ2 and RQ4.
The lifecycle and comparability rows respond to the testing, maintenance, and protocol gaps in RQ3 and RQ4.
The governance row reflects the large difference between risk recognition and implemented controls, while the efficiency row treats cost as a cross-cutting deployment constraint.

\subsection{Execution Contracts for Dependable Interaction}
\label{sec:rq5-dependability}

The first challenge is dependable execution under changing interfaces and uncertain feedback.
Visual observations improve portability, while DOM, XML, and accessibility structures improve precision but introduce platform and privacy assumptions \cite{qin2025ui001, cheng2024seeclick012, lu2024omniparser004, deng2023mind2web127, wen2024autodroid149}.
Neither representation prevents errors from accumulating across long tasks, hidden side effects, dialogs, loading states, and cross-application transitions.
Interactive benchmarks and modular frameworks expose these pressures across web, mobile, and desktop environments \cite{zhou2023webarena128, rawles2024androidworld148, xie2024osworld159, zhang2024ufo109, agashe2024agent101}.
The open problem is to make the full control loop dependable across perception, planning, execution, and recovery.

A promising direction is to define explicit execution contracts for GUI agents.
Such contracts should define valid actions and acceptable state changes.
They should also govern the transition from retry to stopping, confirmation, or recovery.
Current reflection, retry, and replanning mechanisms remain unevenly represented across the literature.
STEVE, VeriSafe, and related verification-oriented studies indicate that trajectory-level reasoning can help diagnose and repair execution failures \cite{lu2025steve243, lee2025verisafe043, chen2025gui316}.
We see an opportunity for transaction-aware GUI execution in which high-risk operations are checked before and after execution, reversible steps are preferred when possible, and irreversible operations are gated by risk-sensitive policies.

Another opportunity is hierarchical state abstraction.
Many agents keep raw screenshots or action histories.
Fewer maintain an explicit belief about task progress and the options available after failure.
A dependable GUI agent should track both the visible interface state and the task-level commitment state.
After filling a form, for example, the system should distinguish local edits from external submission and service confirmation.
This distinction can prevent duplicate actions and make failures easier to explain.
Research on memory-augmented, reflective, and planner--executor systems already points in this direction \cite{zhang2024dynamic028, cheng2025mga025, wu2025auto032, erdogan2025plan242}.
We identify testable and auditable state abstraction as a necessary runtime mechanism.

\subsection{Trajectory-Based Lifecycle Testing}
\label{sec:rq5-lifecycle}

The second challenge is to move GUI-agent quality assurance beyond benchmark execution.
Benchmarks are necessary because they supply tasks, environments, and shared scores.
They are insufficient as the sole quality mechanism because deployed agents must survive continuous change.
Models and prompts evolve alongside wrappers, target interfaces, infrastructure, and user policy.
Benchmark or ablation testing appears in 160 SE-parseable papers, while maintenance-process signals are missing from 279 papers.
We use this gap to motivate a lifecycle quality stack that connects benchmark tasks to software testing, debugging, observability, and maintenance.

A GUI-agent test pyramid would include several layers.
At the bottom, component tests should isolate perception and element localization from action serialization and permission checks.
In the middle, integration tests should validate planner--grounder--executor coordination under controlled UI states.
At the top, end-to-end tests should run realistic tasks in versioned environments.
Existing resources already cover parts of this stack.
ScreenSpot supports perception and localization checks, while Mind2Web provides evidence about action prediction and task decomposition \cite{li2025screenspot051, deng2023mind2web127}.
At the system level, BrowserGym, WABER, AndroidLab, and Windows Agent Arena contribute infrastructure for online evaluation \cite{chezelles2024the168, kara2025waber321, xu2025androidlab152, bonatti2024windows160}.
We see an opportunity to connect these resources into a coherent lifecycle process for deployed agents.

Debugging also needs better abstractions.
A failed trajectory can mix model error with tool misuse or delayed environment feedback.
Policy violations and oracle ambiguity may further obscure the cause.
If the system stores only final success and screenshots, debugging becomes manual and fragile.
Structured traces can turn those failures into reusable regression cases.
The trace must preserve the environment and model configuration together with policy decisions and recovery attempts.
AgentRewardBench and GUI testing work illustrate the value of process-level analysis and reward signals over trajectories \cite{lu2025agentrewardbench261, yang2025gui024}.
We propose treating every failed trajectory as a potential test artifact.
Its metadata should support replay and minimization before the failure enters regression tracking.

Maintenance is the long-term form of the same problem.
GUI agents depend on external software that changes outside the agent developer's control.
A website redesign, a mobile OS update, an API deprecation, or a model-provider change can invalidate assumptions without changing the agent code.
Research opportunities include interface-version monitoring, prompt migration tests, model-version differential testing, wrapper compatibility suites, and deprecation-aware tool schemas.
Modular architectures help because they expose replaceable components, but modularity becomes maintainability only when interfaces are versioned and tested.
This is especially important for systems that advertise cross-platform generality, such as OS-Atlas, UFO, OS-Copilot, and PC Agent \cite{wu2024os213, zhang2024ufo109, wu2024os063, liu2025pc233}.

\subsection{Protocol, Risk, and Drift in Continuous Evaluation}
\label{sec:rq5-evaluation}

The third challenge is to keep evaluation meaningful as models, tasks, and environments change.
Interactive benchmarks now cover web, mobile, desktop, cross-environment, professional, and scientific workflows \cite{zhou2023webarena128, rawles2024androidworld148, xie2024osworld159, xu2025crab167, drouin2024workarena134, sun2025scienceboard183}.
We identify comparability, reproducibility, and risk coverage across benchmark families as the remaining evaluation problem.

Reproducibility requires more than releasing tasks.
Interactive GUI environments contain websites, apps, operating systems, model APIs, authentication states, localization settings, external services, and time-sensitive content.
A reported score can depend on hidden assistance, cached knowledge, action limits, retries, wrappers, browser settings, judge prompts, and environment versions.
Future benchmarks should publish execution protocols as first-class artifacts.
The protocol must define observation and action access together with tool permissions and interaction limits.
Reproduction also requires model and environment versions, judge configuration, failure categories, and intervention rules.
BrowserGym and WABER are useful examples because they treat benchmark infrastructure and evaluation reliability as part of the research object \cite{chezelles2024the168, kara2025waber321}.
The same infrastructure requirements apply to mobile, desktop, and cross-environment benchmarks.

Risk-aware evaluation should also become standard.
A GUI agent that reaches a correct final state after exposing private data, clicking an unsafe confirmation, or repeatedly submitting a form should be judged differently from a safe trajectory.
Safety and privacy benchmarks such as SafeArena, GUIGuard, and VeriSafe begin to supply such evidence by examining unsafe actions, risk detection, and verification mechanisms \cite{tur2025safearena175, wang2026guiguard033, lee2025verisafe043}.
Environmental injection, adversarial web content, and malicious GUI manipulation further show that safety evaluation must be integrated with ordinary task evaluation as part of the same protocol \cite{zhang2025environmental017, liao2024eia322, evtimov2025wasp320}.

Continuous evaluation is the practical extension of reproducibility.
A benchmark result from a fixed date may become stale when the underlying website changes, the base model is updated, or the agent's prompt is modified.
Future leaderboards should therefore distinguish static benchmark snapshots from continuously monitored environments.
They should also report uncertainty, drift, and environmental failures.
In such a setting, a leaderboard would measure task success and performance stability under controlled updates.

\subsection{Bounded Authority and Human Oversight in Deployment}
\label{sec:rq5-deployment}

The final challenge is deployment under real user, organizational, and regulatory constraints.
GUI agents observe user interfaces and perform actions through them, which gives them unusual power compared with agents that only produce text.
They can read private content and access accounts.
Their actions may also alter files, contact other people, or confirm payments and permissions.
Privacy and security are widely recognized, while defense, mitigation, and human oversight mechanisms appear far less often than risk discussion.
We identify a need for deployment architectures that combine capability with least privilege, privacy protection, auditability, and user control.

Permission design is a central problem.
Current agents often operate with broad screen access and broad action authority.
Future systems should expose finer-grained authority.
Read-only observation and reversible editing should be separated from external communication, credential use, and irreversible state changes.
Each capability should have a policy and logging requirement.
Work on GUI safety, proactive verification, and policy compliance provides early building blocks for this direction \cite{wang2026guiguard033, lee2025verisafe043, wu2025verios327, tur2025safearena175}.
We propose making permissions part of the agent architecture and evaluation protocol so that reported performance is interpreted together with action authority.

Privacy engineering is equally important.
Screenshot-based agents, DOM-based agents, and accessibility-based agents expose different kinds of sensitive information.
A deployment-ready system should minimize what is captured, mask or redact sensitive regions when possible, keep local processing local when appropriate, and record data lifecycles for logs and trajectories.
Research on attacks and environmental manipulation shows that privacy and security interact because malicious UI content can influence the agent while it observes private user state \cite{zhang2025attacking333, zharmagambetov2025agentdam179, levy2024st139}.
We propose evaluating privacy leakage, prompt injection, and action risk together.

Human-centered deployment requires a richer account of human participation.
We find sparse HITL evidence in the corpus.
Only three framework papers are explicitly marked as human-in-the-loop, although many real scenarios require confirmation, correction, takeover, or preference judgment.
Existing work on computer-use agents, VeriOS, MobileAgent, and MobileGPT suggests that users can participate through confirmation, SOP integration, repair, and oversight \cite{wang2026computer180, wu2025verios327, ding2024mobileagent343, lee2024mobilegpt335}.
We propose interaction policies for task setup, mid-execution confirmation, failure recovery, and post-hoc audit.
The key question is where human judgment adds safety, accountability, or preference alignment without making the system unusable in routine practice.

Efficiency should be evaluated under the same deployment lens.
GUI-agent cost has both computational and operational components.
Model calls and latency capture only part of it.
Retries, environment setup, human supervision, logging, and evaluation maintenance can dominate in long-running deployments.
MobileAgentBench, Octopus, DistRL, and EcoAgent point to several efficiency paths, including benchmarked mobile cost, compact/on-device models, reinforcement learning efficiency, and economical agent operation \cite{wang2024mobileagentbench154, chen2024octopus342, wang2024distrl344, yi2026ecoagent332}.
We identify cost budgets that include machine and human costs as a research priority.
A safe and useful agent should report what task completion required.
At minimum, this includes model calls, actions, retries, elapsed time, and human confirmation, with sensitive observations reported when relevant.

\textbf{Answer to RQ5.} We identify the transformation of benchmark-performing GUI agents into dependable, maintainable, secure, and deployable systems as the central challenge.
Our synthesis yields four priorities.
Future work should establish dependable execution under dynamic state, lifecycle-centered testing and maintenance, continuous and risk-aware evaluation, and governed human-centered deployment with explicit cost budgets.
These priorities are interdependent.
Execution contracts shape recovery and evaluation, observability supports maintenance and auditing, and permission or privacy controls affect both capability and user burden.
We conclude that GUI-agent research should treat capability, quality, evaluation, and governance as one lifecycle problem.

\section{Discussion}
\label{sec:discussion}

Across the five research questions, we find that the relevant unit of analysis is the complete interactive software system.
Model behavior is mediated by observation and action interfaces and further shaped by memory, runtime checks, environment controllers, evaluation protocols, and user-facing policies.
An improvement to visual grounding can raise task success while leaving recovery or action authority unchanged.
A stronger planner can improve long-horizon control while increasing latency and making failures harder to localize.
GUI-agent progress should consequently be interpreted through the architecture and operating conditions that produce observed behavior.

Observation design is the first system boundary that cuts across model development, architecture, evaluation, and deployment.
Screenshots provide a common interface across web, mobile, and desktop settings, yet they omit stable element identity, hidden state, and explicit interaction semantics.
DOM, XML, and accessibility representations expose more structure while introducing platform dependence and access to potentially sensitive metadata.
OmniParser, SeeClick, ShowUI, and UI-TARS illustrate screenshot-centered perception and action, whereas Mind2Web and AutoDroid show how structured web or mobile state can support grounding and execution \cite{lu2024omniparser004, cheng2024seeclick012, lin2025showui062, qin2025ui001, deng2023mind2web127, wen2024autodroid149}.
We find no single representation that dominates across all operating conditions.
Papers should state the observation channels available to the agent and explain how those channels affect portability, privacy, and diagnostic access.

Agent construction and benchmark construction also remain closely coupled.
WebArena, VisualWebArena, AndroidWorld, OSWorld, WorkArena, ScienceBoard, and CRAB have expanded evaluation from static prediction to interactive tasks across major platforms and professional workflows \cite{zhou2023webarena128, koh2024visualwebarena129, rawles2024androidworld148, xie2024osworld159, drouin2024workarena134, sun2025scienceboard183, xu2025crab167}.
These environments expose capability gaps that offline datasets cannot capture.
They also determine observation privileges, action abstractions, interaction budgets, reset policies, and completion oracles.
We therefore interpret benchmark scores as evidence about an agent stack under a specific protocol.
Cross-benchmark claims require explicit reporting of wrappers, retries, tool access, environment versions, judge settings, and human assistance.
BrowserGym, WABER, AndroidLab, and Windows Agent Arena provide useful foundations by treating evaluation infrastructure and reliability as part of the research contribution \cite{chezelles2024the168, kara2025waber321, xu2025androidlab152, bonatti2024windows160}.

The transition from action prediction to closed-loop control changes the engineering problem.
GUI agents modify external state, and some changes persist after a task fails.
Memory, verification, reflection, and replanning help manage this uncertainty, although their presence does not establish dependable recovery.
STEVE, VeriSafe, AgentRewardBench, and GUI testing studies begin to make trajectories and intermediate outcomes explicit objects of analysis \cite{lu2025steve243, lee2025verisafe043, lu2025agentrewardbench261, yang2025gui024}.
We find that deployed systems need execution contracts defining valid actions, expected state transitions, retry limits, rollback conditions, and escalation rules.
Structured trajectories should record the model and environment configuration together with actions, policy decisions, failures, and recovery attempts.
The same evidence can support debugging, regression testing, continuous evaluation, and audit.

Model and framework design create a related architectural choice.
GUI-native systems such as UI-TARS, CogAgent, ShowUI, and OS-ATLAS demonstrate the value of task-specific data and unified perception-action learning \cite{qin2025ui001, hong2024cogagent003, lin2025showui062, wu2024os063}.
Framework-oriented systems such as AppAgent, UFO, OS-Copilot, CRADLE, and Agent~S expose state construction, planning, execution, and verification as separate responsibilities \cite{li2024appagent206, zhang2024ufo109, wu2024os213, tan2024cradle214, agashe2024agent101}.
Native models can reduce coordination overhead, while modular systems provide clearer locations for testing, replacement, policy enforcement, and failure diagnosis.
The corpus points toward hybrid systems in which learned models provide perception, reasoning, and action priors and explicit runtime modules retain authority over verification, permissions, logging, and recovery.
This design becomes maintainable only when module interfaces, prompts, model versions, and environment adapters are versioned and tested.

Deployment adds coupled constraints involving safety, privacy, usability, and efficiency.
Stronger confirmation policies can reduce harmful actions while increasing user effort.
Privacy filters can limit exposure while removing information needed for grounding or recovery.
Detailed logs improve auditability while creating sensitive data that must be governed.
GUIGuard, SafeArena, VeriSafe, and environmental-injection studies demonstrate that risk can originate in model behavior, interface content, or excessive action authority \cite{wang2026guiguard033, tur2025safearena175, lee2025verisafe043, zhang2025environmental017, liao2024eia322, evtimov2025wasp320}.
VeriOS, MobileAgent, MobileGPT, and human-centered studies further indicate that confirmation, takeover, and preference judgment should be designed as runtime mechanisms \cite{wu2025verios327, ding2024mobileagent343, lee2024mobilegpt335, wang2026computer180}.
We interpret autonomy as bounded delegation with explicit observation access, action authority, and accountability.
Operational cost should include model calls, actions, retries, latency, environment recovery, logging, and human supervision.

These findings have direct implications for research and practice.
Researchers should align each claim with the evidence needed to support it.
Grounding models may emphasize target and action correctness, whereas enterprise or scientific agents require additional evidence on permissions, recovery, privacy, oracle validity, and reproducibility.
System builders should define testable interfaces for observation, state construction, planning, execution, verification, and intervention.
Benchmark designers should version tasks and environments and publish protocols that cover observation access, action permissions, resets, timeouts, authentication, judging, and intervention.
Practitioners should separate read-only and reversible work from external communication, credential use, and irreversible state changes.
We identify system boundaries, lifecycle evidence, action authority, recovery contracts, and human oversight as the shared vocabulary needed to compare GUI-agent research across vision-language modeling, NLP, HCI, mobile computing, security, and software engineering.

\section{Related Surveys and Positioning}
\label{sec:related-work}

Prior surveys have organized the rapidly expanding terminology around GUI agents, multimodal mobile agents, and computer-use systems.
Broad reviews use capabilities such as perception, instruction understanding, planning, grounding, action, and benchmark construction as their main analytical dimensions \cite{zhang2024large005, nguyen2025gui006, wang2024gui008, tang2025a009}.
They establish the pipeline shared by many agents and explain how GUI interaction relates to multimodal foundation models and interactive evaluation.

Specialized surveys narrow this view by platform or technical concern.
Mobile-agent reviews emphasize phone interaction, Android representations, and application automation \cite{wu2024foundations289, liu2025llm065}, while OS-agent reviews focus on desktop control and general computer use \cite{hu2024os285}.
Trustworthiness- and reinforcement-learning-oriented surveys examine safety, robustness, alignment, and sequential decision-making strategies \cite{shi2025towards007, li2025a075}.
Together, these studies demonstrate the field's breadth across NLP, computer vision, HCI, mobile computing, security, reinforcement learning, and systems.

These perspectives are complementary, but their primary unit of analysis is usually a capability, model, platform, benchmark, or risk category.
They provide less synthesis of how these elements interact across requirements, architecture, testing, operation, maintenance, and governance.
The present survey addresses this gap by treating the GUI agent and its surrounding runtime as the unit of engineering analysis.

The distinctive contribution of this survey is its lifecycle-centered software-engineering perspective on GUI agents.
GUI agents are analyzed as systems whose requirements and interfaces connect many components and dependencies.
Execution contracts, tests, and logs determine how those systems are assured, while permissions and maintenance shape their operation over time.
This perspective changes the interpretation of familiar research artifacts.
A benchmark defines an execution protocol, an action space defines an interface contract, a trajectory can serve as debugging and audit evidence, and a model update creates a maintenance event.
The literature contains these connections, but rarely synthesizes them as one lifecycle problem.

The five research questions operationalize this positioning.
They move from field mapping and architectural responsibility to evaluation evidence and lifecycle coverage, then integrate the resulting gaps into a deployment-oriented research agenda.
This sequence distinguishes the survey from reviews that stop at capability taxonomies or benchmark comparison alone in prior work.

The survey also differs from a conventional GUI testing review.
GUI testing has a mature literature on test generation, event exploration, regression, crash detection, and record-and-replay validation.
We include this work only when it directly contributes to agent perception, decision making, execution, evaluation, safety, or lifecycle engineering.
This boundary preserves GUI agents as the object of study while allowing established testing concepts to inform their engineering analysis.

\section{Threats to Validity}
\label{sec:threats}

The first threat concerns corpus coverage and evidence maturity.
GUI-agent research evolves rapidly and uses inconsistent terminology across several communities.
We combined keyword search, snowballing, deduplication, and manual screening to reduce omission risk, although the resulting corpus cannot constitute an exhaustive census.
The high proportion of preprints also means that methods and claims may change after review.
Because the 2026 coverage ends in April, we interpret the 336-paper corpus as a structured snapshot of the field at that time.

The second threat concerns coding and interpretation.
The structured labels simplify papers that often combine several contributions, and lifecycle concerns such as maintainability or human oversight may remain implicit in system descriptions.
We use multi-label and functional coding to reduce this loss and treat absent signals as a lack of explicit evidence under the coding scheme.
The software-engineering synthesis also reflects analytical judgment and may give less attention to fine-grained model or data optimization.
Internal-coherence checks improve consistency, although they do not provide independently measured inter-rater agreement from multiple coders.

The third threat concerns comparability across analytical subsets and evaluation protocols.
RQ1 uses all 336 papers, RQ2 uses 145 framework papers for module statistics, RQ3 uses 252 system-oriented papers, and RQ4 uses 327 SE-parseable records.
Benchmark results also depend on changing models, observation access, action spaces, wrappers, environments, and oracles.
We report the relevant denominators and protocol assumptions with each analysis and avoid direct comparisons across incompatible subsets or benchmark settings.

\section{Conclusion}
\label{sec:conclusion}

GUI agents have evolved from isolated interface-understanding and action-prediction models into closed-loop systems that operate across web, mobile, desktop, and cross-platform environments.
This survey analyzed that evolution from a software-engineering perspective using 336 papers published or posted between 2018 and April 2026.
The five research questions connected the field's growth to its architecture and evaluation evidence.
They then examined lifecycle coverage and future opportunities.
The results show rapid expansion alongside incomplete maturity.
Frameworks and benchmarks dominate the recent literature, screenshots provide a common observation channel, and modular perceive--decide--execute loops have become a recurring architecture.
Evaluation is also more interactive and realistic than earlier static prediction tasks.
However, its evidence remains centered on task success and conditioned by the surrounding protocol.
Across the lifecycle, testing beyond benchmarks and explicit recovery contracts remain limited.
Maintainability, observability, privacy controls, auditability, and systematic human oversight are also underdeveloped.
These findings establish the central conclusion of the survey.
GUI-agent capability and software quality cannot be developed independently.
Dependable progress begins with explicit execution contracts and lifecycle-centered testing.
It also requires reproducible, risk-aware evaluation and observable runtime behavior.
Permission, privacy, operational cost, and human governance must be designed as part of the same system.
Future research should evaluate GUI agents as complete system stacks and report the boundaries under which their capabilities hold.
As GUIs remain an important interface to consequential digital work, building agents that operate through them dependably, securely, and accountably will remain a shared agenda for AI and software-engineering research communities.


\bibliographystyle{ACM-Reference-Format}
\bibliography{main}

@misc{qin2025ui001,
  title = {{UI-TARS}: Pioneering Automated {GUI} Interaction with Native Agents},
  author = {Qin, Yujia and Ye, Yining and Fang, Junjie and Wang, Haoming and Liang, Shihao and Tian, Shizuo and Zhang, Junda and Li, Jiahao and Li, Yunxin and Huang, Shijue and Zhong, Wanjun and Li, Kuanye and Yang, Jiale and Miao, Yu and Lin, Woyu and Liu, Longxiang and Jiang, Xu and Ma, Qianli and Li, Jingyu and Xiao, Xiaojun and Cai, Kai and Li, Chuang and Zheng, Yaowei and Jin, Chaolin and Li, Chen and Zhou, Xiao and Wang, Minchao and Chen, Haoli and Li, Zhaojian and Yang, Haihua and Liu, Haifeng and Lin, Feng and Peng, Tao and Liu, Xin and Shi, Guang},
  year = {2025},
  eprint = {2501.12326},
  archivePrefix = {arXiv},
  primaryClass = {cs.AI},
  doi = {10.48550/arXiv.2501.12326}
}

@misc{gou2024navigating002,
  title = {Navigating the Digital World as Humans Do: Universal Visual Grounding for {GUI} Agents},
  author = {Gou, Boyu and Wang, Ruohan and Zheng, Boyuan and Xie, Yanan and Chang, Cheng and Shu, Yiheng and Sun, Huan and Su, Yu},
  year = {2024},
  eprint = {2410.05243},
  archivePrefix = {arXiv},
  primaryClass = {cs.AI},
  doi = {10.48550/arXiv.2410.05243}
}

@inproceedings{hong2024cogagent003,
  title = {{CogAgent}: A Visual Language Model for {GUI} Agents},
  author = {Hong, Wenyi and Wang, Weihan and Lv, Qingsong and Xu, Jiazheng and Yu, Wenmeng and Ji, Junhui and Wang, Yan and Wang, Zihan and Dong, Yuxiao and Ding, Ming and Tang, Jie},
  booktitle = {2024 IEEE/CVF Conference on Computer Vision and Pattern Recognition (CVPR)},
  year = {2024},
  pages = {14281--14290},
  doi = {10.1109/cvpr52733.2024.01354},
  url = {https://doi.org/10.1109/cvpr52733.2024.01354}
}

@misc{lu2024omniparser004,
  title = {{OmniParser} for Pure Vision Based {GUI} Agent},
  author = {Lü, Yadong and Yang, Jianwei and Shen, Yelong and Awadallah, Ahmed Hassan},
  year = {2024},
  eprint = {2408.00203},
  archivePrefix = {arXiv},
  primaryClass = {cs.CV},
  doi = {10.48550/arXiv.2408.00203}
}

@misc{zhang2024large005,
  title = {Large Language {Model-Brained} {GUI} Agents: A Survey},
  author = {Zhang, Chaoyun and He, Shilin and Qian, Jiaxu and Li, Bowen and Li, Liqun and Qin, Si and Yu, Kang and Ma, Minghua and Liu, G. M. and Lin, Qingwei and Rajmohan, Saravan and Zhang, Dongmei and Zhang, Qi},
  year = {2024},
  eprint = {2411.18279},
  archivePrefix = {arXiv},
  primaryClass = {cs.AI},
  doi = {10.48550/arXiv.2411.18279}
}

@inproceedings{nguyen2025gui006,
  title = {{GUI} Agents: A Survey},
  author = {Nguyen, Dang and Chen, Jian and Wang, Yu and Wu, Gang and Park, Namyong and Hu, Zhengmian and Lyu, Hanjia and Wu, Junda and Aponte, Ryan and Xia, Yu and Li, Xintong and Shi, Jing and Chen, Hongjie and Lai, Viet Dac and Xie, Zhouhang and Kim, Sungchul and Zhang, Ruiyi and Yu, Tong and Tanjim, Mehrab and Ahmed, Nesreen K. and Mathur, Puneet and Yoon, Seunghyun and Yao, Lina and Kveton, Branislav and Kil, Jihyung and Nguyen, Thien Huu and Bui, Trung and Zhou, Tianyi and Rossi, Ryan A. and Dernoncourt, Franck},
  booktitle = {Findings of the Association for Computational Linguistics: ACL 2025},
  year = {2025},
  pages = {22522--22538},
  doi = {10.18653/v1/2025.findings-acl.1158},
  url = {https://doi.org/10.18653/v1/2025.findings-acl.1158}
}

@misc{shi2025towards007,
  title = {Towards Trustworthy {GUI} Agents: A Survey},
  author = {Shi, Yucheng and Yu, Wenhao and Huang, Jingyuan and Yao, Wenlin and Chen, Wenhu and Liu, Ninghao},
  year = {2025},
  eprint = {2503.23434},
  archivePrefix = {arXiv},
  primaryClass = {cs.LG},
  doi = {10.48550/arXiv.2503.23434}
}

@misc{wang2024gui008,
  title = {{GUI} Agents with Foundation Models: A Comprehensive Survey},
  author = {Wang, Shuai and Liu, Weiwen and Chen, J. C. and Zhou, Yuqi and Gan, Weinan and Zeng, X. and Che, Yuhan and Yu, Shicheng and Hao, Xinlong and Kun, Shao and Wang, Bin and Wu, Chuhan and Wang, Y.S. and Tang, Ruiming and Hao, Jianye},
  year = {2024},
  eprint = {2411.04890},
  archivePrefix = {arXiv},
  primaryClass = {cs.AI},
  doi = {10.48550/arXiv.2411.04890}
}

@misc{tang2025a009,
  title = {A Survey on (M){LLM-Based} {GUI} Agents},
  author = {Tang, Fei and Xu, Haolei and Zhang, Hang and Chen, Siqi and Wu, Xingyu and Shen, Yongliang and Zhang, Wenqi and Hou, Guiyang and Tan, Zeqi and Yan, Yuchen and Song, Kaitao and Shao, Jian and Lu, Weiming and Xiao, Jun and Zhuang, Yueting},
  year = {2025},
  eprint = {2504.13865},
  archivePrefix = {arXiv},
  primaryClass = {cs.HC},
  doi = {10.48550/arXiv.2504.13865}
}

@misc{ye2025mobile010,
  title = {{Mobile-Agent-v3}: Fundamental Agents for {GUI} Automation},
  author = {Ye, Jiabo and Zhang, Xi and Xu, Haiyang and Liu, Haowei and Wang, Junyang and Zhu, Zhaoqing and Zheng, Ziwei and Gao, Feiyu and Cao, Junjie and Lu, Zhengxi and Liao, Jitong and Zheng, Qi and Huang, Fei and Zhou, Jingren and Yan, Ming},
  year = {2025},
  eprint = {2508.15144},
  archivePrefix = {arXiv},
  primaryClass = {cs.AI},
  doi = {10.48550/arXiv.2508.15144}
}

@article{chen2025guicourse011,
  title = {{GUICourse}: From General Vision Language Model to Versatile {GUI} Agent},
  author = {Chen, Wentong and Cui, Junbo and Hu, Jinyi and Qin, Yujia and Fang, Junjie and Zhao, Yue and Wang, Chongyi and Liu, Jun and Chen, Guirong and Huo, Yupeng and Yao, Yuan and Lin, Yankai and Liu, Zhiyuan and Sun, Maosong},
  year = {2025},
  pages = {21936--21959},
  doi = {10.18653/v1/2025.acl-long.1065},
  url = {https://doi.org/10.18653/v1/2025.acl-long.1065}
}

@article{cheng2024seeclick012,
  title = {{SeeClick}: Harnessing {GUI} Grounding for Advanced Visual {GUI} Agents},
  author = {Cheng, Kanzhi and Sun, Qiushi and Chu, Yougang and Xu, Fangzhi and Li, Yantao and Zhang, Jianbing and Wu, Zhiyong},
  year = {2024},
  pages = {9313--9332},
  doi = {10.18653/v1/2024.acl-long.505},
  url = {https://doi.org/10.18653/v1/2024.acl-long.505}
}

@article{zhang2026tongui013,
  title = {{TongUI}: {Internet-Scale} Trajectories from Multimodal Web Tutorials for Generalized {GUI} Agents},
  author = {Zhang, Bofei and Shang, Zirui and Gao, Zhi and Zhang, Wang and Xie, Rui and Ma, Xiaojian and Yuan, Tao and Wu, Xinxiao and Zhu, Song-Chun and Li, Qing},
  journal = {Proceedings of the AAAI Conference on Artificial Intelligence},
  year = {2026},
  volume = {40},
  number = {15},
  pages = {12367--12375},
  doi = {10.1609/aaai.v40i15.38229},
  url = {https://doi.org/10.1609/aaai.v40i15.38229}
}

@article{zhang2024android014,
  title = {Android in the Zoo: {Chain-of-Action-Thought} for {GUI} Agents},
  author = {Zhang, Jiwen and Wu, Jihao and Yihua, Teng and Liao, Minghui and Xu, Nuo and Xiao, Xiao and Wei, Zhongyu and Tang, Duyu},
  year = {2024},
  pages = {12016--12031},
  doi = {10.18653/v1/2024.findings-emnlp.702},
  url = {https://doi.org/10.18653/v1/2024.findings-emnlp.702}
}

@misc{shi2025mobilegui015,
  title = {{MobileGUI-RL}: Advancing Mobile {GUI} Agent through Reinforcement Learning in Online Environment},
  author = {Shi, Yucheng and Yu, Wenhao and Li, Zaitang and Wang, Yong‐Lin and Zhang, Hongming and Liu, Ninghao and Mi, Haitao and Yu, Dong},
  year = {2025},
  eprint = {2507.05720},
  archivePrefix = {arXiv},
  primaryClass = {cs.LG},
  doi = {10.48550/arXiv.2507.05720}
}

@misc{nong2024mobileflow016,
  title = {{MobileFlow}: A Multimodal {LLM} For Mobile {GUI} Agent},
  author = {Nong, Songqin and Zhu, Jiali and Wu, Rui and Jin, Jiongchao and Shan, Shuo and Huang, Xiutian and Xu, Wenhao},
  year = {2024},
  eprint = {2407.04346},
  archivePrefix = {arXiv},
  primaryClass = {cs.CV},
  doi = {10.48550/arXiv.2407.04346}
}

@misc{zhang2025environmental017,
  title = {Environmental Injection Attacks against {GUI} Agents in Realistic Dynamic Environments},
  author = {Zhang, Yitong and Li, Xing‐Bin and Cai, Liyi and Li, Jia},
  year = {2025},
  eprint = {2509.11250},
  archivePrefix = {arXiv},
  primaryClass = {cs.CR},
  doi = {10.48550/arXiv.2509.11250}
}

@misc{xie2025gui018,
  title = {{GUI-explorer}: Autonomous Exploration and Mining of Transition-aware Knowledge for {GUI} Agent},
  author = {Xie, Bin and Shao, Rui and Chen, Gongwei and Zhou, Kaiwen and Li, Yinchuan and Liu, Jie and Zhang, Min and Nie, Liqiang},
  year = {2025},
  eprint = {2505.16827},
  archivePrefix = {arXiv},
  primaryClass = {cs.AI},
  doi = {10.48550/arXiv.2505.16827}
}

@misc{wang2025mmbench019,
  title = {{MMBench-GUI}: Hierarchical {Multi-Platform} Evaluation Framework for {GUI} Agents},
  author = {Wang, Xuehui and Wu, Zhenyu and Xie, JingJing and Ding, Zichen and Yang, Bowen and Li, Zehao and Liu, Zhaoyang and Li, Qingyun and Dong, Xuan and Chen, Zhe and Wang, Weiyun and Zhao, Xiangyu and Chen, Jixuan and Duan, Haodong and Xie, Tianbao and Yang, Chenyu and Su, Shiqian and Yu, Yue and Huang, Yuan and Liu, Yiqian and Zhang, Xiao and Zhang, Yanting and Yue, Xiangyu and Su, Weijie and Zhu, Xizhou and Shen, Wei and Dai, Jifeng and Wang, Wenhai},
  year = {2025},
  eprint = {2507.19478},
  archivePrefix = {arXiv},
  primaryClass = {cs.CV},
  doi = {10.48550/arXiv.2507.19478}
}

@inproceedings{sun2025gui020,
  title = {{GUI-Xplore}: Empowering Generalizable {GUI} Agents with One Exploration},
  author = {Sun, Yuchen and Zhao, Shanhui and Yu, Tao and Wen, Hao and Va, Samith and Xu, Mengwei and Li, Yuanchun and Zhang, Chongyang},
  booktitle = {2025 IEEE/CVF Conference on Computer Vision and Pattern Recognition (CVPR)},
  year = {2025},
  pages = {19477--19486},
  doi = {10.1109/cvpr52734.2025.01814},
  url = {https://doi.org/10.1109/cvpr52734.2025.01814}
}

@misc{xiangwu2025auto021,
  title = {{AUTO-Explorer}: Automated Data Collection for {GUI} Agent},
  author = {Guo, Xiangwu and Gao, Difei and Shou, Mike Zheng},
  year = {2025},
  eprint = {2511.06417},
  archivePrefix = {arXiv},
  primaryClass = {cs.AI},
  doi = {10.48550/arXiv.2511.06417}
}

@misc{zhang2025characterizing022,
  title = {Characterizing Unintended Consequences in {Human-GUI} Agent Collaboration for Web Browsing},
  author = {Zhang, Shuning and Chen, Jingruo and Gao, Zhiqi and Gao, Jiajing and Yi, Xin and Li, Hewu},
  year = {2025},
  eprint = {2505.09875},
  archivePrefix = {arXiv},
  primaryClass = {cs.HC},
  doi = {10.48550/arXiv.2505.09875}
}

@article{chen2025less023,
  title = {Less is More: Empowering {GUI} Agent with {Context-Aware} Simplification},
  author = {Chen, Gongwei and Zhou, Xurui and Shao, Rui and Lyu, Yibo and Zhou, Kaiwen and Wang, Shuai and Li, Wentao and Li, Yinchuan and Qi, Zhongang and Nie, Liqiang},
  year = {2025},
  pages = {5901--5911},
  doi = {10.1109/iccv51701.2025.00558},
  url = {https://doi.org/10.1109/iccv51701.2025.00558}
}

@misc{yang2025gui024,
  title = {{GUI-Robust}: A Comprehensive Dataset for Testing {GUI} Agent Robustness in {Real-World} Anomalies},
  author = {Yang, Jingqi and Song, Zeng and Chen, Jiawei and Song, Mingli and Sheng, Zhou and sun, linjun and Ouyang, Xiaogang and Chen, Chun and Wang, Can},
  year = {2025},
  eprint = {2506.14477},
  archivePrefix = {arXiv},
  primaryClass = {cs.AI},
  doi = {10.48550/arXiv.2506.14477}
}

@misc{cheng2025mga025,
  title = {{MGA}: {Memory-Driven} {GUI} Agent for {Observation-Centric} Interaction},
  author = {Cheng, Weihua and Liu, Junming and Sun, Yifei and Shi, Botian and Chen, Yirong and Wang, Ding},
  year = {2025},
  eprint = {2510.24168},
  archivePrefix = {arXiv},
  primaryClass = {cs.AI},
  doi = {10.48550/arXiv.2510.24168}
}

@article{gao2024assistgui026,
  title = {{AssistGUI}: {Task-Oriented} {PC} Graphical User Interface Automation},
  author = {Gao, Difei and Ji, Lei and Bai, Zechen and Ouyang, Mingyu and Li, Peiran and Mao, Donzxing and Wu, Qinchen and Zhang, Weichen and Wang, Peiyi and Guo, Xiangwu and Wang, Hengxu and Zhou, Luowei and Shou, Mike Zheng},
  year = {2024},
  pages = {13289--13298},
  doi = {10.1109/cvpr52733.2024.01262},
  url = {https://doi.org/10.1109/cvpr52733.2024.01262}
}

@misc{hu2024the027,
  title = {The Dawn of {GUI} Agent: A Preliminary Case Study with Claude 3.5 Computer Use},
  author = {Hu, Siyuan and Ouyang, Mingyu and Gao, Difei and Shou, Mike Zheng},
  year = {2024},
  eprint = {2411.10323},
  archivePrefix = {arXiv},
  primaryClass = {cs.AI},
  doi = {10.48550/arXiv.2411.10323}
}

@article{zhang2024dynamic028,
  title = {Dynamic Planning for {LLM-based} Graphical User Interface Automation},
  author = {Zhang, Shaoqing and Zhang, Zhuosheng and Chen, Kehai and Ma, Xinbei and Yang, Muyun and Zhao, Tiejun and Zhang, Min},
  year = {2024},
  pages = {1304--1320},
  doi = {10.18653/v1/2024.findings-emnlp.70},
  url = {https://doi.org/10.18653/v1/2024.findings-emnlp.70}
}

@article{sun2022meta029,
  title = {{META-GUI}: Towards Multi-modal Conversational Agents on Mobile {GUI}},
  author = {Sun, Liangtai and Chen, Xingyu and Chen, Lu and Dai, Tianle and Zhu, Zichen and Yu, Kai},
  year = {2022},
  pages = {6699--6712},
  doi = {10.18653/v1/2022.emnlp-main.449},
  url = {https://doi.org/10.18653/v1/2022.emnlp-main.449}
}

@inproceedings{shaw2023from030,
  title = {From Pixels to {UI} Actions: Learning to Follow Instructions via Graphical User Interfaces},
  author = {Shaw, Peter and Joshi, Mandar and Cohan, James and Berant, Jonathan and Pasupat, Panupong and Hu, Hexiang and Khandelwal, Urvashi and Lee, Kenton and Toutanova, Kristina N},
  booktitle = {Advances in Neural Information Processing Systems 36},
  year = {2023},
  pages = {34354--34370},
  doi = {10.52202/075280-1490},
  url = {https://doi.org/10.52202/075280-1490}
}

@misc{liu2026continual031,
  title = {Continual {GUI} Agents},
  author = {Liu, Ziwei and Kang, Borui and Yuan, Hangjie and Zhao, Zixiang and Li, Wei and Zhu, Yifan and Feng, Tao},
  year = {2026},
  eprint = {2601.20732},
  archivePrefix = {arXiv},
  primaryClass = {cs.LG},
  doi = {10.48550/arXiv.2601.20732}
}

@misc{wu2025auto032,
  title = {Auto-scaling Continuous Memory for {GUI} Agent},
  author = {Wu, Wenyi and Zhou, Kun and Yuan, Ruoxin and Yu, Vivian and Wang, Stephen and Hu, Zhiting and Huang, Biwei},
  year = {2025},
  eprint = {2510.09038},
  archivePrefix = {arXiv},
  primaryClass = {cs.AI},
  doi = {10.48550/arXiv.2510.09038}
}

@misc{wang2026guiguard033,
  title = {{GUIGuard}: Toward a General Framework for {Privacy-Preserving} {GUI} Agents},
  author = {Wang, Yanxi and Zhang, Zhiling and Zhou, Wenbo and Zhang, Weiming and Zhang, Jie and Zhu, Qiannan and Shi, Yu and Zheng, Shuxin and He, Jiyan},
  year = {2026},
  eprint = {2601.18842},
  archivePrefix = {arXiv},
  primaryClass = {cs.CR},
  doi = {10.48550/arXiv.2601.18842}
}

@misc{li2025uipro034,
  title = {{UIPro}: Unleashing Superior Interaction Capability For {GUI} Agents},
  author = {Li, Hongxin and Su, Jingran and Chen, Jingfan and Ju, Zheng and Chen, Yuntao and Li, Qing and Zhang, Zhaoxiang},
  year = {2025},
  eprint = {2509.17328},
  archivePrefix = {arXiv},
  primaryClass = {cs.CV},
  doi = {10.48550/arXiv.2509.17328}
}

@inproceedings{huang2025spiritsight035,
  title = {{SpiritSight} Agent: Advanced {GUI} Agent with One Look},
  author = {Huang, Zhiyuan and Cheng, Ziming and Pan, Junting and Hou, Zhaohui and Zhan, Mingjie},
  booktitle = {2025 IEEE/CVF Conference on Computer Vision and Pattern Recognition (CVPR)},
  year = {2025},
  pages = {29490--29500},
  doi = {10.1109/cvpr52734.2025.02746},
  url = {https://doi.org/10.1109/cvpr52734.2025.02746}
}

@misc{chen2025toward036,
  title = {Toward a {Human-Centered} Evaluation Framework for Trustworthy {LLM-Powered} {GUI} Agents},
  author = {Chen, Chaoran and Zhang, Zhiping and Khalilov, Ibrahim and Guo, Bingcan and Gebreegziabher, Simret A and Ye, Yanfang and Xiao, Ziang and Yao, Yaxing and Li, Tianshi and Li, Toby Jia-Jun},
  year = {2025},
  eprint = {2504.17934},
  archivePrefix = {arXiv},
  primaryClass = {cs.HC},
  doi = {10.48550/arXiv.2504.17934}
}

@misc{lyu2026personalalign037,
  title = {{PersonalAlign}: Hierarchical Implicit Intent Alignment for Personalized {GUI} Agent with {Long-Term} {User-Centric} Records},
  author = {Lyu, Yibo and Chen, Gongwei and Shao, Rui and Guan, Weili and Nie, Liqiang},
  year = {2026},
  eprint = {2601.09636},
  archivePrefix = {arXiv},
  primaryClass = {cs.AI},
  doi = {10.48550/arXiv.2601.09636}
}

@misc{xu2025deskvision038,
  title = {{DeskVision}: Large Scale Desktop Region Captioning for Advanced {GUI} Agents},
  author = {Xu, Yibin and Yang, Liang and Chen, Hao and Wang, Hua and Chen, Zhi and Tang, Yaohua},
  year = {2025},
  eprint = {2503.11170},
  archivePrefix = {arXiv},
  primaryClass = {cs.CL},
  doi = {10.48550/arXiv.2503.11170}
}

@misc{chen2025mpr039,
  title = {{MPR-GUI}: Benchmarking and Enhancing Multilingual Perception and Reasoning in {GUI} Agents},
  author = {Chen, Ruihan and Li, Qiming and Feng, Xiaocheng and Zhong, Weihong and Yang, Xiaoliang and Gu, Yuxuan and Zhou, Zekun and Lu, Yunfei and Ren, Haoyu and Chen, Kun and Tu, Dandan and Qin, Bing},
  year = {2025},
  eprint = {2512.00756},
  archivePrefix = {arXiv},
  primaryClass = {cs.AI},
  doi = {10.48550/arXiv.2512.00756}
}

@misc{tang2026agent040,
  title = {Agent Alpha: Tree Search Unifying Generation, Exploration and Evaluation for {Computer-Use} Agents},
  author = {Tang, Sizhe and Chen, Rongqian and Lan, Tian},
  year = {2026},
  eprint = {2602.02995},
  archivePrefix = {arXiv},
  primaryClass = {cs.AI},
  doi = {10.48550/arXiv.2602.02995}
}

@misc{yan2025lasm041,
  title = {{LaSM}: Layer-wise Scaling Mechanism for Defending Pop-up Attack on {GUI} Agents},
  author = {Yan, Zihe and Gui, Jiaping and Zhang, Zhuosheng and Liu, Gongshen},
  year = {2025},
  eprint = {2507.10610},
  archivePrefix = {arXiv},
  primaryClass = {cs.CR},
  doi = {10.48550/arXiv.2507.10610}
}

@misc{zhang2025does042,
  title = {Does {Chain-of-Thought} Reasoning Help Mobile {GUI} Agent? An Empirical Study},
  author = {Zhang, Li and Gao, Longxi and Xu, Mengwei},
  year = {2025},
  eprint = {2503.16788},
  archivePrefix = {arXiv},
  primaryClass = {cs.AI},
  doi = {10.48550/arXiv.2503.16788}
}

@inproceedings{lee2025verisafe043,
  title = {{VeriSafe} Agent: Safeguarding Mobile {GUI} Agent via Logic-based Action Verification},
  author = {Lee, Jungjae and Lee, Dongjae and Choi, Chihun and Im, Youngmin and Wi, Jaeyoung and Heo, Kihong and Oh, Sangeun and Lee, Sunjae and Shin, Insik},
  booktitle = {Proceedings of the 31st Annual International Conference on Mobile Computing and Networking},
  year = {2025},
  pages = {817--831},
  doi = {10.1145/3680207.3765248},
  url = {https://doi.org/10.1145/3680207.3765248}
}

@misc{liu2025learnact044,
  title = {{LearnAct}: {Few-Shot} Mobile {GUI} Agent with a Unified Demonstration Benchmark},
  author = {Liu, Guangyi and Zhao, Pengxiang and Liu, Liang and Chen, Zhiming and Chai, Yuxiang and Ren, Shuai and Wang, Hao and He, Shibo and Meng, Wenchao},
  year = {2025},
  eprint = {2504.13805},
  archivePrefix = {arXiv},
  primaryClass = {cs.HC},
  doi = {10.48550/arXiv.2504.13805}
}

@misc{kwak2025mega045,
  title = {{MEGA-GUI}: Multi-stage Enhanced Grounding Agents for {GUI} Elements},
  author = {Kwak, SeokJoo and Kim, Jihoon and Kim, Boyoun and Yoon, Jung Jae and Jang, Wooseok and Hong, Jeonghoon and Yang, Jaeho and Kwon, Yeong-Dae},
  year = {2025},
  eprint = {2511.13087},
  archivePrefix = {arXiv},
  primaryClass = {cs.AI},
  doi = {10.48550/arXiv.2511.13087}
}

@article{du2026test046,
  title = {{Test-Time} Reinforcement Learning for {GUI} Grounding via Region Consistency},
  author = {Du, Yong and Yan, Yuchen and Tang, Fei and Lu, Zhengxi and Zong, Chang and Lu, Weiming and Jiang, Shengpei and Shen, Yongliang},
  journal = {Proceedings of the AAAI Conference on Artificial Intelligence},
  year = {2026},
  volume = {40},
  number = {36},
  pages = {30593--30601},
  doi = {10.1609/aaai.v40i36.40314},
  url = {https://doi.org/10.1609/aaai.v40i36.40314}
}

@misc{chen2025the047,
  title = {The Obvious Invisible Threat: {LLM-Powered} {GUI} Agents' Vulnerability to {Fine-Print} Injections},
  author = {Chen, Chaoran and Zhang, Zhiping and Guo, Bingcan and Ma, Shang and Khalilov, Ibrahim and Gebreegziabher, Simret A and Ye, Yanfang and Xiao, Ziang and Yao, Yaxing and Li, Tianshi and Li, Toby Jia-Jun},
  year = {2025},
  eprint = {2504.11281},
  archivePrefix = {arXiv},
  primaryClass = {cs.HC},
  doi = {10.48550/arXiv.2504.11281}
}

@inproceedings{liu2026infiguiagent048,
  title = {{InfiGUIAgent}: A Multimodal Generalist {GUI} Agent with Native Reasoning and Reflection},
  author = {Liu, Yuhang and Li, Pengxiang and Wei, Zishu and Xie, Congkai and Hu, Xueyu and Xu, Xinchen and Zhang, Shengyu and Han, Xiaotian and Yang, Hongxia and Wu, Fei},
  booktitle = {Proceedings of the 19th Conference of the European Chapter of the Association for Computational Linguistics (Volume 1: Long Papers)},
  year = {2026},
  pages = {1035--1051},
  doi = {10.18653/v1/2026.eacl-long.45},
  url = {https://doi.org/10.18653/v1/2026.eacl-long.45}
}

@misc{nong2025craft049,
  title = {{CRAFT-GUI}: {Curriculum-Reinforced} Agent For {GUI} Tasks},
  author = {Nong, Songqin and Tang, Xiaoxuan and Xu, Jingxuan and Zhou, Sheng and Chen, Jianfeng and Jiang, Tao and Xu, Wenhao},
  year = {2025},
  eprint = {2508.11360},
  archivePrefix = {arXiv},
  primaryClass = {cs.AI},
  doi = {10.48550/arXiv.2508.11360}
}

@misc{dai2025advancing050,
  title = {Advancing Mobile {GUI} Agents: A {Verifier-Driven} Approach to Practical Deployment},
  author = {Dai, Gaole and Jiang, Shiqi and Cao, Ting and Li, Yuanchun and Yang, Yuqing and Tan, Rui and Li, Mo and Qiu, Lili},
  year = {2025},
  eprint = {2503.15937},
  archivePrefix = {arXiv},
  primaryClass = {cs.AI},
  doi = {10.48550/arXiv.2503.15937}
}

@inproceedings{li2025screenspot051,
  title = {{ScreenSpot-Pro}: {GUI} Grounding for Professional {High-Resolution} Computer Use},
  author = {Li, Kaixin and Meng, Ziyang and Lin, Hongzhan and Luo, Ziyang and Tian, Yuchen and Ma, Jing and Huang, Zhiyong and Chua, Tat-Seng},
  booktitle = {Proceedings of the 33rd ACM International Conference on Multimedia},
  year = {2025},
  pages = {8778--8786},
  doi = {10.1145/3746027.3755688},
  url = {https://doi.org/10.1145/3746027.3755688}
}

@inproceedings{chen2025ccagent052,
  title = {{CCAgent}: Coordinating Collaborative Data Scaling for Operating System Agents via {Web3}},
  author = {Chen, Liang and Zhao, Haozhe and Huang, Yinzhen and Luo, Yang and Lin, Tsekai and Xie, Weichu and Wu, Ruoyu and Wang, Peiyi and Xu, Runxin and Wu, Ming and Chang, Baobao},
  booktitle = {Proceedings of the 34th ACM International Conference on Information and Knowledge Management},
  year = {2025},
  pages = {280--290},
  doi = {10.1145/3746252.3761392},
  url = {https://doi.org/10.1145/3746252.3761392}
}

@inproceedings{chen2025pg053,
  title = {{PG-Agent}: An Agent Powered by Page Graph},
  author = {Chen, Weizhi and Wang, Ziwei and Yang, Leyang and Zhou, Sheng and Tang, Xiaoxuan and Bu, Jiajun and Li, Yong and Jiang, Wei},
  booktitle = {Proceedings of the 33rd ACM International Conference on Multimedia},
  year = {2025},
  pages = {6878--6887},
  doi = {10.1145/3746027.3755189},
  url = {https://doi.org/10.1145/3746027.3755189}
}

@inproceedings{wu2025mmpro054,
  title = {{MMPro}: A Decoupled {Perception-Thinking-Execution} Framework for Secure {GUI} Agent},
  author = {Wu, Benlong and Qi, Yuang and Shang, Xiuwei and Zhang, Weiming and Yu, Nenghai and Chen, Kejiang},
  booktitle = {Proceedings of the 33rd ACM International Conference on Multimedia},
  year = {2025},
  pages = {4679--4687},
  doi = {10.1145/3746027.3755553},
  url = {https://doi.org/10.1145/3746027.3755553}
}

@inproceedings{zhao2025appagent055,
  title = {{AppAgent-Pro}: A Proactive {GUI} Agent System for Multidomain Information Integration and User Assistance},
  author = {Zhao, Yuyang and Shi, Wentao and Feng, Fuli and He, Xiangnan},
  booktitle = {Proceedings of the 34th ACM International Conference on Information and Knowledge Management},
  year = {2025},
  pages = {6767--6771},
  doi = {10.1145/3746252.3761473},
  url = {https://doi.org/10.1145/3746252.3761473}
}

@inproceedings{gao2024assisteditor056,
  title = {{AssistEditor}: {Multi-Agent} Collaboration for {GUI} Workflow Automation in Video Creation},
  author = {Gao, Difei and Hu, Siyuan and Bai, Zechen and Lin, Qinghong and Shou, Mike Zheng},
  booktitle = {Proceedings of the 32nd ACM International Conference on Multimedia},
  year = {2024},
  pages = {11255--11257},
  doi = {10.1145/3664647.3684998},
  url = {https://doi.org/10.1145/3664647.3684998}
}

@inproceedings{liu2025hijacking058,
  title = {Hijacking {JARVIS}: Benchmarking Mobile {GUI} Agents against Unprivileged Third Parties},
  author = {Liu, Guohong and Ye, Jialei and Liu, Jiacheng and Li, Yuanchun and Liu, Wei and Gao, Pengzhi and Luan, Jian and Liu, Yunxin},
  booktitle = {Proceedings of the 2nd International Workshop on Edge and Mobile Foundation Models},
  year = {2025},
  pages = {12--18},
  doi = {10.1145/3737902.3768354},
  url = {https://doi.org/10.1145/3737902.3768354}
}

@inproceedings{jin2025scp060,
  title = {<scp>{ScreenLLM}:</scp> Stateful Screen Schema for Efficient Action Understanding and Prediction},
  author = {Jin, Yiqiao and Petrangeli, Stefano and Shen, Yu and Wu, Gang},
  booktitle = {Companion Proceedings of the ACM on Web Conference 2025},
  year = {2025},
  pages = {2008--2013},
  doi = {10.1145/3701716.3718379},
  url = {https://doi.org/10.1145/3701716.3718379}
}

@misc{zeng2025uitron061,
  title = {{UItron}: Foundational {GUI} Agent with Advanced Perception and Planning},
  author = {Zeng, Zhixiong and Huang, Jing and Zheng, Liming and Han, Wenkang and Zhong, Yufeng and Chen, Lei and Yang, Longrong and Chu, Yingjie and He, Yuzhi and Ma, Lin},
  year = {2025},
  eprint = {2508.21767},
  archivePrefix = {arXiv},
  primaryClass = {cs.CV},
  doi = {10.48550/arXiv.2508.21767}
}

@inproceedings{lin2025showui062,
  title = {{ShowUI}: One {Vision-Language-Action} Model for {GUI} Visual Agent},
  author = {Lin, Kevin Qinghong and Li, Linjie and Gao, Difei and Yang, Zhengyuan and Wu, Shiwei and Bai, Zechen and Lei, Stan Weixian and Wang, Lijuan and Shou, Mike Zheng},
  booktitle = {2025 IEEE/CVF Conference on Computer Vision and Pattern Recognition (CVPR)},
  year = {2025},
  pages = {19498--19508},
  doi = {10.1109/cvpr52734.2025.01816},
  url = {https://doi.org/10.1109/cvpr52734.2025.01816}
}

@misc{wu2024os063,
  title = {{OS-ATLAS}: A Foundation Action Model for Generalist {GUI} Agents},
  author = {Wu, Zhiyong and Wu, Zhenyu and Xu, Fangzhi and Wang, Yian and Sun, Qiushi and Jia, Chengyou and Cheng, Kanzhi and Ding, Zichen and Chen, Liheng and Liang, Paul Pu and Qiao, Yu},
  year = {2024},
  eprint = {2410.23218},
  archivePrefix = {arXiv},
  primaryClass = {cs.CL},
  doi = {10.48550/arXiv.2410.23218}
}

@inproceedings{wang2025ponder064,
  title = {Ponder \&amp; Press: Advancing Visual {GUI} Agent towards General Computer Control},
  author = {Wang, Yiqin and Zhang, Haoji and Tian, Jingqi and Tang, Yansong},
  booktitle = {Findings of the Association for Computational Linguistics: ACL 2025},
  year = {2025},
  pages = {1461--1473},
  doi = {10.18653/v1/2025.findings-acl.76},
  url = {https://doi.org/10.18653/v1/2025.findings-acl.76}
}

@misc{liu2025llm065,
  title = {{LLM-Powered} {GUI} Agents in Phone Automation: Surveying Progress and Prospects},
  author = {Liu, Guangyi and Zhao, Pengxiang and Liang, Yaozhen and Liu, Liang and Guo, Yaxuan and Xiao, Han and Lin, Weifeng and Chai, Yuxiang and Han, Yue and Ren, Shuai and Wang, Hao and Liang, Xiaoyu and Wang, WenHao and Wu, Tianze and Lu, Zhengxi and Chen, Siheng and LiLinghao and Wang, Hao and Xiong, Guanjing and Liu, Yong and Li, Hongsheng},
  year = {2025},
  eprint = {2504.19838},
  archivePrefix = {arXiv},
  primaryClass = {cs.HC},
  doi = {10.48550/arXiv.2504.19838}
}

@misc{luo2025gui066,
  title = {{GUI-R1} : A Generalist {R1-Style} {Vision-Language} Action Model For {GUI} Agents},
  author = {Luo, Run and Wang, Lu and He, Wanwei and Chen, Longze and Li, Jiaming and Xia, Xiaobo},
  year = {2025},
  eprint = {2504.10458},
  archivePrefix = {arXiv},
  primaryClass = {cs.CV},
  doi = {10.48550/arXiv.2504.10458}
}

@misc{xu2024aguvis067,
  title = {Aguvis: Unified Pure Vision Agents for Autonomous {GUI} Interaction},
  author = {Xu, Yiheng and Wang, Zekun and Wang, Junli and Lu, Dunjie and Xie, Tianbao and Saha, Amrita and Sahoo, Doyen and Yu, Tao and Xiong, Caiming},
  year = {2024},
  eprint = {2412.04454},
  archivePrefix = {arXiv},
  primaryClass = {cs.CL},
  doi = {10.48550/arXiv.2412.04454}
}

@misc{sun2024os068,
  title = {{OS-Genesis}: Automating {GUI} Agent Trajectory Construction via Reverse Task Synthesis},
  author = {Sun, Qiushi and Cheng, Kanzhi and Ding, Zichen and Jin, Chuanyang and Wang, Yian and Xu, Fangzhi and Wu, Zhenyu and Jia, Chengyou and Chen, Liheng and Liu, Zhoumianze and Kao, Ben and Li, Guohao and He, Junxian and Qiao, Yu and Wu, Zhiyong},
  year = {2024},
  eprint = {2412.19723},
  archivePrefix = {arXiv},
  primaryClass = {cs.AI},
  doi = {10.48550/arXiv.2412.19723}
}

@misc{wu2025gui069,
  title = {{GUI-Actor}: {Coordinate-Free} Visual Grounding for {GUI} Agents},
  author = {Wu, Qianhui and Cheng, Kanzhi and Yang, Rui and Zhang, Chaoyun and Yang, Jianwei and Jiang, Huiqiang and Mu, Jian and Peng, Baolin and Qiao, Bo and Tan, Reuben and Qin, Si and Liden, Lars and Lin, Qingwei and Zhang, Huan and Zhang, Tong and Zhang, Jianbing and Zhang, Dongmei and Gao, Jianfeng},
  year = {2025},
  eprint = {2506.03143},
  archivePrefix = {arXiv},
  primaryClass = {cs.CL},
  doi = {10.48550/arXiv.2506.03143}
}

@misc{zheng2025vem070,
  title = {{VEM}: {Environment-Free} Exploration for Training {GUI} Agent with Value Environment Model},
  author = {Zheng, Jiani and Wang, Lu and Yang, Fangkai and Zhang, Chaoyun and Mei, Lingrui and Yin, Wenjie and Lin, Qingwei and Zhang, Dongmei and Rajmohan, Saravan and Zhang, Qi},
  year = {2025},
  eprint = {2502.18906},
  archivePrefix = {arXiv},
  primaryClass = {cs.LG},
  doi = {10.48550/arXiv.2502.18906}
}

@misc{tang2025magicgui071,
  title = {{MagicGUI}: A Foundational Mobile {GUI} Agent with Scalable Data Pipeline and Reinforcement Fine-tuning},
  author = {Tang, Liujian and Dong, Shaokang and Huang, Yijia and Xiang, Minqi and Ruan, Hongtao and Wang, Bin and Li, Shuo and Xi, Zhiheng and Cao, Zhihui and Pang, Hailiang and Kong, Heng and Yang, He and Chai, Mingxu and Gao, Zhilin and Liu, Xingyu and Fu, Yingnan and Liu, Jiaming and Huang, Xuanjing and Jiang, Yu-Gang and Gui, Tao and Zhang, Qi and Wang, Kang and Zhang, Yunke and Wang, Yuran},
  year = {2025},
  eprint = {2508.03700},
  archivePrefix = {arXiv},
  primaryClass = {cs.HC},
  doi = {10.48550/arXiv.2508.03700}
}

@inproceedings{cheng2025os072,
  title = {{OS-Kairos}: Adaptive Interaction for {MLLM-Powered} {GUI} Agents},
  author = {Cheng, Pengzhou and Wu, Zheng and Wu, Zongru and Ju, Tianjie and Zhang, Aston and Zhang, Zhuosheng and Liu, Gongshen},
  booktitle = {Findings of the Association for Computational Linguistics: ACL 2025},
  year = {2025},
  pages = {6701--6725},
  doi = {10.18653/v1/2025.findings-acl.348},
  url = {https://doi.org/10.18653/v1/2025.findings-acl.348}
}

@misc{jiang2025appagentx073,
  title = {{AppAgentX}: Evolving {GUI} Agents as Proficient Smartphone Users},
  author = {Jiang, Wenjia and Zhuang, Yangyang and Song, Chenxi and Yang, Xu and Zhou, Joey Tianyi and Zhang, Chi},
  year = {2025},
  eprint = {2503.02268},
  archivePrefix = {arXiv},
  primaryClass = {cs.AI},
  doi = {10.48550/arXiv.2503.02268}
}

@misc{chai2025a3074,
  title = {{A3}: Android Agent Arena for Mobile {GUI} Agents with {Essential-State} Procedural Evaluation},
  author = {Chai, Yuxiang and Tang, Shunye and Xiao, Han and Lin, Weifeng and Li, Hanhao and Zhang, Jiayu and Liu, Liang and Zhao, Pengxiang and Liu, Guangyi and Wang, Guozhi and Ren, Shuai and Han, Rongduo and Zhang, Haining and Huang, Siyuan and Li, Hongsheng},
  year = {2025},
  eprint = {2501.01149},
  archivePrefix = {arXiv},
  primaryClass = {cs.AI},
  doi = {10.48550/arXiv.2501.01149}
}

@misc{li2025a075,
  title = {A Survey on {GUI} Agents with Foundation Models Enhanced by Reinforcement Learning},
  author = {Li, Jiahao and Huang, Kaer},
  year = {2025},
  eprint = {2504.20464},
  archivePrefix = {arXiv},
  primaryClass = {cs.AI},
  doi = {10.48550/arXiv.2504.20464}
}

@misc{huang2025scaletrack076,
  title = {{ScaleTrack}: Scaling and back-tracking Automated {GUI} Agents},
  author = {Huang, Jing and Zeng, Zhixiong and Han, Wenkang and Zhong, Yufeng and Zheng, Liming and Fu, Shuai and Chen, Jingyuan and Ma, Lin},
  year = {2025},
  eprint = {2505.00416},
  archivePrefix = {arXiv},
  primaryClass = {cs.AI},
  doi = {10.48550/arXiv.2505.00416}
}

@misc{liu2025infigui077,
  title = {{InfiGUI-R1}: Advancing Multimodal {GUI} Agents from Reactive Actors to Deliberative Reasoners},
  author = {Liu, Yuhang and Li, Pengxiang and Xie, Congkai and Hu, Xavier and Han, Xiaotian and Zhang, Shengyu and Yang, Hongxia and Wu, Fei},
  year = {2025},
  eprint = {2504.14239},
  archivePrefix = {arXiv},
  primaryClass = {cs.AI},
  doi = {10.48550/arXiv.2504.14239}
}

@article{chen2026gui078,
  title = {{GUI-Eyes}: {Tool-Augmented} Perception for Visual Grounding in {GUI} Agents},
  author = {Chen, Chen and Shao, Jiawei and Lu, Dakuan and Hu, Haoyi and Liu, Xiangcheng and Yao, Hantao and Liu, Wu},
  journal = {Proceedings of the AAAI Conference on Artificial Intelligence},
  year = {2026},
  volume = {40},
  number = {35},
  pages = {29350--29358},
  doi = {10.1609/aaai.v40i35.40175},
  url = {https://doi.org/10.1609/aaai.v40i35.40175}
}

@misc{yang2025macosworld079,
  title = {macOSWorld: A Multilingual Interactive Benchmark for {GUI} Agents},
  author = {Yang, Pei and Ci, Hai and Shou, Mike Zheng},
  year = {2025},
  eprint = {2506.04135},
  archivePrefix = {arXiv},
  primaryClass = {cs.AI},
  doi = {10.48550/arXiv.2506.04135}
}

@article{dong2025mt081,
  title = {{MT-Agent}: Constructing a {GUI} Agent via Modality Enhancement and {Text-Guided} Fusion},
  author = {Dong, Jinhan and Jin, Lei and Zhang, Zhihong and Tang, Wei and Zhang, Runqing and Xu, Liqiang and Xing, Junliang},
  journal = {IEEE Internet of Things Journal},
  year = {2025},
  pages = {1},
  doi = {10.1109/jiot.2025.3600573},
  url = {https://doi.org/10.1109/jiot.2025.3600573}
}

@article{lu2026ui082,
  title = {{UI-R1}: Enhancing Efficient Action Prediction of {GUI} Agents by Reinforcement Learning},
  author = {Lu, Zhengxi and Chai, Yuxiang and Guo, Yaxuan and Yin, Xi and Liu, Liang and Wang, Hao and Xiao, Han and Ren, Shuai and Zhao, Pengxiang and Liu, Guangyi and Xiong, Guanjing and Li, Hongsheng},
  journal = {Proceedings of the AAAI Conference on Artificial Intelligence},
  year = {2026},
  volume = {40},
  number = {21},
  pages = {17608--17616},
  doi = {10.1609/aaai.v40i21.38816},
  url = {https://doi.org/10.1609/aaai.v40i21.38816}
}

@misc{zhang2026omegause083,
  title = {{OmegaUse}: Building a {General-Purpose} {GUI} Agent for Autonomous Task Execution},
  author = {Zhang, Le and Xiao, Yixiong and Lu, Xinjiang and Cao, Jingjia and Zhao, Yusai and Zhou, Jingbo and An, Lang and Feng, Zikan and Sha, Wanxiang and Shi, Yu and Xiao, Congxi and Xiong, Jian and Zhang, Yankai and Wu, Hua and Wang, Haifeng},
  year = {2026},
  eprint = {2601.20380},
  archivePrefix = {arXiv},
  primaryClass = {cs.AI},
  doi = {10.48550/arXiv.2601.20380}
}

@inproceedings{xia2025g084,
  title = {{G-TADS}: {GUI} {Task-Ability} Decoupling Strategy for {High-Adaptability} Multimodal Intelligent Agents},
  author = {Xia, Zhiqiang and Zhang, Xinyuan and Li, Yang and Liu, Yuchen and Shi, Runyu and Xu, Jiaming},
  booktitle = {2025 IEEE International Conference on Multimedia and Expo (ICME)},
  year = {2025},
  pages = {1--6},
  doi = {10.1109/icme59968.2025.11209200},
  url = {https://doi.org/10.1109/icme59968.2025.11209200}
}

@inproceedings{singh2025trishul085,
  title = {Trishul: Towards Region Identification and Screen Hierarchy Understanding for Large {VLM} Based {GUI} Agents},
  author = {Singh, Kunal and Singh, Shreyas and Khanna, Mukund},
  booktitle = {2025 IEEE/CVF Conference on Computer Vision and Pattern Recognition Workshops (CVPRW)},
  year = {2025},
  pages = {170--179},
  doi = {10.1109/cvprw67362.2025.00022},
  url = {https://doi.org/10.1109/cvprw67362.2025.00022}
}

@inproceedings{ye2025gg086,
  title = {{GG-R1}: A {Fine-Grained} {RFT} Approach for {GUI} Grounding},
  author = {Ye, Jian and Zhao, Xin and Feng, Xuanzhen and Gao, Xiang and Zhang, Xin and Li, Lanting and Hong, Wentao},
  booktitle = {2025 9th IEEE International Conference on Network Intelligence and Digital Content (IC-NIDC)},
  year = {2025},
  pages = {16--20},
  doi = {10.1109/ic-nidc67200.2025.11390507},
  url = {https://doi.org/10.1109/ic-nidc67200.2025.11390507}
}

@inproceedings{wen2025autodroid087,
  title = {{AutoDroid-V2}: Boosting {SLM-based} {GUI} Agents via Code Generation},
  author = {Wen, Hao and Tian, Shizuo and Pavlov, Borislav and Du, Wenjie and Li, Yixuan and Chang, Ge and Zhao, Shanhui and Liu, Jiacheng and Liu, Yunxin and Zhang, Ya-Qin and Li, Yuanchun},
  booktitle = {Proceedings of the 23rd Annual International Conference on Mobile Systems, Applications and Services},
  year = {2025},
  pages = {223--235},
  doi = {10.1145/3711875.3729134},
  url = {https://doi.org/10.1145/3711875.3729134}
}

@misc{xiong2025gui088,
  title = {{GUI-PRA}: Process Reward Agent for {GUI} Tasks},
  author = {Xiong, Tao and Hu, Xavier and Chen, Yurun and Liu, Yuhang and Wu, Changqiao and Gao, Pengzhi and Liu, Wei and Luan, Jian and Zhang, Shengyu},
  year = {2025},
  eprint = {2509.23263},
  archivePrefix = {arXiv},
  primaryClass = {cs.AI},
  doi = {10.48550/arXiv.2509.23263}
}

@inproceedings{wu2025backtrackagent089,
  title = {{BacktrackAgent}: Enhancing {GUI} Agent with Error Detection and Backtracking Mechanism},
  author = {Wu, Qinzhuo and Gao, Pengzhi and Liu, Wei and Luan, Jian},
  booktitle = {Proceedings of the 2025 Conference on Empirical Methods in Natural Language Processing},
  year = {2025},
  pages = {4250--4272},
  doi = {10.18653/v1/2025.emnlp-main.212},
  url = {https://doi.org/10.18653/v1/2025.emnlp-main.212}
}

@misc{hao2025uncertainty090,
  title = {{Uncertainty-Aware} {GUI} Agent: Adaptive Perception through Component Recommendation and {Human-in-the-Loop} Refinement},
  author = {Hao, Chao and Wang, Shuai and Zhou, Kaiwen},
  year = {2025},
  eprint = {2508.04025},
  archivePrefix = {arXiv},
  primaryClass = {cs.AI},
  doi = {10.48550/arXiv.2508.04025}
}

@misc{zhang2025breaking091,
  title = {Breaking the Data Barrier -- Building {GUI} Agents Through Task Generalization},
  author = {Zhang, Junlei and Ding, Zichen and Ma, Chang and Chen, Zijie and Sun, Qiushi and Lan, Zhenzhong and He, Junxian},
  year = {2025},
  eprint = {2504.10127},
  archivePrefix = {arXiv},
  primaryClass = {cs.AI},
  doi = {10.48550/arXiv.2504.10127}
}

@misc{yuan2025enhancing092,
  title = {Enhancing Visual Grounding for {GUI} Agents via {Self-Evolutionary} Reinforcement Learning},
  author = {Yuan, Xinbin and Zhang, Jian and Li, Kaixin and Cai, Zhuoxuan and Yao, Lujian and Chen, Jie and Wang, Enguang and Hou, Qibin and Chen, Jinwei and Jiang, Peng-Tao and Li, Bo},
  year = {2025},
  eprint = {2505.12370},
  archivePrefix = {arXiv},
  primaryClass = {cs.AI},
  doi = {10.48550/arXiv.2505.12370}
}

@inproceedings{zhang2025ui093,
  title = {{UI-Hawk}: Unleashing the Screen Stream Understanding for Mobile {GUI} Agents},
  author = {Zhang, Jiwen and Yu, Ya-Qi and Liao, Minghui and Li, WenTao and Wu, Jihao and Wei, Zhongyu},
  booktitle = {Proceedings of the 2025 Conference on Empirical Methods in Natural Language Processing},
  year = {2025},
  pages = {18228--18247},
  doi = {10.18653/v1/2025.emnlp-main.920},
  url = {https://doi.org/10.18653/v1/2025.emnlp-main.920}
}

@misc{xiao2025ui094,
  title = {{UI-Genie}: A {Self-Improving} Approach for Iteratively Boosting {MLLM-based} Mobile {GUI} Agents},
  author = {Xiao, Han and Wang, Guozhi and Chai, Yuxiang and Lu, Zimu and Lin, Weifeng and He, Hao and Fan, Lue and Bian, Liuyang and Hu, Rui and Liu, Liang and Ren, Shuai and Wen, Yafei and Chen, Xiaoxin and Zhou, Aojun and Li, Hongsheng},
  year = {2025},
  eprint = {2505.21496},
  archivePrefix = {arXiv},
  primaryClass = {cs.CL},
  doi = {10.48550/arXiv.2505.21496}
}

@article{ma2026beyond095,
  title = {Beyond element-level understanding: Explicit relational understanding for {GUI} agents},
  author = {Ma, Longhui and Zhao, Di and Wang, Siwei and Lv, Zhao and Wang, Miao},
  journal = {Pattern Recognition},
  year = {2026},
  volume = {176},
  pages = {113262},
  doi = {10.1016/j.patcog.2026.113262},
  url = {https://doi.org/10.1016/j.patcog.2026.113262}
}

@misc{lu2025arpo096,
  title = {{ARPO}: {End-to-End} Policy Optimization for {GUI} Agents with Experience Replay},
  author = {Lu, Fanbin and Zhong, Zhisheng and Liu, Shu and Fu, Chi-Wing and Jia, Jiaya},
  year = {2025},
  eprint = {2505.16282},
  archivePrefix = {arXiv},
  primaryClass = {cs.CV},
  doi = {10.48550/arXiv.2505.16282}
}

@misc{han2025uitron097,
  title = {{UITron-Speech}: Towards Automated {GUI} Agents Based on Speech Instructions},
  author = {Han, Wenkang and Zeng, Zhixiong and Huang, Jing and Jiang, Shu and Zheng, Liming and Yang, Longrong and Qiu, Haibo and Yao, Chang and Chen, Jingyuan and Ma, Lin},
  year = {2025},
  eprint = {2506.11127},
  archivePrefix = {arXiv},
  primaryClass = {cs.CL},
  doi = {10.48550/arXiv.2506.11127}
}

@misc{xie2025mirage098,
  title = {{Mirage-1}: Augmenting and Updating {GUI} Agent with Hierarchical Multimodal Skills},
  author = {Xie, Yuquan and Li, Zaijing and Shao, Rui and Chen, Gongwei and Zhou, Kaiwen and Li, Yinchuan and Jiang, Dongmei and Nie, Liqiang},
  year = {2025},
  eprint = {2506.10387},
  archivePrefix = {arXiv},
  primaryClass = {cs.AI},
  doi = {10.48550/arXiv.2506.10387}
}

@misc{xu2025mobilerl099,
  title = {{MobileRL}: Online Agentic Reinforcement Learning for Mobile {GUI} Agents},
  author = {Xu, Yifan and Liu, Xiao and Liu, Xinghan and Fu, Jiaqi and Zhang, Hanchen and Jing, Bohao and Zhang, Shudan and Wang, Yuting and Zhao, Wenyi and Dong, Yuxiao},
  year = {2025},
  eprint = {2509.18119},
  archivePrefix = {arXiv},
  primaryClass = {cs.LG},
  doi = {10.48550/arXiv.2509.18119}
}

@misc{zhou2025hiconagent100,
  title = {{HiconAgent}: History Context-aware Policy Optimization for {GUI} Agents},
  author = {Zhou, Xurui and Chen, Gongwei and Xie, Yuquan and Li, Zaijing and Zhou, Kaiwen and Wang, Shuai and Yang, Shuo and Tian, Zhuotao and Shao, Rui},
  year = {2025},
  eprint = {2512.01763},
  archivePrefix = {arXiv},
  primaryClass = {cs.CV},
  doi = {10.48550/arXiv.2512.01763}
}

@misc{agashe2024agent101,
  title = {Agent S: An Open Agentic Framework that Uses Computers Like a Human},
  author = {Agashe, Saaket and Han, Jiuzhou and Gan, Shuyu and Yang, Jiachen and Li, Ang and Wang, Xin Eric},
  year = {2024},
  eprint = {2410.08164},
  archivePrefix = {arXiv},
  primaryClass = {cs.AI},
  doi = {10.48550/arXiv.2410.08164}
}

@inproceedings{xu2025retrieval102,
  title = {Retrieval-augmented {GUI} Agents with Generative Guidelines},
  author = {Xu, Ran and Ma, Kaixin and Yu, Wenhao and Zhang, Hongming and Ho, Joyce C. and Yang, Carl and Yu, Dong},
  booktitle = {Proceedings of the 2025 Conference on Empirical Methods in Natural Language Processing},
  year = {2025},
  pages = {17877--17886},
  doi = {10.18653/v1/2025.emnlp-main.902},
  url = {https://doi.org/10.18653/v1/2025.emnlp-main.902}
}

@misc{yang2026gui103,
  title = {{GUI-Libra}: Training Native {GUI} Agents to Reason and Act with Action-aware Supervision and Partially Verifiable {RL}},
  author = {Yang, Rui and Wu, Qianhui and Wang, Zhaoyang and Chen, Hanyang and Yang, Ke and Cheng, Hao and Yao, Huaxiu and Peng, Baolin and Zhang, Huan and Gao, Jianfeng and Zhang, Tong},
  year = {2026},
  eprint = {2602.22190},
  archivePrefix = {arXiv},
  primaryClass = {cs.LG},
  doi = {10.48550/arXiv.2602.22190}
}

@misc{tang2025lpo104,
  title = {{LPO}: Towards Accurate {GUI} Agent Interaction via Location Preference Optimization},
  author = {Tang, Jiaqi and Xia, Yu and Wu, Yi-Feng and Hu, Yuwei and Chen, Yuhui and Chen, Qing-Guo and Xu, Xiaogang and Wu, Xiangyu and Lu, Hao and Ma, Yanqing and Lu, Shiyin and Chen, Qifeng},
  year = {2025},
  eprint = {2506.09373},
  archivePrefix = {arXiv},
  primaryClass = {cs.LG},
  doi = {10.48550/arXiv.2506.09373}
}

@misc{lian2025ui105,
  title = {{UI-AGILE}: Advancing {GUI} Agents with Effective Reinforcement Learning and Precise {Inference-Time} Grounding},
  author = {Lian, Shuquan and Wu, Yuhang and Ma, Jia and Ding, Yifan and Song, Zihan and Chen, Bingqi and Zheng, Xiawu and Li, Hui and Ji, Rongrong},
  year = {2025},
  eprint = {2507.22025},
  archivePrefix = {arXiv},
  primaryClass = {cs.AI},
  doi = {10.48550/arXiv.2507.22025}
}

@misc{tian2025agentprog106,
  title = {{AgentProg}: Empowering {Long-Horizon} {GUI} Agents with {Program-Guided} Context Management},
  author = {Tian, Shizuo and Wen, Hao and Chen, Yuxuan and Liu, Jiacheng and Zhao, Shanhui and Liu, Guohong and Ren, Ju and Liu, Yunxin and Li, Yuanchun},
  year = {2025},
  eprint = {2512.10371},
  archivePrefix = {arXiv},
  primaryClass = {cs.AI},
  doi = {10.48550/arXiv.2512.10371}
}

@misc{huang2025gui107,
  title = {{GUI-KV}: Efficient {GUI} Agents via {KV} Cache with {Spatio-Temporal} Awareness},
  author = {Huang, Kung-Hsiang and Qiu, Haoyi and Dai, Yutong and Xiong, Caiming and Wu, Chien-Sheng},
  year = {2025},
  eprint = {2510.00536},
  archivePrefix = {arXiv},
  primaryClass = {cs.CL},
  doi = {10.48550/arXiv.2510.00536}
}

@misc{anand2025afragent108,
  title = {{AFRAgent} : An Adaptive Feature Renormalization Based High Resolution Aware {GUI} agent},
  author = {Anand, Neeraj and Jain, Rishabh and Patnaik, Sohan and Krishnamurthy, Balaji and Sarkar, Mausoom},
  year = {2025},
  eprint = {2512.00846},
  archivePrefix = {arXiv},
  primaryClass = {cs.CV},
  doi = {10.48550/arXiv.2512.00846}
}

@misc{zhang2024ufo109,
  title = {{UFO}: A {UI-Focused} Agent for Windows {OS} Interaction},
  author = {Zhang, Chaoyun and Li, Liqun and He, Shilin and Zhang, Xu and Qiao, Bo and Qin, Si and Ma, Minghua and Kang, Yu and Lin, Qingwei and Rajmohan, Saravan and Zhang, Dongmei and Zhang, Qi},
  year = {2024},
  eprint = {2402.07939},
  archivePrefix = {arXiv},
  primaryClass = {cs.HC},
  doi = {10.48550/arXiv.2402.07939}
}

@inproceedings{wang2025inreact110,
  title = {{INREACT}: An {Inspire-Then-Reinforce} Training Framework For Multimodal {GUI} Agent},
  author = {Wang, Yuanlei and Zhang, Liuzhou and Luo, Haohao and Shen, Ying},
  booktitle = {Findings of the Association for Computational Linguistics: EMNLP 2025},
  year = {2025},
  pages = {9148--9160},
  doi = {10.18653/v1/2025.findings-emnlp.486},
  url = {https://doi.org/10.18653/v1/2025.findings-emnlp.486}
}

@inproceedings{tao2025understanding111,
  title = {Understanding {GUI} Agent Localization Biases through Logit Sharpness},
  author = {Tao, Xingjian and Wang, Yiwei and Cai, Yujun and Yang, Zhicheng and Tang, Jing},
  booktitle = {Findings of the Association for Computational Linguistics: EMNLP 2025},
  year = {2025},
  pages = {23361--23374},
  doi = {10.18653/v1/2025.findings-emnlp.1268},
  url = {https://doi.org/10.18653/v1/2025.findings-emnlp.1268}
}

@inproceedings{guan2025kg112,
  title = {{KG-RAG}: Enhancing {GUI} Agent {Decision-Making} via Knowledge {Graph-Driven} {Retrieval-Augmented} Generation},
  author = {Guan, Ziyi and Li, Jason Chun Lok and Hou, Zhijian and Zhang, Pingping and Xu, Donglai and Zhao, Yuzhi and Wu, Mengyang and Chen, Jinpeng and Nguyen, Thanh-Toan and Xian, Pengfei and Ma, Wenao and Qin, Shengchao and Chesi, Graziano and Wong, Ngai},
  booktitle = {Proceedings of the 2025 Conference on Empirical Methods in Natural Language Processing},
  year = {2025},
  pages = {5396--5405},
  doi = {10.18653/v1/2025.emnlp-main.274},
  url = {https://doi.org/10.18653/v1/2025.emnlp-main.274}
}

@inproceedings{chai2025amex113,
  title = {{AMEX}: Android Multi-annotation Expo Dataset for Mobile {GUI} Agents},
  author = {Chai, Yuxiang and Huang, Siyuan and Niu, Yazhe and Xiao, Han and Liu, Liang and Wang, Guozhi and Zhang, Dingyu and Ren, Shuai and Li, Hongsheng},
  booktitle = {Findings of the Association for Computational Linguistics: ACL 2025},
  year = {2025},
  pages = {2138--2156},
  doi = {10.18653/v1/2025.findings-acl.110},
  url = {https://doi.org/10.18653/v1/2025.findings-acl.110}
}

@inproceedings{zhao2025cola114,
  title = {{COLA}: Collaborative {Multi-Agent} Framework with Dynamic Task Scheduling for {GUI} Automation},
  author = {Zhao, Di and Ma, Longhui and Wang, Siwei and Wang, Miao and Lv, Zhao},
  booktitle = {Proceedings of the 2025 Conference on Empirical Methods in Natural Language Processing},
  year = {2025},
  pages = {4570--4593},
  doi = {10.18653/v1/2025.emnlp-main.227},
  url = {https://doi.org/10.18653/v1/2025.emnlp-main.227}
}

@inproceedings{wang2025fedmabench115,
  title = {{FedMABench}: Benchmarking Mobile {GUI} Agents on Decentralized Heterogeneous User Data},
  author = {Wang, WenHao and Yu, Zijie and Ye, Rui and Zhang, Jianqing and Liu, Guangyi and Liu, Liang and Chen, Siheng and Wang, Yanfeng},
  booktitle = {Proceedings of the 2025 Conference on Empirical Methods in Natural Language Processing},
  year = {2025},
  pages = {26398--26419},
  doi = {10.18653/v1/2025.emnlp-main.1341},
  url = {https://doi.org/10.18653/v1/2025.emnlp-main.1341}
}

@inproceedings{ma2024coco116,
  title = {{CoCo-Agent}: A Comprehensive Cognitive {MLLM} Agent for Smartphone {GUI} Automation},
  author = {Ma, Xinbei and Zhang, Zhuosheng and Zhao, Hai},
  booktitle = {Findings of the Association for Computational Linguistics ACL 2024},
  year = {2024},
  pages = {9097--9110},
  doi = {10.18653/v1/2024.findings-acl.539},
  url = {https://doi.org/10.18653/v1/2024.findings-acl.539}
}

@inproceedings{ma2025caution117,
  title = {Caution for the Environment: Multimodal {LLM} Agents are Susceptible to Environmental Distractions},
  author = {Ma, Xinbei and Wang, Yiting and Yao, Yao and Yuan, Tongxin and Zhang, Aston and Zhang, Zhuosheng and Zhao, Hai},
  booktitle = {Proceedings of the 63rd Annual Meeting of the Association for Computational Linguistics (Volume 1: Long Papers)},
  year = {2025},
  pages = {22324--22339},
  doi = {10.18653/v1/2025.acl-long.1087},
  url = {https://doi.org/10.18653/v1/2025.acl-long.1087}
}

@inproceedings{cheng2025hidden118,
  title = {Hidden Ghost Hand: Unveiling Backdoor Vulnerabilities in {MLLM-Powered} Mobile {GUI} Agents},
  author = {Cheng, Pengzhou and Hu, Haowen and Wu, Zheng and Wu, Zongru and Ju, Tianjie and Ding, Daizong and Zhang, Zhuosheng and Liu, Gongshen},
  booktitle = {Findings of the Association for Computational Linguistics: EMNLP 2025},
  year = {2025},
  pages = {7781--7805},
  doi = {10.18653/v1/2025.findings-emnlp.411},
  url = {https://doi.org/10.18653/v1/2025.findings-emnlp.411}
}

@inproceedings{park2025r119,
  title = {{R-VLM}: {Region-Aware} Vision Language Model for Precise {GUI} Grounding},
  author = {Park, Joonhyung and Tang, Peng and Das, Sagnik and Appalaraju, Srikar and Singh, Kunwar Yashraj and Manmatha, R. and Ghadar, Shabnam},
  booktitle = {Findings of the Association for Computational Linguistics: ACL 2025},
  year = {2025},
  pages = {9669--9685},
  doi = {10.18653/v1/2025.findings-acl.501},
  url = {https://doi.org/10.18653/v1/2025.findings-acl.501}
}

@inproceedings{ahn2025flashadventure120,
  title = {{FlashAdventure}: A Benchmark for {GUI} Agents Solving Full Story Arcs in Diverse Adventure Games},
  author = {Ahn, Jaewoo and Kim, Junseo and Yun, Heeseung and Son, Jaehyeon and Park, Dongmin and Cho, Jaewoong and Kim, Gunhee},
  booktitle = {Proceedings of the 2025 Conference on Empirical Methods in Natural Language Processing},
  year = {2025},
  pages = {23365--23395},
  doi = {10.18653/v1/2025.emnlp-main.1192},
  url = {https://doi.org/10.18653/v1/2025.emnlp-main.1192}
}

@misc{fan2024read121,
  title = {Read Anywhere Pointed: Layout-aware {GUI} Screen Reading with {Tree-of-Lens} Grounding},
  author = {Fan, Yue and Ding, Lei and Kuo, Ching-Chen and Jiang, Shan and Zhao, Yang and Guan, Xinze and Yang, Jie and Zhang, Yi and Wang, Xin Eric},
  year = {2024},
  eprint = {2406.19263},
  archivePrefix = {arXiv},
  primaryClass = {cs.CL},
  doi = {10.48550/arXiv.2406.19263}
}

@inproceedings{zhang2024you122,
  title = {You Only Look at Screens: Multimodal {Chain-of-Action} Agents},
  author = {Zhang, Zhuosheng and Zhang, Aston},
  booktitle = {Findings of the Association for Computational Linguistics ACL 2024},
  year = {2024},
  pages = {3132--3149},
  doi = {10.18653/v1/2024.findings-acl.186},
  url = {https://doi.org/10.18653/v1/2024.findings-acl.186}
}

@inproceedings{lu2025transbench123,
  title = {{TransBench}: Breaking Barriers for Transferable Graphical User Interface Agents in Dynamic Digital Environments},
  author = {Lu, Yuheng and Yu, Qian and Wang, Hongru and Liu, Zeming and Su, Wei and Liu, Yanping and Guo, Yuhang and Liang, Maocheng and Wang, Yunhong and Wang, Haifeng},
  booktitle = {Findings of the Association for Computational Linguistics: ACL 2025},
  year = {2025},
  pages = {12464--12478},
  doi = {10.18653/v1/2025.findings-acl.645},
  url = {https://doi.org/10.18653/v1/2025.findings-acl.645}
}

@misc{liu2018reinforcement124,
  title = {Reinforcement Learning on Web Interfaces Using {Workflow-Guided} Exploration},
  author = {Liu, Evan Zheran and Guu, Kelvin and Pasupat, Panupong and Shi, Tianlin and Liang, Percy},
  year = {2018},
  eprint = {1802.08802},
  archivePrefix = {arXiv},
  primaryClass = {cs.AI},
  doi = {10.48550/arXiv.1802.08802}
}

@misc{xu2021grounding125,
  title = {Grounding {Open-Domain} Instructions to Automate Web Support Tasks},
  author = {Xu, Nancy and Masling, Sam and Du, Michael and Campagna, Giovanni and Heck, Larry and Landay, James and Lam, Monica S},
  year = {2021},
  eprint = {2103.16057},
  archivePrefix = {arXiv},
  primaryClass = {cs.CL},
  doi = {10.48550/arXiv.2103.16057}
}

@inproceedings{yao2022webshop126,
  title = {{WebShop}: Towards Scalable {Real-World} Web Interaction with Grounded Language Agents},
  author = {Yao, Shunyu and Chen, Howard and Yang, John and Narasimhan, Karthik},
  booktitle = {Advances in Neural Information Processing Systems 35},
  year = {2022},
  pages = {20744--20757},
  doi = {10.52202/068431-1508},
  url = {https://doi.org/10.52202/068431-1508}
}

@inproceedings{deng2023mind2web127,
  title = {{Mind2Web}: Towards a Generalist Agent for the Web},
  author = {Deng, Xiang and Gu, Yu and Zheng, Boyuan and Chen, Shijie and Stevens, Sam and Wang, Boshi and Sun, Huan and Su, Yu},
  booktitle = {Advances in Neural Information Processing Systems 36},
  year = {2023},
  pages = {28091--28114},
  doi = {10.52202/075280-1220},
  url = {https://doi.org/10.52202/075280-1220}
}

@misc{zhou2023webarena128,
  title = {{WebArena}: A Realistic Web Environment for Building Autonomous Agents},
  author = {Zhou, Shuyan and Xu, Frank F. and Zhu, Hao and Zhou, Xuhui and Lo, Robert and Sridhar, Abishek and Cheng, Xianyi and Ou, Tianyue and Bisk, Yonatan and Fried, Daniel and Alon, Uri and Neubig, Graham},
  year = {2023},
  eprint = {2307.13854},
  archivePrefix = {arXiv},
  primaryClass = {cs.AI},
  doi = {10.48550/arXiv.2307.13854}
}

@inproceedings{koh2024visualwebarena129,
  title = {{VisualWebArena}: Evaluating Multimodal Agents on Realistic Visual Web Tasks},
  author = {Koh, Jing Yu and Lo, Robert and Jang, Lawrence and Duvvur, Vikram and Lim, Ming and Huang, Po-Yu and Neubig, Graham and Zhou, Shuyan and Salakhutdinov, Russ and Fried, Daniel},
  booktitle = {Proceedings of the 62nd Annual Meeting of the Association for Computational Linguistics (Volume 1: Long Papers)},
  year = {2024},
  pages = {881--905},
  doi = {10.18653/v1/2024.acl-long.50},
  url = {https://doi.org/10.18653/v1/2024.acl-long.50}
}

@inproceedings{deng2024on130,
  title = {On the Multi-turn Instruction Following for Conversational Web Agents},
  author = {Deng, Yang and Zhang, Xuan and Zhang, Wenxuan and Yuan, Yifei and Ng, See-Kiong and Chua, Tat-Seng},
  booktitle = {Proceedings of the 62nd Annual Meeting of the Association for Computational Linguistics (Volume 1: Long Papers)},
  year = {2024},
  pages = {8795--8812},
  doi = {10.18653/v1/2024.acl-long.477},
  url = {https://doi.org/10.18653/v1/2024.acl-long.477}
}

@inproceedings{tian2025mmina131,
  title = {{MMInA}: Benchmarking Multihop Multimodal Internet Agents},
  author = {Tian, Shulin and Zhang, Ziniu and Chen, Liangyu and Liu, Ziwei},
  booktitle = {Findings of the Association for Computational Linguistics: ACL 2025},
  year = {2025},
  pages = {13682--13697},
  doi = {10.18653/v1/2025.findings-acl.703},
  url = {https://doi.org/10.18653/v1/2025.findings-acl.703}
}

@misc{pan2024webcanvas132,
  title = {{WebCanvas}: Benchmarking Web Agents in Online Environments},
  author = {Pan, Yichen and Kong, Dehan and Zhou, Sida and Cui, Cheng and Leng, Yifei and Jiang, Bing and Liu, Hangyu and Shang, Yanyi and Zhou, Shuyan and Wu, Tongshuang and Wu, Zhengyang},
  year = {2024},
  eprint = {2406.12373},
  archivePrefix = {arXiv},
  primaryClass = {cs.CL},
  doi = {10.48550/arXiv.2406.12373}
}

@misc{drouin2024workarena134,
  title = {{WorkArena}: How Capable Are Web Agents at Solving Common Knowledge Work Tasks?},
  author = {Drouin, Alexandre and Gasse, Maxime and Caccia, Massimo and Laradji, Issam H. and Verme, Manuel Del and Marty, Tom and Boisvert, Léo and Thakkar, Megh and Cappart, Quentin and Vazquez, David and Chapados, Nicolas and Lacoste, Alexandre},
  year = {2024},
  eprint = {2403.07718},
  archivePrefix = {arXiv},
  primaryClass = {cs.LG},
  doi = {10.48550/arXiv.2403.07718}
}

@misc{jang2024videowebarena135,
  title = {{VideoWebArena}: Evaluating Long Context Multimodal Agents with Video Understanding Web Tasks},
  author = {Jang, Lawrence and Li, Yinheng and Zhao, Dan and Ding, Charles and Lin, Justin and Liang, Paul Pu and Bonatti, Rogerio and Koishida, Kazuhito},
  year = {2024},
  eprint = {2410.19100},
  archivePrefix = {arXiv},
  primaryClass = {cs.CV},
  doi = {10.48550/arXiv.2410.19100}
}

@inproceedings{ma2025caution136,
  title = {Caution for the Environment: Multimodal {LLM} Agents are Susceptible to Environmental Distractions},
  author = {Ma, Xinbei and Wang, Yiting and Yao, Yao and Yuan, Tongxin and Zhang, Aston and Zhang, Zhuosheng and Zhao, Hai},
  booktitle = {Proceedings of the 63rd Annual Meeting of the Association for Computational Linguistics (Volume 1: Long Papers)},
  year = {2025},
  pages = {22324--22339},
  doi = {10.18653/v1/2025.acl-long.1087},
  url = {https://doi.org/10.18653/v1/2025.acl-long.1087}
}

@article{chen2024webvln137,
  title = {{WebVLN}: {Vision-and-Language} Navigation on Websites},
  author = {Chen, Qi and Pitawela, Dileepa and Zhao, Chongyang and Zhou, Gengze and Chen, Hsiang-Ting and Wu, Qi},
  journal = {Proceedings of the AAAI Conference on Artificial Intelligence},
  year = {2024},
  volume = {38},
  number = {2},
  pages = {1165--1173},
  doi = {10.1609/aaai.v38i2.27878},
  url = {https://doi.org/10.1609/aaai.v38i2.27878}
}

@misc{lu2024weblinx138,
  title = {{WebLINX}: {Real-World} Website Navigation with {Multi-Turn} Dialogue},
  author = {Lù, Xing Han and Kasner, Zdeněk and Reddy, Siva},
  year = {2024},
  eprint = {2402.05930},
  archivePrefix = {arXiv},
  primaryClass = {cs.CL},
  doi = {10.48550/arXiv.2402.05930}
}

@misc{levy2024st139,
  title = {{ST-WebAgentBench}: A Benchmark for Evaluating Safety and Trustworthiness in Web Agents},
  author = {Levy, Ido and Wiesel, Ben and Marreed, Sami and Oved, Alon and Yaeli, Avi and Shlomov, Segev},
  year = {2024},
  eprint = {2410.06703},
  archivePrefix = {arXiv},
  primaryClass = {cs.AI},
  doi = {10.48550/arXiv.2410.06703}
}

@misc{furuta2023exposing140,
  title = {Exposing Limitations of Language Model Agents in {Sequential-Task} Compositions on the Web},
  author = {Furuta, Hiroki and Matsuo, Yutaka and Faust, Aleksandra and Gur, Izzeddin},
  year = {2023},
  eprint = {2311.18751},
  archivePrefix = {arXiv},
  primaryClass = {cs.LG},
  doi = {10.48550/arXiv.2311.18751}
}

@inproceedings{xu2025turkingbench141,
  title = {{TurkingBench}: A Challenge Benchmark for Web Agents},
  author = {Xu, Kevin and Kordi, Yeganeh and Nayak, Tanay and Asija, Adi and Wang, Yizhong and Sanders, Kate and Byerly, Adam and Zhang, Jingyu and Durme, Benjamin Van and Khashabi, Daniel},
  booktitle = {Proceedings of the 2025 Conference of the Nations of the Americas Chapter of the Association for Computational Linguistics: Human Language Technologies (Volume 1: Long Papers)},
  year = {2025},
  pages = {3694--3710},
  doi = {10.18653/v1/2025.naacl-long.188},
  url = {https://doi.org/10.18653/v1/2025.naacl-long.188}
}

@misc{shahbandeh2024naviqate142,
  title = {{NaviQAte}: {Functionality-Guided} Web Application Navigation},
  author = {Shahbandeh, Mobina and Alian, Parsa and Nashid, Noor and Mesbah, Ali},
  year = {2024},
  eprint = {2409.10741},
  archivePrefix = {arXiv},
  primaryClass = {cs.SE},
  doi = {10.48550/arXiv.2409.10741}
}

@misc{liu2024visualwebbench143,
  title = {{VisualWebBench}: How Far Have Multimodal {LLMs} Evolved in Web Page Understanding and Grounding?},
  author = {Liu, Junpeng and Song, Yifan and Lin, Bill Yuchen and Lam, Wai and Neubig, Graham and Li, Yuanzhi and Yue, Xiang},
  year = {2024},
  eprint = {2404.05955},
  archivePrefix = {arXiv},
  primaryClass = {cs.CL},
  doi = {10.48550/arXiv.2404.05955}
}

@inproceedings{wornow2024wonderbread144,
  title = {{WONDERBREAD}: A Benchmark for Evaluating Multimodal Foundation Models on Business Process Management Tasks},
  author = {Wornow, Michael and Narayan, Avanika and Viggiano, Ben and Khare, Ishan and Verma, Tathagat and Thompson, Tibor and Hernandez, Miguel and Sundar, Sudharsan and Trujillo, Chloe and Chawla, Krrish and Lu, Rongfei and Shen, Justin and Nagaraj, Divya and Martinez, Joshua and Agrawal, Vardhan and Hudson, Althea and Shah, Nigam and Ré, Christopher},
  booktitle = {Advances in Neural Information Processing Systems 37},
  year = {2024},
  pages = {115963--116021},
  doi = {10.52202/079017-3682},
  url = {https://doi.org/10.52202/079017-3682}
}

@inproceedings{zheng2024webolympus145,
  title = {{WebOlympus}: An Open Platform for Web Agents on Live Websites},
  author = {Zheng, Boyuan and Gou, Boyu and Salisbury, Scott and Du, Zheng and Sun, Huan and Su, Yu},
  booktitle = {Proceedings of the 2024 Conference on Empirical Methods in Natural Language Processing: System Demonstrations},
  year = {2024},
  pages = {187--197},
  doi = {10.18653/v1/2024.emnlp-demo.20},
  url = {https://doi.org/10.18653/v1/2024.emnlp-demo.20}
}

@misc{zhang2023mobile146,
  title = {{Mobile-Env}: Building Qualified Evaluation Benchmarks for {LLM-GUI} Interaction},
  author = {Zhang, Danyang and Shen, Zhennan and Xie, Rui and Zhang, Situo and Xie, Tianbao and Zhao, Zihan and Chen, Siyuan and Chen, Lu and Xu, Hongshen and Cao, Ruisheng and Yu, Kai},
  year = {2023},
  eprint = {2305.08144},
  archivePrefix = {arXiv},
  primaryClass = {cs.AI},
  doi = {10.48550/arXiv.2305.08144}
}

@misc{lee2024benchmarking147,
  title = {Benchmarking Mobile Device Control Agents across Diverse Configurations},
  author = {Lee, Juyong and Min, Taywon and An, Minyong and Hahm, Dongyoon and Lee, Haeone and Kim, Changyeon and Lee, Kimin},
  year = {2024},
  eprint = {2404.16660},
  archivePrefix = {arXiv},
  primaryClass = {cs.HC},
  doi = {10.48550/arXiv.2404.16660}
}

@misc{rawles2024androidworld148,
  title = {{AndroidWorld}: A Dynamic Benchmarking Environment for Autonomous Agents},
  author = {Rawles, Christopher and Clinckemaillie, Sarah and Chang, Yifan and Waltz, Jonathan and Lau, Gabrielle and Fair, Marybeth and Li, Alice and Bishop, William and Li, Wei and Campbell-Ajala, Folawiyo and Toyama, Daniel and Berry, Robert and Tyamagundlu, Divya and Lillicrap, Timothy and Riva, Oriana},
  year = {2024},
  eprint = {2405.14573},
  archivePrefix = {arXiv},
  primaryClass = {cs.AI},
  doi = {10.48550/arXiv.2405.14573}
}

@inproceedings{wen2024autodroid149,
  title = {{AutoDroid}: {LLM-powered} Task Automation in Android},
  author = {Wen, Hao and Li, Yuanchun and Liu, Guohong and Zhao, Shanhui and Yu, Tao and Li, Toby Jia-Jun and Jiang, Shiqi and Liu, Yunhao and Zhang, Yaqin and Liu, Yunxin},
  booktitle = {Proceedings of the 30th Annual International Conference on Mobile Computing and Networking},
  year = {2024},
  pages = {543--557},
  doi = {10.1145/3636534.3649379},
  url = {https://doi.org/10.1145/3636534.3649379}
}

@misc{rawles2023android150,
  title = {Android in the Wild: A {Large-Scale} Dataset for Android Device Control},
  author = {Rawles, Christopher and Li, Alice and Rodriguez, Daniel and Riva, Oriana and Lillicrap, Timothy},
  year = {2023},
  eprint = {2307.10088},
  archivePrefix = {arXiv},
  primaryClass = {cs.LG},
  doi = {10.48550/arXiv.2307.10088}
}

@inproceedings{xing2024understanding151,
  title = {Understanding the Weakness of Large Language Model Agents within a Complex Android Environment},
  author = {Xing, Mingzhe and Zhang, Rongkai and Xue, Hui and Chen, Qi and Yang, Fan and Xiao, Zhen},
  booktitle = {Proceedings of the 30th ACM SIGKDD Conference on Knowledge Discovery and Data Mining},
  year = {2024},
  pages = {6061--6072},
  doi = {10.1145/3637528.3671650},
  url = {https://doi.org/10.1145/3637528.3671650}
}

@inproceedings{xu2025androidlab152,
  title = {{AndroidLab}: Training and Systematic Benchmarking of Android Autonomous Agents},
  author = {Xu, Yifan and Liu, Xiao and Sun, Xueqiao and Cheng, Siyi and Yu, Hao and Lai, Hanyu and Zhang, Shudan and Zhang, Dan and Tang, Jie and Dong, Yuxiao},
  booktitle = {Proceedings of the 63rd Annual Meeting of the Association for Computational Linguistics (Volume 1: Long Papers)},
  year = {2025},
  pages = {2144--2166},
  doi = {10.18653/v1/2025.acl-long.107},
  url = {https://doi.org/10.18653/v1/2025.acl-long.107}
}

@inproceedings{zhang2024llamatouch153,
  title = {{LlamaTouch}: A Faithful and Scalable Testbed for Mobile {UI} Task Automation},
  author = {Zhang, Li and Wang, Shihe and Jia, Xianqing and Zheng, Zhihan and Yan, Yunhe and Gao, Longxi and Li, Yuanchun and Xu, Mengwei},
  booktitle = {Proceedings of the 37th Annual ACM Symposium on User Interface Software and Technology},
  year = {2024},
  pages = {1--13},
  doi = {10.1145/3654777.3676382},
  url = {https://doi.org/10.1145/3654777.3676382}
}

@misc{wang2024mobileagentbench154,
  title = {{MobileAgentBench}: An Efficient and {User-Friendly} Benchmark for Mobile {LLM} Agents},
  author = {Wang, Luyuan and Deng, Yongyu and Zha, Yiwei and Mao, Guodong and Wang, Qinmin and Min, Tianchen and Chen, Wei and Chen, Shoufa},
  year = {2024},
  eprint = {2406.08184},
  archivePrefix = {arXiv},
  primaryClass = {cs.AI},
  doi = {10.48550/arXiv.2406.08184}
}

@inproceedings{deng2024mobile155,
  title = {{Mobile-Bench}: An Evaluation Benchmark for {LLM-based} Mobile Agents},
  author = {Deng, Shihan and Xu, Weikai and Sun, Hongda and Liu, Wei and Tan, Tao and Liujianfeng, Liujianfeng and Li, Ang and Luan, Jian and Wang, Bin and Yan, Rui and Shang, Shuo},
  booktitle = {Proceedings of the 62nd Annual Meeting of the Association for Computational Linguistics (Volume 1: Long Papers)},
  year = {2024},
  pages = {8813--8831},
  doi = {10.18653/v1/2024.acl-long.478},
  url = {https://doi.org/10.18653/v1/2024.acl-long.478}
}

@article{lee2026mobilesafetybench156,
  title = {{MobileSafetyBench}: Evaluating Safety of Autonomous Agents in Mobile Device Control},
  author = {Lee, Juyong and Hahm, Dongyoon and Choi, June Suk and Knox, W. Bradley and Lee, Kimin},
  journal = {Proceedings of the AAAI Conference on Artificial Intelligence},
  year = {2026},
  volume = {40},
  number = {44},
  pages = {37565--37573},
  doi = {10.1609/aaai.v40i44.41090},
  url = {https://doi.org/10.1609/aaai.v40i44.41090}
}

@misc{chen2024spa157,
  title = {{SPA-Bench}: A Comprehensive Benchmark for {SmartPhone} Agent Evaluation},
  author = {Chen, Jingxuan and Yuen, Derek and Xie, Bin and Yang, Yuhao and Chen, Gongwei and Wu, Zhihao and Yixing, Li and Zhou, Xurui and Liu, Weiwen and Wang, Shuai and Zhou, Kaiwen and Shao, Rui and Nie, Liqiang and Wang, Yasheng and Hao, Jianye and Wang, Jun and Shao, Kun},
  year = {2024},
  eprint = {2410.15164},
  archivePrefix = {arXiv},
  primaryClass = {cs.AI},
  doi = {10.48550/arXiv.2410.15164}
}

@misc{liu2024visualagentbench158,
  title = {{VisualAgentBench}: Towards Large Multimodal Models as Visual Foundation Agents},
  author = {Liu, Xiao and Zhang, Tianjie and Gu, Yu and Iong, Iat Long and Xu, Yifan and Song, Xixuan and Zhang, Shudan and Lai, Hanyu and Liu, Xinyi and Zhao, Hanlin and Sun, Jiadai and Yang, Xinyue and Yang, Yu and Qi, Zehan and Yao, Shuntian and Sun, Xueqiao and Cheng, Siyi and Zheng, Qinkai and Yu, Hao and Zhang, Hanchen and Hong, Wenyi and Ding, Ming and Pan, Lihang and Gu, Xiaotao and Zeng, Aohan and Du, Zhengxiao and Song, Chan Hee and Su, Yu and Dong, Yuxiao and Tang, Jie},
  year = {2024},
  eprint = {2408.06327},
  archivePrefix = {arXiv},
  primaryClass = {cs.AI},
  doi = {10.48550/arXiv.2408.06327}
}

@inproceedings{xie2024osworld159,
  title = {{OSWorld}: Benchmarking Multimodal Agents for {Open-Ended} Tasks in Real Computer Environments},
  author = {Xie, Tianbao and Zhang, Danyang and Chen, Jixuan and Li, Xiaochuan and Zhao, Siheng and Cao, Ruisheng and Hua, Toh and Cheng, Zhoujun and Shin, Dongchan and Lei, Fangyu and Liu, Yitao and Xu, Yiheng and Zhou, Shuyan and Savarese, Silvio and Xiong, Caiming and Zhong, Victor and Yu, Tao},
  booktitle = {Advances in Neural Information Processing Systems 37},
  year = {2024},
  pages = {52040--52094},
  doi = {10.52202/079017-1650},
  url = {https://doi.org/10.52202/079017-1650}
}

@misc{bonatti2024windows160,
  title = {Windows Agent Arena: Evaluating {Multi-Modal} {OS} Agents at Scale},
  author = {Bonatti, Rogerio and Zhao, Dan and Bonacci, Francesco and Dupont, Dillon and Abdali, Sara and Li, Yinheng and Lu, Yadong and Wagle, Justin and Koishida, Kazuhito and Bucker, Arthur and Jang, Lawrence and Hui, Zack},
  year = {2024},
  eprint = {2409.08264},
  archivePrefix = {arXiv},
  primaryClass = {cs.AI},
  doi = {10.48550/arXiv.2409.08264}
}

@article{kapoor2024omniact161,
  title = {{OmniACT}: A Dataset and Benchmark for Enabling Multimodal Generalist Autonomous Agents for Desktop and Web},
  author = {Kapoor, Raghav and Butala, Yash Parag and Russak, Melisa and Koh, Jing Yu and Kamble, Kiran and AlShikh, Waseem and Salakhutdinov, Ruslan},
  journal = {Lecture Notes in Computer Science},
  year = {2024},
  pages = {161--178},
  doi = {10.1007/978-3-031-73113-6\_10},
  url = {https://doi.org/10.1007/978-3-031-73113-6\_10}
}

@inproceedings{lin2024videogui162,
  title = {{VideoGUI}: A Benchmark for {GUI} Automation from Instructional Videos},
  author = {Lin, Kevin and Li, Linjie and Gao, Difei and Wu, Qinchen and Yan, Mingyi and Yang, Zhengyuan and Wang, Lijuan and Shou, Mike},
  booktitle = {Advances in Neural Information Processing Systems 37},
  year = {2024},
  pages = {69329--69360},
  doi = {10.52202/079017-2214},
  url = {https://doi.org/10.52202/079017-2214}
}

@inproceedings{cao2024spider2163,
  title = {{Spider2-V}: How Far Are Multimodal Agents From Automating Data Science and Engineering Workflows?},
  author = {Cao, Ruisheng and Lei, Fangyu and Wu, Haoyuan and Chen, Jixuan and Fu, Yeqiao and Gao, Hongcheng and Xiong, Xinzhuang and Zhang, Hanchong and Mao, Yuchen and Hu, Wenjing and Xie, Tianbao and Xu, Hongshen and Zhang, Danyang and Wang, Sida and Sun, Ruoxi and Yin, Pengcheng and Xiong, Caiming and Ni, Ansong and Liu, Qian and Zhong, Victor and Chen, Lu and Yu, Kai and Yu, Tao},
  booktitle = {Advances in Neural Information Processing Systems 37},
  year = {2024},
  pages = {107703--107744},
  doi = {10.52202/079017-3421},
  url = {https://doi.org/10.52202/079017-3421}
}

@misc{wu2024gui164,
  title = {{GUI} Action Narrator: Where and When Did That Action Take Place?},
  author = {Wu, Qinchen and Gao, Difei and Lin, Kevin Qinghong and Wu, Zhuoyu and Guo, Xiangwu and Li, Peiran and Zhang, Weichen and Wang, Hengxu and Shou, Mike Zheng},
  year = {2024},
  eprint = {2406.13719},
  archivePrefix = {arXiv},
  primaryClass = {cs.CV},
  doi = {10.48550/arXiv.2406.13719}
}

@misc{wang2024officebench165,
  title = {{OfficeBench}: Benchmarking Language Agents across Multiple Applications for Office Automation},
  author = {Wang, Zilong and Cui, Yuedong and Zhong, Li and Zhang, Zimin and Yin, Da and Lin, Bill Yuchen and Shang, Jingbo},
  year = {2024},
  eprint = {2407.19056},
  archivePrefix = {arXiv},
  primaryClass = {cs.CL},
  doi = {10.48550/arXiv.2407.19056}
}

@misc{zheng2024agentstudio166,
  title = {{AgentStudio}: A Toolkit for Building General Virtual Agents},
  author = {Zheng, Longtao and Huang, Zhiyuan and Xue, Zhenghai and Wang, Xinrun and An, Bo and Yan, Shuicheng},
  year = {2024},
  eprint = {2403.17918},
  archivePrefix = {arXiv},
  primaryClass = {cs.AI},
  doi = {10.48550/arXiv.2403.17918}
}

@inproceedings{xu2025crab167,
  title = {{CRAB}: Cross-environment Agent Benchmark for Multimodal Language Model Agents},
  author = {Xu, Tianqi and Chen, Linyao and Wu, Dai-Jie and Chen, Yanjun and Zhang, Zecheng and Yao, Xiang and Xie, Zhiqiang and Chen, Yongchao and Liu, Shilong and Qian, Bochen and Yang, Anjie and Jin, Zhaoxuan and Deng, Jianbo and Torr, Philip and Ghanem, Bernard and Li, Guohao},
  booktitle = {Findings of the Association for Computational Linguistics: ACL 2025},
  year = {2025},
  pages = {21607--21647},
  doi = {10.18653/v1/2025.findings-acl.1113},
  url = {https://doi.org/10.18653/v1/2025.findings-acl.1113}
}

@misc{chezelles2024the168,
  title = {The {BrowserGym} Ecosystem for Web Agent Research},
  author = {Chezelles, Thibault Le Sellier De and Gasse, Maxime and Drouin, Alexandre and Caccia, Massimo and Boisvert, Léo and Thakkar, Megh and Marty, Tom and Assouel, Rim and Shayegan, Sahar Omidi and Jang, Lawrence Keunho and Lù, Xing Han and Yoran, Ori and Kong, Dehan and Xu, Frank F. and Reddy, Siva and Cappart, Quentin and Neubig, Graham and Salakhutdinov, Ruslan and Chapados, Nicolas and Lacoste, Alexandre},
  year = {2024},
  eprint = {2412.05467},
  archivePrefix = {arXiv},
  primaryClass = {cs.LG},
  doi = {10.48550/arXiv.2412.05467}
}

@misc{wu2025webwalker170,
  title = {{WebWalker}: Benchmarking {LLMs} in Web Traversal},
  author = {Wu, Jialong and Yin, Wenbiao and Jiang, Yong and Wang, Zhenglin and Xi, Zekun and Fang, Runnan and Zhang, Linhai and He, Yulan and Zhou, Deyu and Xie, Pengjun and Huang, Fei},
  year = {2025},
  eprint = {2501.07572},
  archivePrefix = {arXiv},
  primaryClass = {cs.CL},
  doi = {10.48550/arXiv.2501.07572}
}

@misc{ran2025beyond171,
  title = {Beyond Pass or Fail: {Multi-Dimensional} Benchmarking of Foundation Models for Goal-based Mobile {UI} Navigation},
  author = {Ran, Dezhi and Wu, Mengzhou and Yu, Hao and Li, Yuetong and Ren, Jun and Cao, Yuan and Zeng, Xia and Lu, Haochuan and Xu, Zexin and Xu, Mengqian and Su, Ting and Yao, Liangchao and Xiong, Ting and Yang, Wei and Deng, Yuetang and Marron, Assaf and Harel, David and Xie, Tao},
  year = {2025},
  eprint = {2501.02863},
  archivePrefix = {arXiv},
  primaryClass = {cs.SE},
  doi = {10.48550/arXiv.2501.02863}
}

@misc{thomas2025webgames173,
  title = {{WebGames}: Challenging {General-Purpose} {Web-Browsing} {AI} Agents},
  author = {Thomas, George and Chan, Alex J. and Kang, Jikun and Wu, Wenqi and Christianos, Filippos and Greenlee, Fraser and Toulis, Andy and Purtorab, Marvin},
  year = {2025},
  eprint = {2502.18356},
  archivePrefix = {arXiv},
  primaryClass = {cs.LG},
  doi = {10.48550/arXiv.2502.18356}
}

@misc{sun2025autoeval174,
  title = {{AutoEval}: A Practical Framework for Autonomous Evaluation of Mobile Agents},
  author = {Sun, Jiahui and Hua, Zhichao and Xia, Yubin},
  year = {2025},
  eprint = {2503.02403},
  archivePrefix = {arXiv},
  primaryClass = {cs.AI},
  doi = {10.48550/arXiv.2503.02403}
}

@misc{tur2025safearena175,
  title = {{SafeArena}: Evaluating the Safety of Autonomous Web Agents},
  author = {Tur, Ada Defne and Meade, Nicholas and Lù, Xing Han and Zambrano, Alejandra and Patel, Arkil and Durmus, Esin and Gella, Spandana and Stańczak, Karolina and Reddy, Siva},
  year = {2025},
  eprint = {2503.04957},
  archivePrefix = {arXiv},
  primaryClass = {cs.LG},
  doi = {10.48550/arXiv.2503.04957}
}

@misc{nayak2025ui177,
  title = {{UI-Vision}: A Desktop-centric {GUI} Benchmark for Visual Perception and Interaction},
  author = {Nayak, Shravan and Jian, Xiangru and Lin, Kevin Qinghong and Rodriguez, Juan A. and Kalsi, Montek and Awal, Rabiul and Chapados, Nicolas and Özsu, M. Tamer and Agrawal, Aishwarya and Vazquez, David and Pal, Christopher and Taslakian, Perouz and Gella, Spandana and Rajeswar, Sai},
  year = {2025},
  eprint = {2503.15661},
  archivePrefix = {arXiv},
  primaryClass = {cs.CV},
  doi = {10.48550/arXiv.2503.15661}
}

@misc{xue2025an178,
  title = {An Illusion of Progress? Assessing the Current State of Web Agents},
  author = {Xue, Tianci and Qi, Weijian and Shi, Tianneng and Song, Chan Hee and Gou, Boyu and Song, Dawn and Sun, Huan and Su, Yu},
  year = {2025},
  eprint = {2504.01382},
  archivePrefix = {arXiv},
  primaryClass = {cs.AI},
  doi = {10.48550/arXiv.2504.01382}
}

@misc{zharmagambetov2025agentdam179,
  title = {{AgentDAM}: Privacy Leakage Evaluation for Autonomous Web Agents},
  author = {Zharmagambetov, Arman and Guo, Chuan and Evtimov, Ivan and Pavlova, Maya and Salakhutdinov, Ruslan and Chaudhuri, Kamalika},
  year = {2025},
  eprint = {2503.09780},
  archivePrefix = {arXiv},
  primaryClass = {cs.AI},
  doi = {10.48550/arXiv.2503.09780}
}

@inproceedings{wang2026computer180,
  title = {Computer Agent Arena: Toward {Human-Centric} Evaluation and Analysis of {Computer-Use} Agents},
  author = {Wang, Bowen and Wang, Xinyuan and Deng, Jiaqi and Xie, Tianbao and Li, Ryan and Zhang, Yanzhe and Wang, Junli and Lu, Dunjie and Gong, Zicheng and Li, Gavin and Hua, Toh Jing and Chiang, Wei-Lin and Stoica, Ion and Yang, Diyi and Su, Yu and Zhang, Yi and Wang, Zhiguo and Zhong, Victor and Yu, Tao},
  booktitle = {International Conference on Learning Representations},
  year = {2026},
  url = {https://openreview.net/forum?id=3x4SDbXbgl}
}

@misc{garg2025real181,
  title = {{REAL}: Benchmarking Autonomous Agents on Deterministic Simulations of Real Websites},
  author = {Garg, Divyansh and VanWeelden, Shaun and Caples, Diego and Draguns, Andis and Ravi, Nikil and Putta, Pranav and Garg, Naman and Abraham, Tomas and Lara, Michael and Lopez, Federico and Liu, James and Gundawar, Atharva and Hebbar, Prannay and Joo, Youngchul and Gu, Jindong and London, Charles and Witt, Christian Schroeder de and Motwani, Sumeet},
  year = {2025},
  eprint = {2504.11543},
  archivePrefix = {arXiv},
  primaryClass = {cs.AI},
  doi = {10.48550/arXiv.2504.11543}
}

@misc{song2025bearcubs182,
  title = {{BEARCUBS}: A benchmark for computer-using web agents},
  author = {Song, Yixiao and Thai, Katherine and Pham, Chau Minh and Chang, Yapei and Nadaf, Mazin and Iyyer, Mohit},
  year = {2025},
  eprint = {2503.07919},
  archivePrefix = {arXiv},
  primaryClass = {cs.AI},
  doi = {10.48550/arXiv.2503.07919}
}

@misc{sun2025scienceboard183,
  title = {{ScienceBoard}: Evaluating Multimodal Autonomous Agents in Realistic Scientific Workflows},
  author = {Sun, Qiushi and Liu, Zhoumianze and Ma, Chang and Ding, Zichen and Xu, Fangzhi and Yin, Zhangyue and Zhao, Haiteng and Wu, Zhenyu and Cheng, Kanzhi and Liu, Zhaoyang and Wang, Jianing and Li, Qintong and Tang, Xiangru and Xie, Tianbao and Feng, Xiachong and Li, Xiang and Kao, Ben and Wang, Wenhai and Qi, Biqing and Kong, Lingpeng and Wu, Zhiyong},
  year = {2025},
  eprint = {2505.19897},
  archivePrefix = {arXiv},
  primaryClass = {cs.AI},
  doi = {10.48550/arXiv.2505.19897}
}

@misc{chae2024web184,
  title = {Web Agents with World Models: Learning and Leveraging Environment Dynamics in Web Navigation},
  author = {Chae, Hyungjoo and Kim, Namyoung and Ong, Kai Tzu-iunn and Gwak, Minju and Song, Gwanwoo and Kim, Jihoon and Kim, Sunghwan and Lee, Dongha and Yeo, Jinyoung},
  year = {2024},
  eprint = {2410.13232},
  archivePrefix = {arXiv},
  primaryClass = {cs.CL},
  doi = {10.48550/arXiv.2410.13232}
}

@misc{gur2023a185,
  title = {A {Real-World} {WebAgent} with Planning, Long Context Understanding, and Program Synthesis},
  author = {Gur, Izzeddin and Furuta, Hiroki and Huang, Austin and Safdari, Mustafa and Matsuo, Yutaka and Eck, Douglas and Faust, Aleksandra},
  year = {2023},
  eprint = {2307.12856},
  archivePrefix = {arXiv},
  primaryClass = {cs.LG},
  doi = {10.48550/arXiv.2307.12856}
}

@misc{ma2023laser186,
  title = {{LASER}: {LLM} Agent with {State-Space} Exploration for Web Navigation},
  author = {Ma, Kaixin and Zhang, Hongming and Wang, Hongwei and Pan, Xiaoman and Yu, Wenhao and Yu, Dong},
  year = {2023},
  eprint = {2309.08172},
  archivePrefix = {arXiv},
  primaryClass = {cs.CL},
  doi = {10.48550/arXiv.2309.08172}
}

@inproceedings{he2024webvoyager187,
  title = {{WebVoyager}: Building an {End-to-End} Web Agent with Large Multimodal Models},
  author = {He, Hongliang and Yao, Wenlin and Ma, Kaixin and Yu, Wenhao and Dai, Yong and Zhang, Hongming and Lan, Zhenzhong and Yu, Dong},
  booktitle = {Proceedings of the 62nd Annual Meeting of the Association for Computational Linguistics (Volume 1: Long Papers)},
  year = {2024},
  pages = {6864--6890},
  doi = {10.18653/v1/2024.acl-long.371},
  url = {https://doi.org/10.18653/v1/2024.acl-long.371}
}

@inproceedings{lai2024autowebglm188,
  title = {{AutoWebGLM}: A Large Language Model-based Web Navigating Agent},
  author = {Lai, Hanyu and Liu, Xiao and Iong, Iat Long and Yao, Shuntian and Chen, Yuxuan and Shen, Pengbo and Yu, Hao and Zhang, Hanchen and Zhang, Xiaohan and Dong, Yuxiao and Tang, Jie},
  booktitle = {Proceedings of the 30th ACM SIGKDD Conference on Knowledge Discovery and Data Mining},
  year = {2024},
  pages = {5295--5306},
  doi = {10.1145/3637528.3671620},
  url = {https://doi.org/10.1145/3637528.3671620}
}

@misc{xie2023openagents189,
  title = {{OpenAgents}: An Open Platform for Language Agents in the Wild},
  author = {Xie, Tianbao and Zhou, Fan and Cheng, Zhoujun and Shi, Peng and Weng, Luoxuan and Liu, Yitao and Hua, Toh Jing and Zhao, Junning and Liu, Qian and Liu, Che and Liu, Leo Z. and Xu, Yiheng and Su, Hongjin and Shin, Dongchan and Xiong, Caiming and Yu, Tao},
  year = {2023},
  eprint = {2310.10634},
  archivePrefix = {arXiv},
  primaryClass = {cs.CL},
  doi = {10.48550/arXiv.2310.10634}
}

@misc{zheng2024gpt190,
  title = {{GPT-4V}(ision) is a Generalist Web Agent, if Grounded},
  author = {Zheng, Boyuan and Gou, Boyu and Kil, Jihyung and Sun, Huan and Su, Yu},
  year = {2024},
  eprint = {2401.01614},
  archivePrefix = {arXiv},
  primaryClass = {cs.IR},
  doi = {10.48550/arXiv.2401.01614}
}

@misc{kil2024dual191,
  title = {{Dual-View} Visual Contextualization for Web Navigation},
  author = {Kil, Jihyung and Song, Chan Hee and Zheng, Boyuan and Deng, Xiang and Su, Yu and Chao, Wei-Lun},
  year = {2024},
  eprint = {2402.04476},
  archivePrefix = {arXiv},
  primaryClass = {cs.CV},
  doi = {10.48550/arXiv.2402.04476}
}

@misc{abuelsaad2024agent192,
  title = {{Agent-E}: From Autonomous Web Navigation to Foundational Design Principles in Agentic Systems},
  author = {Abuelsaad, Tamer and Akkil, Deepak and Dey, Prasenjit and Jagmohan, Ashish and Vempaty, Aditya and Kokku, Ravi},
  year = {2024},
  eprint = {2407.13032},
  archivePrefix = {arXiv},
  primaryClass = {cs.AI},
  doi = {10.48550/arXiv.2407.13032}
}

@misc{koh2024tree193,
  title = {Tree Search for Language Model Agents},
  author = {Koh, Jing Yu and McAleer, Stephen and Fried, Daniel and Salakhutdinov, Ruslan},
  year = {2024},
  eprint = {2407.01476},
  archivePrefix = {arXiv},
  primaryClass = {cs.AI},
  doi = {10.48550/arXiv.2407.01476}
}

@article{zhang2025webpilot194,
  title = {{WebPilot}: A Versatile and Autonomous {Multi-Agent} System for Web Task Execution with Strategic Exploration},
  author = {Zhang, Yao and Ma, Zijian and Ma, Yunpu and Han, Zhen and Wu, Yu and Tresp, Volker},
  journal = {Proceedings of the AAAI Conference on Artificial Intelligence},
  year = {2025},
  volume = {39},
  number = {22},
  pages = {23378--23386},
  doi = {10.1609/aaai.v39i22.34505},
  url = {https://doi.org/10.1609/aaai.v39i22.34505}
}

@misc{song2024beyond195,
  title = {Beyond Browsing: {API-Based} Web Agents},
  author = {Song, Yueqi and Xu, Frank and Zhou, Shuyan and Neubig, Graham},
  year = {2024},
  eprint = {2410.16464},
  archivePrefix = {arXiv},
  primaryClass = {cs.CL},
  doi = {10.48550/arXiv.2410.16464}
}

@misc{yang2024agentoccam196,
  title = {{AgentOccam}: A Simple Yet Strong Baseline for {LLM-Based} Web Agents},
  author = {Yang, Ke and Liu, Yao and Chaudhary, Sapana and Fakoor, Rasool and Chaudhari, Pratik and Karypis, George and Rangwala, Huzefa},
  year = {2024},
  eprint = {2410.13825},
  archivePrefix = {arXiv},
  primaryClass = {cs.AI},
  doi = {10.48550/arXiv.2410.13825}
}

@misc{murty2024nnetnav197,
  title = {{NNetNav}: Unsupervised Learning of Browser Agents Through Environment Interaction in the Wild},
  author = {Murty, Shikhar and Zhu, Hao and Bahdanau, Dzmitry and Manning, Christopher D.},
  year = {2024},
  eprint = {2410.02907},
  archivePrefix = {arXiv},
  primaryClass = {cs.CL},
  doi = {10.48550/arXiv.2410.02907}
}

@inproceedings{iong2024openwebagent198,
  title = {{OpenWebAgent}: An Open Toolkit to Enable Web Agents on Large Language Models},
  author = {Iong, Iat Long and Liu, Xiao and Chen, Yuxuan and Lai, Hanyu and Yao, Shuntian and Shen, Pengbo and Yu, Hao and Dong, Yuxiao and Tang, Jie},
  booktitle = {Proceedings of the 62nd Annual Meeting of the Association for Computational Linguistics (Volume 3: System Demonstrations)},
  year = {2024},
  pages = {72--81},
  doi = {10.18653/v1/2024.acl-demos.8},
  url = {https://doi.org/10.18653/v1/2024.acl-demos.8}
}

@misc{tang2024steward199,
  title = {Steward: Natural Language Web Automation},
  author = {Tang, Brian and Shin, Kang G.},
  year = {2024},
  eprint = {2409.15441},
  archivePrefix = {arXiv},
  primaryClass = {cs.AI},
  doi = {10.48550/arXiv.2409.15441}
}

@misc{gu2024is200,
  title = {Is Your {LLM} Secretly a World Model of the Internet? {Model-Based} Planning for Web Agents},
  author = {Gu, Yu and Zhang, Kai and Ning, Yuting and Zheng, Boyuan and Gou, Boyu and Xue, Tianci and Chang, Cheng and Srivastava, Sanjari and Xie, Yanan and Qi, Peng and Sun, Huan and Su, Yu},
  year = {2024},
  eprint = {2411.06559},
  archivePrefix = {arXiv},
  primaryClass = {cs.AI},
  doi = {10.48550/arXiv.2411.06559}
}

@misc{putta2024agent201,
  title = {Agent Q: Advanced Reasoning and Learning for Autonomous {AI} Agents},
  author = {Putta, Pranav and Mills, Edmund and Garg, Naman and Motwani, Sumeet and Finn, Chelsea and Garg, Divyansh and Rafailov, Rafael},
  year = {2024},
  eprint = {2408.07199},
  archivePrefix = {arXiv},
  primaryClass = {cs.AI},
  doi = {10.48550/arXiv.2408.07199}
}

@inproceedings{song2024visiontasker202,
  title = {{VisionTasker}: Mobile Task Automation Using Vision Based {UI} Understanding and {LLM} Task Planning},
  author = {Song, Yunpeng and Bian, Yiheng and Tang, Yongtao and Ma, Guiyu and Cai, Zhongmin},
  booktitle = {Proceedings of the 37th Annual ACM Symposium on User Interface Software and Technology},
  year = {2024},
  pages = {1--17},
  doi = {10.1145/3654777.3676386},
  url = {https://doi.org/10.1145/3654777.3676386}
}

@misc{wen2023droidbot203,
  title = {{DroidBot-GPT}: {GPT-powered} {UI} Automation for Android},
  author = {Wen, Hao and Wang, Hongming and Liu, Jiaxuan and Li, Yuanchun},
  year = {2023},
  eprint = {2304.07061},
  archivePrefix = {arXiv},
  primaryClass = {cs.SE},
  doi = {10.48550/arXiv.2304.07061}
}

@misc{yan2023gpt204,
  title = {{GPT-4V} in Wonderland: Large Multimodal Models for {Zero-Shot} Smartphone {GUI} Navigation},
  author = {Yan, An and Yang, Zhengyuan and Zhu, Wanrong and Lin, Kevin and Li, Linjie and Wang, Jianfeng and Yang, Jianwei and Zhong, Yiwu and McAuley, Julian and Gao, Jianfeng and Liu, Zicheng and Wang, Lijuan},
  year = {2023},
  eprint = {2311.07562},
  archivePrefix = {arXiv},
  primaryClass = {cs.CV},
  doi = {10.48550/arXiv.2311.07562}
}

@inproceedings{zhang2025appagent205,
  title = {{AppAgent}: Multimodal Agents as Smartphone Users},
  author = {Zhang, Chi and Yang, Zhao and Liu, Jiaxuan and Li, Yanda and Han, Yucheng and Chen, Xin and Huang, Zebiao and Fu, Bin and Yu, Gang},
  booktitle = {Proceedings of the 2025 CHI Conference on Human Factors in Computing Systems},
  year = {2025},
  pages = {1--20},
  doi = {10.1145/3706598.3713600},
  url = {https://doi.org/10.1145/3706598.3713600}
}

@misc{li2024appagent206,
  title = {{AppAgent} v2: Advanced Agent for Flexible Mobile Interactions},
  author = {Li, Yanda and Zhang, Chi and Jiang, Wenjia and Yang, Wanqi and Fu, Bin and Cheng, Pei and Chen, Xin and Chen, Ling and Wei, Yunchao},
  year = {2024},
  eprint = {2408.11824},
  archivePrefix = {arXiv},
  primaryClass = {cs.HC},
  doi = {10.48550/arXiv.2408.11824}
}

@misc{niu2024screenagent207,
  title = {{ScreenAgent}: A Vision Language Model-driven Computer Control Agent},
  author = {Niu, Runliang and Li, Jindong and Wang, Shiqi and Fu, Yali and Hu, Xiyu and Leng, Xueyuan and Kong, He and Chang, Yi and Wang, Qi},
  year = {2024},
  eprint = {2402.07945},
  archivePrefix = {arXiv},
  primaryClass = {cs.HC},
  doi = {10.48550/arXiv.2402.07945}
}

@misc{wang2024mobile208,
  title = {{Mobile-Agent}: Autonomous {Multi-Modal} Mobile Device Agent with Visual Perception},
  author = {Wang, Junyang and Xu, Haiyang and Ye, Jiabo and Yan, Ming and Shen, Weizhou and Zhang, Ji and Huang, Fei and Sang, Jitao},
  year = {2024},
  eprint = {2401.16158},
  archivePrefix = {arXiv},
  primaryClass = {cs.CL},
  doi = {10.48550/arXiv.2401.16158}
}

@misc{wang2024mobile209,
  title = {{Mobile-Agent-v2}: Mobile Device Operation Assistant with Effective Navigation via {Multi-Agent} Collaboration},
  author = {Wang, Junyang and Xu, Haiyang and Jia, Haitao and Zhang, Xi and Yan, Ming and Shen, Weizhou and Zhang, Ji and Huang, Fei and Sang, Jitao},
  year = {2024},
  eprint = {2406.01014},
  archivePrefix = {arXiv},
  primaryClass = {cs.CL},
  doi = {10.48550/arXiv.2406.01014}
}

@misc{zhang2024mobileexperts210,
  title = {{MobileExperts}: A Dynamic {Tool-Enabled} Agent Team in Mobile Devices},
  author = {Zhang, Jiayi and Zhao, Chuang and Zhao, Yihan and Yu, Zhaoyang and He, Ming and Fan, Jianping},
  year = {2024},
  eprint = {2407.03913},
  archivePrefix = {arXiv},
  primaryClass = {cs.AI},
  doi = {10.48550/arXiv.2407.03913}
}

@misc{christianos2024lightweight211,
  title = {Lightweight Neural App Control},
  author = {Christianos, Filippos and Papoudakis, Georgios and Coste, Thomas and Hao, Jianye and Wang, Jun and Shao, Kun},
  year = {2024},
  eprint = {2410.17883},
  archivePrefix = {arXiv},
  primaryClass = {cs.AI},
  doi = {10.48550/arXiv.2410.17883}
}

@inproceedings{zhu2025moba212,
  title = {{MobA}: Multifaceted {Memory-Enhanced} Adaptive Planning for Efficient Mobile Task Automation},
  author = {Zhu, Zichen and Tang, Hao and Li, Yansi and Liu, Dingye and Xu, Hongshen and Lan, Kunyao and Zhang, Danyang and Jiang, Yixuan and Zhou, Hao and Wang, Chenrun and Zhang, Situo and Sun, Liangtai and Wang, Yixiao and Sun, Yuheng and Chen, Lu and Yu, Kai},
  booktitle = {Proceedings of the 2025 Conference of the Nations of the Americas Chapter of the Association for Computational Linguistics: Human Language Technologies (System Demonstrations)},
  year = {2025},
  pages = {535--549},
  doi = {10.18653/v1/2025.naacl-demo.43},
  url = {https://doi.org/10.18653/v1/2025.naacl-demo.43}
}

@misc{wu2024os213,
  title = {{OS-Copilot}: Towards Generalist Computer Agents with {Self-Improvement}},
  author = {Wu, Zhiyong and Han, Chengcheng and Ding, Zichen and Weng, Zhenmin and Liu, Zhoumianze and Yao, Shunyu and Yu, Tao and Kong, Lingpeng},
  year = {2024},
  eprint = {2402.07456},
  archivePrefix = {arXiv},
  primaryClass = {cs.AI},
  doi = {10.48550/arXiv.2402.07456}
}

@misc{tan2024cradle214,
  title = {Cradle: Empowering Foundation Agents Towards General Computer Control},
  author = {Tan, Weihao and Zhang, Wentao and Xu, Xinrun and Xia, Haochong and Ding, Ziluo and Li, Boyu and Zhou, Bohan and Yue, Junpeng and Jiang, Jiechuan and Li, Yewen and An, Ruyi and Qin, Molei and Zong, Chuqiao and Zheng, Longtao and Wu, Yujie and Chai, Xiaoqiang and Bi, Yifei and Xie, Tianbao and Gu, Pengjie and Li, Xiyun and Zhang, Ceyao and Tian, Long and Wang, Chaojie and Wang, Xinrun and Karlsson, Börje F. and An, Bo and Yan, Shuicheng and Lu, Zongqing},
  year = {2024},
  eprint = {2403.03186},
  archivePrefix = {arXiv},
  primaryClass = {cs.AI},
  doi = {10.48550/arXiv.2403.03186}
}

@inproceedings{li2023a215,
  title = {A {Zero-Shot} Language Agent for Computer Control with Structured Reflection},
  author = {Li, Tao and Li, Gang and Deng, Zhiwei and Wang, Bryan and Li, Yang},
  booktitle = {Findings of the Association for Computational Linguistics: EMNLP 2023},
  year = {2023},
  pages = {11261--11274},
  doi = {10.18653/v1/2023.findings-emnlp.753},
  url = {https://doi.org/10.18653/v1/2023.findings-emnlp.753}
}

@misc{liu2024autoglm216,
  title = {{AutoGLM}: Autonomous Foundation Agents for {GUIs}},
  author = {Liu, Xiao and Qin, Bo and Liang, Dongzhu and Dong, Guang and Lai, Hanyu and Zhang, Hanchen and Zhao, Hanlin and Iong, Iat Long and Sun, Jiadai and Wang, Jiaqi and Gao, Junjie and Shan, Junjun and Liu, Kangning and Zhang, Shudan and Yao, Shuntian and Cheng, Siyi and Yao, Wentao and Zhao, Wenyi and Liu, Xinghan and Liu, Xinyi and Chen, Xinying and Yang, Xinyue and Yang, Yang and Xu, Yifan and Yang, Yu and Wang, Yujia and Xu, Yulin and Qi, Zehan and Dong, Yuxiao and Tang, Jie},
  year = {2024},
  eprint = {2411.00820},
  archivePrefix = {arXiv},
  primaryClass = {cs.HC},
  doi = {10.48550/arXiv.2411.00820}
}

@inproceedings{pawlowski2025tinyclick217,
  title = {{TinyClick}: {Single-Turn} Agent for Empowering {GUI} Automation},
  author = {Pawlowski, Pawel and Zawistowski, Krystian and Lapacz, Wojciech and Wiacek, Adam and Skorupa, Marcin and Postansque, Sebastien and Hoscilowicz, Jakub},
  booktitle = {Interspeech 2025},
  year = {2025},
  pages = {3035--3039},
  doi = {10.21437/interspeech.2025-176},
  url = {https://doi.org/10.21437/interspeech.2025-176}
}

@misc{wang2024oscar218,
  title = {{OSCAR}: Operating System Control via {State-Aware} Reasoning and {Re-Planning}},
  author = {Wang, Xiaoqiang and Liu, Bang},
  year = {2024},
  eprint = {2410.18963},
  archivePrefix = {arXiv},
  primaryClass = {cs.AI},
  doi = {10.48550/arXiv.2410.18963}
}

@inproceedings{jia2025agentstore219,
  title = {{AgentStore}: Scalable Integration of Heterogeneous Agents As Specialized Generalist Computer Assistant},
  author = {Jia, Chengyou and Luo, Minnan and Dang, Zhuohang and Sun, Qiushi and Xu, Fangzhi and Hu, Junlin and Xie, Tianbao and Wu, Zhiyong},
  booktitle = {Findings of the Association for Computational Linguistics: ACL 2025},
  year = {2025},
  pages = {8908--8934},
  doi = {10.18653/v1/2025.findings-acl.466},
  url = {https://doi.org/10.18653/v1/2025.findings-acl.466}
}

@misc{song2024mmac220,
  title = {{MMAC-Copilot}: Multi-modal Agent Collaboration Operating System Copilot},
  author = {Song, Zirui and Li, Yaohang and Fang, Meng and Chen, Zhenhao and Shi, Zecheng and Huang, Yuan and Chen, Ling},
  year = {2024},
  eprint = {2404.18074},
  archivePrefix = {arXiv},
  primaryClass = {cs.AI},
  doi = {10.48550/arXiv.2404.18074}
}

@misc{shen2024scribeagent221,
  title = {{ScribeAgent}: Towards Specialized Web Agents Using {Production-Scale} Workflow Data},
  author = {Shen, Junhong and Jain, Atishay and Xiao, Zedian and Amlekar, Ishan and Hadji, Mouad and Podolny, Aaron and Talwalkar, Ameet},
  year = {2024},
  eprint = {2411.15004},
  archivePrefix = {arXiv},
  primaryClass = {cs.CL},
  doi = {10.48550/arXiv.2411.15004}
}

@misc{zhou2024proposer222,
  title = {{Proposer-Agent-Evaluator}({PAE}): Autonomous Skill Discovery For Foundation Model Internet Agents},
  author = {Zhou, Yifei and Yang, Qianlan and Lin, Kaixiang and Bai, Min and Zhou, Xiong and Wang, Yu-Xiong and Levine, Sergey and Li, Erran},
  year = {2024},
  eprint = {2412.13194},
  archivePrefix = {arXiv},
  primaryClass = {cs.LG},
  doi = {10.48550/arXiv.2412.13194}
}

@misc{he2024pc223,
  title = {{PC} Agent: While You Sleep, {AI} Works -- A Cognitive Journey into Digital World},
  author = {He, Yanheng and Jin, Jiahe and Xia, Shijie and Su, Jiadi and Fan, Runze and Zou, Haoyang and Hu, Xiangkun and Liu, Pengfei},
  year = {2024},
  eprint = {2412.17589},
  archivePrefix = {arXiv},
  primaryClass = {cs.AI},
  doi = {10.48550/arXiv.2412.17589}
}

@article{liu2025wepo224,
  title = {{WEPO}: Web Element Preference Optimization for {LLM-based} Web Navigation},
  author = {Liu, Jiarun and Hao, Jia and Zhang, Chunhong and Hu, Zheng},
  journal = {Proceedings of the AAAI Conference on Artificial Intelligence},
  year = {2025},
  volume = {39},
  number = {25},
  pages = {26614--26622},
  doi = {10.1609/aaai.v39i25.34863},
  url = {https://doi.org/10.1609/aaai.v39i25.34863}
}

@misc{wang2025mobile225,
  title = {{Mobile-Agent-E}: {Self-Evolving} Mobile Assistant for Complex Tasks},
  author = {Wang, Zhenhailong and Xu, Haiyang and Wang, Junyang and Zhang, Xi and Yan, Ming and Zhang, Ji and Huang, Fei and Ji, Heng},
  year = {2025},
  eprint = {2501.11733},
  archivePrefix = {arXiv},
  primaryClass = {cs.CL},
  doi = {10.48550/arXiv.2501.11733}
}

@misc{su2025learn226,
  title = {Learn-by-interact: A {Data-Centric} Framework for {Self-Adaptive} Agents in Realistic Environments},
  author = {Su, Hongjin and Sun, Ruoxi and Yoon, Jinsung and Yin, Pengcheng and Yu, Tao and Arık, Sercan Ö.},
  year = {2025},
  eprint = {2501.10893},
  archivePrefix = {arXiv},
  primaryClass = {cs.LG},
  doi = {10.48550/arXiv.2501.10893}
}

@inproceedings{huang2025r2d2227,
  title = {{R2D2}: Remembering, Replaying and Dynamic Decision Making with a Reflective Agentic Memory},
  author = {Huang, Tenghao and Basu, Kinjal and Abdelaziz, Ibrahim and Kapanipathi, Pavan and May, Jonathan and Chen, Muhao},
  booktitle = {Proceedings of the 63rd Annual Meeting of the Association for Computational Linguistics (Volume 1: Long Papers)},
  year = {2025},
  pages = {30318--30330},
  doi = {10.18653/v1/2025.acl-long.1464},
  url = {https://doi.org/10.18653/v1/2025.acl-long.1464}
}

@misc{hoscilowicz2024clickagent228,
  title = {{ClickAgent}: Enhancing {UI} Location Capabilities of Autonomous Agents},
  author = {Hoscilowicz, Jakub and Maj, Bartosz and Kozakiewicz, Bartosz and Tymoshchuk, Oleksii and Janicki, Artur},
  year = {2024},
  eprint = {2410.11872},
  archivePrefix = {arXiv},
  primaryClass = {cs.HC},
  doi = {10.48550/arXiv.2410.11872}
}

@inproceedings{wu2025reachagent229,
  title = {{ReachAgent}: Enhancing Mobile Agent via Page Reaching and Operation},
  author = {Wu, Qinzhuo and Liu, Wei and Luan, Jian and Wang, Bin},
  booktitle = {Proceedings of the 2025 Conference of the Nations of the Americas Chapter of the Association for Computational Linguistics: Human Language Technologies (Volume 1: Long Papers)},
  year = {2025},
  pages = {4760--4775},
  doi = {10.18653/v1/2025.naacl-long.244},
  url = {https://doi.org/10.18653/v1/2025.naacl-long.244}
}

@inproceedings{wang2025mobilea3gent230,
  title = {{MobileA3gent}: Training Mobile {GUI} Agents Using Decentralized {Self-Sourced} Data from Diverse Users},
  author = {Wang, WenHao and Yuan, Mengying and Yu, Zijie and Liu, Guangyi and Ye, Rui and Jin, Tian and Chen, Siheng and Wang, Yanfeng},
  booktitle = {Proceedings of the Fourth Workshop on Bridging Human-Computer Interaction and Natural Language Processing (HCI+NLP)},
  year = {2025},
  pages = {79--112},
  doi = {10.18653/v1/2025.hcinlp-1.8},
  url = {https://doi.org/10.18653/v1/2025.hcinlp-1.8}
}

@article{huang2025prompt2task231,
  title = {{Prompt2Task}: Automating {UI} Tasks on Smartphones from Textual Prompts},
  author = {Huang, Tian and Yu, Chun and Shi, Weinan and Peng, Zijian and Yang, David and Sun, Weiqi and Shi, Yuanchun},
  journal = {ACM Transactions on Computer-Human Interaction},
  year = {2025},
  volume = {32},
  number = {3},
  pages = {1--45},
  doi = {10.1145/3716132},
  url = {https://doi.org/10.1145/3716132}
}

@misc{zhang2025symbiotic232,
  title = {Symbiotic Cooperation for Web Agents: Harnessing Complementary Strengths of Large and Small {LLMs}},
  author = {Zhang, Ruichen and Qiu, Mufan and Tan, Zhen and Zhang, Mohan and Lu, Vincent and Peng, Jie and Xu, Kaidi and Agudelo, Leandro Z. and Qian, Peter and Chen, Tianlong},
  year = {2025},
  eprint = {2502.07942},
  archivePrefix = {arXiv},
  primaryClass = {cs.MA},
  doi = {10.48550/arXiv.2502.07942}
}

@misc{liu2025pc233,
  title = {{PC-Agent}: A Hierarchical {Multi-Agent} Collaboration Framework for Complex Task Automation on {PC}},
  author = {Liu, Haowei and Zhang, Xi and Xu, Haiyang and Wanyan, Yuyang and Wang, Junyang and Yan, Ming and Zhang, Ji and Yuan, Chunfeng and Xu, Changsheng and Hu, Weiming and Huang, Fei},
  year = {2025},
  eprint = {2502.14282},
  archivePrefix = {arXiv},
  primaryClass = {cs.CV},
  doi = {10.48550/arXiv.2502.14282}
}

@misc{wang2025mobile234,
  title = {{Mobile-Agent-V}: Learning Mobile Device Operation Through {Video-Guided} {Multi-Agent} Collaboration},
  author = {Wang, Junyang and Xu, Haiyang and Zhang, Xi and Yan, Ming and Zhang, Ji and Huang, Fei and Sang, Jitao},
  year = {2025},
  eprint = {2502.17110},
  archivePrefix = {arXiv},
  primaryClass = {cs.CL},
  doi = {10.48550/arXiv.2502.17110}
}

@inproceedings{liu2025mobilesteward235,
  title = {{MobileSteward}: Integrating Multiple {App-Oriented} Agents with {Self-Evolution} to Automate {Cross-App} Instructions},
  author = {Liu, Yuxuan and Sun, Hongda and Liu, Wei and Luan, Jian and Du, Bo and Yan, Rui},
  booktitle = {Proceedings of the 31st ACM SIGKDD Conference on Knowledge Discovery and Data Mining V.1},
  year = {2025},
  pages = {883--893},
  doi = {10.1145/3690624.3709171},
  url = {https://doi.org/10.1145/3690624.3709171}
}

@misc{aggarwal2025programming236,
  title = {Programming with Pixels: {Computer-Use} Meets Software Engineering},
  author = {Aggarwal, Pranjal and Welleck, Sean},
  year = {2025},
  eprint = {2502.18525},
  archivePrefix = {arXiv},
  primaryClass = {cs.SE},
  doi = {10.48550/arXiv.2502.18525}
}

@misc{zhang2025litewebagent237,
  title = {{LiteWebAgent}: The {Open-Source} Suite for {VLM-Based} {Web-Agent} Applications},
  author = {Zhang, Danqing and Rama, Balaji and Ni, Jingyi and He, Shiying and Zhao, Fu and Chen, Kunyu and Chen, Arnold and Cao, Junyu},
  year = {2025},
  eprint = {2503.02950},
  archivePrefix = {arXiv},
  primaryClass = {cs.AI},
  doi = {10.48550/arXiv.2503.02950}
}

@misc{zhou2025chop238,
  title = {{CHOP}: Mobile Operating Assistant with Constrained High-frequency Optimized Subtask Planning},
  author = {Zhou, Yuqi and Wang, Shuai and Dai, Sunhao and Jia, Qinglin and Du, Zhaocheng and Dong, Zhenhua and Xu, Jun},
  year = {2025},
  eprint = {2503.03743},
  archivePrefix = {arXiv},
  primaryClass = {cs.AI},
  doi = {10.48550/arXiv.2503.03743}
}

@misc{he2025advancing239,
  title = {Advancing Language {Multi-Agent} Learning with Credit {Re-Assignment} for Interactive Environment Generalization},
  author = {He, Zhitao and Liu, Zijun and Li, Peng and Fung, Yi R. and Yan, Ming and Zhang, Ji and Huang, Fei and Liu, Yang},
  year = {2025},
  eprint = {2502.14496},
  archivePrefix = {arXiv},
  primaryClass = {cs.CL},
  doi = {10.48550/arXiv.2502.14496}
}

@article{wornow2024automating240,
  title = {Automating the Enterprise with Foundation Models},
  author = {Wornow, Michael and Narayan, Avanika and Opsahl-Ong, Krista and McIntyre, Quinn and Shah, Nigam and Ré, Christopher},
  journal = {Proceedings of the VLDB Endowment},
  year = {2024},
  volume = {17},
  number = {11},
  pages = {2805--2812},
  doi = {10.14778/3681954.3681964},
  url = {https://doi.org/10.14778/3681954.3681964}
}

@inproceedings{dammu2025towards241,
  title = {Towards Ethical and Personalized Web Navigation Agents: A Framework for {User-Aligned} Task Execution},
  author = {Dammu, Preetam Prabhu Srikar},
  booktitle = {Proceedings of the Eighteenth ACM International Conference on Web Search and Data Mining},
  year = {2025},
  pages = {1074--1076},
  doi = {10.1145/3701551.3707420},
  url = {https://doi.org/10.1145/3701551.3707420}
}

@misc{erdogan2025plan242,
  title = {{Plan-and-Act}: Improving Planning of Agents for {Long-Horizon} Tasks},
  author = {Erdogan, Lutfi Eren and Lee, Nicholas and Kim, Sehoon and Moon, Suhong and Furuta, Hiroki and Anumanchipalli, Gopala and Keutzer, Kurt and Gholami, Amir},
  year = {2025},
  eprint = {2503.09572},
  archivePrefix = {arXiv},
  primaryClass = {cs.CL},
  doi = {10.48550/arXiv.2503.09572}
}

@misc{lu2025steve243,
  title = {{STEVE}: A Step Verification Pipeline for Computer-use Agent Training},
  author = {Lu, Fanbin and Zhong, Zhisheng and Wei, Ziqin and Liu, Shu and Fu, Chi-Wing and Jia, Jiaya},
  year = {2025},
  eprint = {2503.12532},
  archivePrefix = {arXiv},
  primaryClass = {cs.CV},
  doi = {10.48550/arXiv.2503.12532}
}

@misc{zheng2025skillweaver244,
  title = {{SkillWeaver}: Web Agents can {Self-Improve} by Discovering and Honing Skills},
  author = {Zheng, Boyuan and Fatemi, Michael Y. and Jin, Xiaolong and Wang, Zora Zhiruo and Gandhi, Apurva and Song, Yueqi and Gu, Yu and Srinivasa, Jayanth and Liu, Gaowen and Neubig, Graham and Su, Yu},
  year = {2025},
  eprint = {2504.07079},
  archivePrefix = {arXiv},
  primaryClass = {cs.AI},
  doi = {10.48550/arXiv.2504.07079}
}

@misc{wang2025inducing245,
  title = {Inducing Programmatic Skills for Agentic Tasks},
  author = {Wang, Zora Zhiruo and Gandhi, Apurva and Neubig, Graham and Fried, Daniel},
  year = {2025},
  eprint = {2504.06821},
  archivePrefix = {arXiv},
  primaryClass = {cs.CL},
  doi = {10.48550/arXiv.2504.06821}
}

@misc{zhang2025webrollback246,
  title = {{WebRollback}: Enhancing Web Agents with Explicit Rollback Mechanisms},
  author = {Zhang, Zhisong and Fang, Tianqing and Ma, Kaixin and Yu, Wenhao and Zhang, Hongming and Mi, Haitao and Yu, Dong},
  year = {2025},
  eprint = {2504.11788},
  archivePrefix = {arXiv},
  primaryClass = {cs.CL},
  doi = {10.48550/arXiv.2504.11788}
}

@misc{agashe2025agent247,
  title = {Agent {S2}: A Compositional {Generalist-Specialist} Framework for Computer Use Agents},
  author = {Agashe, Saaket and Wong, Kyle and Tu, Vincent and Yang, Jiachen and Li, Ang and Wang, Xin Eric},
  year = {2025},
  eprint = {2504.00906},
  archivePrefix = {arXiv},
  primaryClass = {cs.AI},
  doi = {10.48550/arXiv.2504.00906}
}

@misc{hu2025guiding248,
  title = {Guiding {VLM} Agents with Process Rewards at Inference Time for {GUI} Navigation},
  author = {Hu, Zhiyuan and Xiong, Shiyun and Zhang, Yifan and Ng, See-Kiong and Luu, Anh Tuan and An, Bo and Yan, Shuicheng and Hooi, Bryan},
  year = {2025},
  eprint = {2504.16073},
  archivePrefix = {arXiv},
  primaryClass = {cs.CL},
  doi = {10.48550/arXiv.2504.16073}
}

@misc{zhang2025ufo2249,
  title = {{UFO2}: The Desktop {AgentOS}},
  author = {Zhang, Chaoyun and Huang, He and Ni, Chiming and Mu, Jian and Qin, Si and He, Shilin and Wang, Lu and Yang, Fangkai and Zhao, Pu and Du, Chao and Li, Liqun and Kang, Yu and Jiang, Zhao and Zheng, Suzhen and Wang, Rujia and Qian, Jiaxu and Ma, Minghua and Lou, Jian-Guang and Lin, Qingwei and Rajmohan, Saravan and Zhang, Dongmei},
  year = {2025},
  eprint = {2504.14603},
  archivePrefix = {arXiv},
  primaryClass = {cs.AI},
  doi = {10.48550/arXiv.2504.14603}
}

@misc{ye2023proagent250,
  title = {{ProAgent}: From Robotic Process Automation to Agentic Process Automation},
  author = {Ye, Yining and Cong, Xin and Tian, Shizuo and Cao, Jiannan and Wang, Hao and Qin, Yujia and Lu, Yaxi and Yu, Heyang and Wang, Huadong and Lin, Yankai and Liu, Zhiyuan and Sun, Maosong},
  year = {2023},
  eprint = {2311.10751},
  archivePrefix = {arXiv},
  primaryClass = {cs.RO},
  doi = {10.48550/arXiv.2311.10751}
}

@inproceedings{guan2024intelligent251,
  title = {Intelligent Agents with {LLM-based} Process Automation},
  author = {Guan, Yanchu and Wang, Dong and Chu, Zhixuan and Wang, Shiyu and Ni, Feiyue and Song, Ruihua and Zhuang, Chenyi},
  booktitle = {Proceedings of the 30th ACM SIGKDD Conference on Knowledge Discovery and Data Mining},
  year = {2024},
  pages = {5018--5027},
  doi = {10.1145/3637528.3671646},
  url = {https://doi.org/10.1145/3637528.3671646}
}

@misc{vu2024gptvoicetasker253,
  title = {{GPTVoiceTasker}: {LLM-Powered} Virtual Assistant for Smartphone},
  author = {Vu, Minh Duc and Wang, Han and Li, Zhuang and Chen, Jieshan and Zhao, Shengdong and Xing, Zhenchang and Chen, Chunyang},
  year = {2024},
  eprint = {2401.14268},
  archivePrefix = {arXiv},
  primaryClass = {cs.HC},
  doi = {10.48550/arXiv.2401.14268}
}

@misc{pan2023autotask254,
  title = {{AutoTask}: Executing Arbitrary Voice Commands by Exploring and Learning from Mobile {GUI}},
  author = {Pan, Lihang and Wang, Bowen and Yu, Chun and Chen, Yuxuan and Zhang, Xiangyu and Shi, Yuanchun},
  year = {2023},
  eprint = {2312.16062},
  archivePrefix = {arXiv},
  primaryClass = {cs.HC},
  doi = {10.48550/arXiv.2312.16062}
}

@misc{huang2024promptrpa255,
  title = {{PromptRPA}: Generating Robotic Process Automation on Smartphones from Textual Prompts},
  author = {Huang, Tian and Yu, Chun and Shi, Weinan and Peng, Zijian and Yang, David and Sun, Weiqi and Shi, Yuanchun},
  year = {2024},
  eprint = {2404.02475},
  archivePrefix = {arXiv},
  primaryClass = {cs.HC},
  doi = {10.48550/arXiv.2404.02475}
}

@misc{srinivasan2025webnav257,
  title = {{WebNav}: An Intelligent Agent for {Voice-Controlled} Web Navigation},
  author = {Srinivasan, Trisanth and Patapati, Santosh},
  year = {2025},
  eprint = {2503.13843},
  archivePrefix = {arXiv},
  primaryClass = {cs.AI},
  doi = {10.48550/arXiv.2503.13843}
}

@misc{chaimalas2025explorer258,
  title = {Explorer: Robust Collection of Interactable {GUI} Elements},
  author = {Chaimalas, Iason and Vyšniauskas, Arnas and Brostow, Gabriel},
  year = {2025},
  eprint = {2504.09352},
  archivePrefix = {arXiv},
  primaryClass = {cs.HC},
  doi = {10.48550/arXiv.2504.09352}
}

@misc{yang2025pixelweb259,
  title = {{PixelWeb}: The First Web {GUI} Dataset with {Pixel-Wise} Labels},
  author = {Yang, Qi and Bi, Weichen and Shen, Haiyang and Guo, Yaoqi and Ma, Yun},
  year = {2025},
  eprint = {2504.16419},
  archivePrefix = {arXiv},
  primaryClass = {cs.CV},
  doi = {10.48550/arXiv.2504.16419}
}

@misc{zhang2025breaking260,
  title = {Breaking the Data Barrier -- Building {GUI} Agents Through Task Generalization},
  author = {Zhang, Junlei and Ding, Zichen and Ma, Chang and Chen, Zijie and Sun, Qiushi and Lan, Zhenzhong and He, Junxian},
  year = {2025},
  eprint = {2504.10127},
  archivePrefix = {arXiv},
  primaryClass = {cs.AI},
  doi = {10.48550/arXiv.2504.10127}
}

@misc{lu2025agentrewardbench261,
  title = {{AgentRewardBench}: Evaluating Automatic Evaluations of Web Agent Trajectories},
  author = {Lù, Xing Han and Kazemnejad, Amirhossein and Meade, Nicholas and Patel, Arkil and Shin, Dongchan and Zambrano, Alejandra and Stańczak, Karolina and Shaw, Peter and Pal, Christopher J. and Reddy, Siva},
  year = {2025},
  eprint = {2504.08942},
  archivePrefix = {arXiv},
  primaryClass = {cs.LG},
  doi = {10.48550/arXiv.2504.08942}
}

@article{ye2026realwebassist262,
  title = {{RealWebAssist}: A Benchmark for {Long-Horizon} Web Assistance with {Real-World} Users},
  author = {Ye, Suyu and Shi, Haojun and Shih, Darren and Yun, Hyokun and Roosta, Tanya G. and Shu, Tianmin},
  journal = {Proceedings of the AAAI Conference on Artificial Intelligence},
  year = {2026},
  volume = {40},
  number = {40},
  pages = {34441--34449},
  doi = {10.1609/aaai.v40i40.40742},
  url = {https://doi.org/10.1609/aaai.v40i40.40742}
}

@inproceedings{liu2025ui263,
  title = {{UI-E2I-Synth}: Advancing {GUI} Grounding with {Large-Scale} Instruction Synthesis},
  author = {Liu, Xinyi and Zhang, Xiaoyi and Zhang, Ziyun and Lu, Yan},
  booktitle = {Findings of the Association for Computational Linguistics: ACL 2025},
  year = {2025},
  pages = {15668--15684},
  doi = {10.18653/v1/2025.findings-acl.809},
  url = {https://doi.org/10.18653/v1/2025.findings-acl.809}
}

@misc{luo2025vimo264,
  title = {{ViMo}: A Generative Visual {GUI} World Model for App Agents},
  author = {Luo, Dezhao and Tang, Bohan and Li, Kang and Papoudakis, Georgios and Song, Jifei and Gong, Shaogang and Hao, Jianye and Wang, Jun and Shao, Kun},
  year = {2025},
  eprint = {2504.13936},
  archivePrefix = {arXiv},
  primaryClass = {cs.HC},
  doi = {10.48550/arXiv.2504.13936}
}

@misc{wang2025fedmabench267,
  title = {{FedMABench}: Benchmarking Mobile Agents on Decentralized Heterogeneous User Data},
  author = {Wang, Wenhao and Yu, Zijie and Ye, Rui and Zhang, Jianqing and Chen, Siheng and Wang, Yanfeng},
  year = {2025},
  eprint = {2503.05143},
  archivePrefix = {arXiv},
  primaryClass = {cs.AI},
  doi = {10.48550/arXiv.2503.05143}
}

@misc{cheng2025navi268,
  title = {Navi-plus: Managing Ambiguous {GUI} Navigation Tasks with Follow-up Questions},
  author = {Cheng, Ziming and Huang, Zhiyuan and Pan, Junting and Hou, Zhaohui and Zhan, Mingjie},
  year = {2025},
  eprint = {2503.24180},
  archivePrefix = {arXiv},
  primaryClass = {cs.CV},
  doi = {10.48550/arXiv.2503.24180}
}

@misc{wu2025smoothing269,
  title = {Smoothing Grounding and Reasoning for {MLLM-Powered} {GUI} Agents with {Query-Oriented} Pivot Tasks},
  author = {Wu, Zongru and Cheng, Pengzhou and Wu, Zheng and Ju, Tianjie and Zhang, Zhuosheng and Liu, Gongshen},
  year = {2025},
  eprint = {2503.00401},
  archivePrefix = {arXiv},
  primaryClass = {cs.CL},
  doi = {10.48550/arXiv.2503.00401}
}

@misc{hui2025winclick270,
  title = {{WinClick}: {GUI} Grounding with Multimodal Large Language Models},
  author = {Hui, Zheng and Li, Yinheng and zhao, Dan and Chen, Tianyi and Banbury, Colby and Koishida, Kazuhito},
  year = {2025},
  eprint = {2503.04730},
  archivePrefix = {arXiv},
  primaryClass = {cs.CL},
  doi = {10.48550/arXiv.2503.04730}
}

@misc{tang2025think271,
  title = {Think Twice, Click Once: Enhancing {GUI} Grounding via Fast and Slow Systems},
  author = {Tang, Fei and Shen, Yongliang and Zhang, Hang and Chen, Siqi and Hou, Guiyang and Zhang, Wenqi and Zhang, Wenqiao and Song, Kaitao and Lu, Weiming and Zhuang, Yueting},
  year = {2025},
  eprint = {2503.06470},
  archivePrefix = {arXiv},
  primaryClass = {cs.AI},
  doi = {10.48550/arXiv.2503.06470}
}

@inproceedings{wang2025mp272,
  title = {{MP-GUI}: Modality Perception with {MLLMs} for {GUI} Understanding},
  author = {Wang, Ziwei and Chen, Weizhi and Yang, Leyang and Zhou, Sheng and Zhao, Shengchu and Zhan, Hanbei and Jin, Jiongchao and Li, Liangcheng and Shao, Zirui and Bu, Jiajun},
  booktitle = {2025 IEEE/CVF Conference on Computer Vision and Pattern Recognition (CVPR)},
  year = {2025},
  pages = {29711--29721},
  doi = {10.1109/cvpr52734.2025.02766},
  url = {https://doi.org/10.1109/cvpr52734.2025.02766}
}

@article{lu2026ui273,
  title = {{UI-R1}: Enhancing Efficient Action Prediction of {GUI} Agents by Reinforcement Learning},
  author = {Lu, Zhengxi and Chai, Yuxiang and Guo, Yaxuan and Yin, Xi and Liu, Liang and Wang, Hao and Xiao, Han and Ren, Shuai and Zhao, Pengxiang and Liu, Guangyi and Xiong, Guanjing and Li, Hongsheng},
  journal = {Proceedings of the AAAI Conference on Artificial Intelligence},
  year = {2026},
  volume = {40},
  number = {21},
  pages = {17608--17616},
  doi = {10.1609/aaai.v40i21.38816},
  url = {https://doi.org/10.1609/aaai.v40i21.38816}
}

@misc{zhao2025cola274,
  title = {{COLA}: A Scalable {Multi-Agent} Framework For Windows {UI} Task Automation},
  author = {Zhao, Di and Ma, Longhui and Wang, Siwei and Wang, Miao and Lv, Zhao},
  year = {2025},
  eprint = {2503.09263},
  archivePrefix = {arXiv},
  primaryClass = {cs.MA},
  doi = {10.48550/arXiv.2503.09263}
}

@inproceedings{pahuja2025explorer276,
  title = {Explorer: Scaling Exploration-driven Web Trajectory Synthesis for Multimodal Web Agents},
  author = {Pahuja, Vardaan and Lu, Yadong and Rosset, Corby and Gou, Boyu and Mitra, Arindam and Whitehead, Spencer and Su, Yu and Awadallah, Ahmed Hassan},
  booktitle = {Findings of the Association for Computational Linguistics: ACL 2025},
  year = {2025},
  pages = {6300--6323},
  doi = {10.18653/v1/2025.findings-acl.326},
  url = {https://doi.org/10.18653/v1/2025.findings-acl.326}
}

@misc{trabucco2025insta277,
  title = {{InSTA}: Towards {Internet-Scale} Training For Agents},
  author = {Trabucco, Brandon and Sigurdsson, Gunnar and Piramuthu, Robinson and Salakhutdinov, Ruslan},
  year = {2025},
  eprint = {2502.06776},
  archivePrefix = {arXiv},
  primaryClass = {cs.LG},
  doi = {10.48550/arXiv.2502.06776}
}

@misc{papoudakis2025appvlm278,
  title = {{AppVLM}: A Lightweight Vision Language Model for Online App Control},
  author = {Papoudakis, Georgios and Coste, Thomas and Wu, Zhihao and Hao, Jianye and Wang, Jun and Shao, Kun},
  year = {2025},
  eprint = {2502.06395},
  archivePrefix = {arXiv},
  primaryClass = {cs.AI},
  doi = {10.48550/arXiv.2502.06395}
}

@inproceedings{yang2025rwkv279,
  title = {{RWKV-UI}: {UI} Understanding with Enhanced Perception and Reasoning},
  author = {Yang, Jiaxi and Hou, Haowen},
  booktitle = {2025 IEEE International Conference on Multimedia and Expo (ICME)},
  year = {2025},
  pages = {1--6},
  doi = {10.1109/icme59968.2025.11210007},
  url = {https://doi.org/10.1109/icme59968.2025.11210007}
}

@misc{wu2025advancing280,
  title = {Advancing Autonomous {VLM} Agents via Variational {Subgoal-Conditioned} Reinforcement Learning},
  author = {Wu, Qingyuan and Liu, Jianheng and Hao, Jianye and Wang, Jun and Shao, Kun},
  year = {2025},
  eprint = {2502.07949},
  archivePrefix = {arXiv},
  primaryClass = {cs.LG},
  doi = {10.48550/arXiv.2502.07949}
}

@inproceedings{yang2025magma281,
  title = {Magma: A Foundation Model for Multimodal {AI} Agents},
  author = {Yang, Jianwei and Tan, Reuben and Wu, Qianhui and Zheng, Ruijie and Peng, Baolin and Liang, Yongyuan and Gu, Yu and Cai, Mu and Ye, Seonghyeon and Jang, Joel and Deng, Yuquan and Gao, Jianfeng},
  booktitle = {2025 IEEE/CVF Conference on Computer Vision and Pattern Recognition (CVPR)},
  year = {2025},
  pages = {14203--14214},
  doi = {10.1109/cvpr52734.2025.01325},
  url = {https://doi.org/10.1109/cvpr52734.2025.01325}
}

@inproceedings{bai2025digi282,
  title = {{Digi-Q}: Learning {VLM} {Q-Value} Functions for Training {Device-Control} Agents},
  author = {Bai, Hao and Zhou, Yifei and Li, Li and Levine, Sergey and Kumar, Aviral},
  booktitle = {International Conference on Learning Representations},
  year = {2025},
  url = {https://proceedings.iclr.cc/paper\_files/paper/2025/hash/519abe71ee55aac4fe821bbd731c6645-Abstract-Conference.html}
}

@inproceedings{li2025autogui283,
  title = {{AutoGUI}: Scaling {GUI} Grounding with Automatic Functionality Annotations from {LLMs}},
  author = {Li, Hongxin and Chen, Jingfan and Su, Jingran and Chen, Yuntao and Qing, Li and Zhang, Zhaoxiang},
  booktitle = {Proceedings of the 63rd Annual Meeting of the Association for Computational Linguistics (Volume 1: Long Papers)},
  year = {2025},
  pages = {10323--10358},
  doi = {10.18653/v1/2025.acl-long.510},
  url = {https://doi.org/10.18653/v1/2025.acl-long.510}
}

@article{hu2024os285,
  title = {{OS} Agents: A Survey on {MLLM-Based} Agents for General Computing Devices Use},
  author = {Hu, Xueyu and Xiong, Tao and Yi, Biao and Wei, Zishu and Xiao, Ruixuan and Chen, Yurun and Ye, Jiasheng and Tao, Meiling and Zhou, Xiangxin and Zhao, Ziyu and Li, Yuhuai and Xu, Shengze and Wang, Shawn and Xu, Xinchen and Qiao, Shuofei and Kuang, Kun and Zeng, Tieyong and Wang, Liang and Li, Jiwei and Jiang, Yuchen Eleanor and Zhou, Wangchunshu and Wang, Guoyin and Yin, Keting and Zhao, Zhou and Yang, Hongxia and Wu, Fan and Zhang, Shengyu and Wu, Fei},
  year = {2024},
  doi = {10.20944/preprints202412.2294.v1},
  url = {https://doi.org/10.20944/preprints202412.2294.v1}
}

@misc{xu2024agenttrek286,
  title = {{AgentTrek}: Agent Trajectory Synthesis via Guiding Replay with Web Tutorials},
  author = {Xu, Yiheng and Lu, Dunjie and Shen, Zhennan and Wang, Junli and Wang, Zekun and Mao, Yuchen and Xiong, Caiming and Yu, Tao},
  year = {2024},
  eprint = {2412.09605},
  archivePrefix = {arXiv},
  primaryClass = {cs.CL},
  doi = {10.48550/arXiv.2412.09605}
}

@misc{shen2024falcon287,
  title = {{Falcon-UI}: Understanding {GUI} Before Following User Instructions},
  author = {Shen, Huawen and Liu, Chang and Li, Gengluo and Wang, Xinlong and Zhou, Yu and Ma, Can and Ji, Xiangyang},
  year = {2024},
  eprint = {2412.09362},
  archivePrefix = {arXiv},
  primaryClass = {cs.CL},
  doi = {10.48550/arXiv.2412.09362}
}

@article{xu2025attention288,
  title = {{Attention-Driven} {GUI} Grounding: Leveraging Pretrained Multimodal Large Language Models Without {Fine-Tuning}},
  author = {Xu, Hai-Ming and Chen, Qi and Wang, Lei and Liu, Lingqiao},
  journal = {Proceedings of the AAAI Conference on Artificial Intelligence},
  year = {2025},
  volume = {39},
  number = {8},
  pages = {8851--8859},
  doi = {10.1609/aaai.v39i8.32957},
  url = {https://doi.org/10.1609/aaai.v39i8.32957}
}

@misc{wu2024foundations289,
  title = {Foundations and Recent Trends in Multimodal Mobile Agents: A Survey},
  author = {Wu, Biao and Li, Yanda and Zhang, Zhiwei and Wei, Yunchao and Fang, Meng and Chen, Ling},
  year = {2024},
  eprint = {2411.02006},
  archivePrefix = {arXiv},
  primaryClass = {cs.AI},
  doi = {10.48550/arXiv.2411.02006}
}

@misc{qi2024webrl290,
  title = {{WebRL}: Training {LLM} Web Agents via {Self-Evolving} Online Curriculum Reinforcement Learning},
  author = {Qi, Zehan and Liu, Xiao and Iong, Iat Long and Lai, Hanyu and Sun, Xueqiao and Zhao, Wenyi and Yang, Yu and Yang, Xinyue and Sun, Jiadai and Yao, Shuntian and Zhang, Tianjie and Xu, Wei and Tang, Jie and Dong, Yuxiao},
  year = {2024},
  eprint = {2411.02337},
  archivePrefix = {arXiv},
  primaryClass = {cs.CL},
  doi = {10.48550/arXiv.2411.02337}
}

@inproceedings{he2025openwebvoyager291,
  title = {{OpenWebVoyager}: Building Multimodal Web Agents via Iterative {Real-World} Exploration, Feedback and Optimization},
  author = {He, Hongliang and Yao, Wenlin and Ma, Kaixin and Yu, Wenhao and Zhang, Hongming and Fang, Tianqing and Lan, Zhenzhong and Yu, Dong},
  booktitle = {Proceedings of the 63rd Annual Meeting of the Association for Computational Linguistics (Volume 1: Long Papers)},
  year = {2025},
  pages = {27545--27564},
  doi = {10.18653/v1/2025.acl-long.1336},
  url = {https://doi.org/10.18653/v1/2025.acl-long.1336}
}

@misc{li2024ferret292,
  title = {{Ferret-UI} 2: Mastering Universal User Interface Understanding Across Platforms},
  author = {Li, Zhangheng and You, Keen and Zhang, Haotian and Feng, Di and Agrawal, Harsh and Li, Xiujun and Moorthy, Mohana Prasad Sathya and Nichols, Jeff and Yang, Yinfei and Gan, Zhe},
  year = {2024},
  eprint = {2410.18967},
  archivePrefix = {arXiv},
  primaryClass = {cs.CV},
  doi = {10.48550/arXiv.2410.18967}
}

@inproceedings{wu2024mobilevlm293,
  title = {{MobileVLM}: A {Vision-Language} Model for Better Intra- and {Inter-UI} Understanding},
  author = {Wu, Qinzhuo and Xu, Weikai and Liu, Wei and Tan, Tao and Liujianfeng, Liujian and Li, Ang and Luan, Jian and Wang, Bin and Shang, Shuo},
  booktitle = {Findings of the Association for Computational Linguistics: EMNLP 2024},
  year = {2024},
  pages = {10231--10251},
  doi = {10.18653/v1/2024.findings-emnlp.599},
  url = {https://doi.org/10.18653/v1/2024.findings-emnlp.599}
}

@misc{gao2024mobileviews294,
  title = {{MobileViews}: A Million-scale and Diverse Mobile {GUI} Dataset},
  author = {Gao, Longxi and Zhang, Li and Wang, Shihe and Gao, Pengzhi and Liu, Wei and Luan, Jian and Wang, Shangguang and Li, Yuanchun and Xu, Mengwei},
  year = {2024},
  eprint = {2409.14337},
  archivePrefix = {arXiv},
  primaryClass = {cs.HC},
  doi = {10.48550/arXiv.2409.14337}
}

@inproceedings{ran2024guardian295,
  title = {Guardian: A Runtime Framework for {LLM-Based} {UI} Exploration},
  author = {Ran, Dezhi and Wang, Hao and Song, Zihe and Wu, Mengzhou and Cao, Yuan and Zhang, Ying and Yang, Wei and Xie, Tao},
  booktitle = {Proceedings of the 33rd ACM SIGSOFT International Symposium on Software Testing and Analysis},
  year = {2024},
  pages = {958--970},
  doi = {10.1145/3650212.3680334},
  url = {https://doi.org/10.1145/3650212.3680334}
}

@inproceedings{ziyang2024vga296,
  title = {{VGA}: Vision {GUI} Assistant - Minimizing Hallucinations through {Image-Centric} {Fine-Tuning}},
  author = {Ziyang, Meng and Dai, Yu and Gong, Zezheng and Guo, Shaoxiong and Tang, Minglong and Wei, Tongquan},
  booktitle = {Findings of the Association for Computational Linguistics: EMNLP 2024},
  year = {2024},
  pages = {1261--1279},
  doi = {10.18653/v1/2024.findings-emnlp.68},
  url = {https://doi.org/10.18653/v1/2024.findings-emnlp.68}
}

@misc{lu2024guiodyssey297,
  title = {{GUIOdyssey}: A Comprehensive Dataset for {Cross-App} {GUI} Navigation on Mobile Devices},
  author = {Lu, Quanfeng and Shao, Wenqi and Liu, Zitao and Du, Lingxiao and Meng, Fanqing and Li, Boxuan and Chen, Botong and Huang, Siyuan and Zhang, Kaipeng and Luo, Ping},
  year = {2024},
  eprint = {2406.08451},
  archivePrefix = {arXiv},
  primaryClass = {cs.CV},
  doi = {10.48550/arXiv.2406.08451}
}

@article{chen2026octo298,
  title = {{Octo-Planner}: {On-Device} Language Model for {Planner-Action} Agents},
  author = {Chen, Wei and Li, Zhiyuan and Guo, Zhen and Shen, Yikang},
  journal = {Lecture Notes in Computer Science},
  year = {2026},
  pages = {141--156},
  doi = {10.1007/978-3-032-18011-7\_9},
  url = {https://doi.org/10.1007/978-3-032-18011-7\_9}
}

@misc{wang2024e299,
  title = {{E-ANT}: A {Large-Scale} Dataset for Efficient Automatic {GUI} {NavigaTion}},
  author = {Wang, Ke and Xia, Tianyu and Gu, Zhangxuan and Zhao, Yi and Shen, Shuheng and Meng, Changhua and Wang, Weiqiang and Xu, Ke},
  year = {2024},
  eprint = {2406.14250},
  archivePrefix = {arXiv},
  primaryClass = {cs.CV},
  doi = {10.48550/arXiv.2406.14250}
}

@misc{chen2024gui300,
  title = {{GUI-World}: A Video Benchmark and Dataset for Multimodal {GUI-oriented} Understanding},
  author = {Chen, Dongping and Huang, Yue and Wu, Siyuan and Tang, Jingyu and Chen, Liuyi and Bai, Yilin and He, Zhigang and Wang, Chenlong and Zhou, Huichi and Li, Yiqiang and Zhou, Tianshuo and Yu, Yue and Gao, Chujie and Zhang, Qihui and Gui, Yi and Li, Zhen and Wan, Yao and Zhou, Pan and Gao, Jianfeng and Sun, Lichao},
  year = {2024},
  eprint = {2406.10819},
  archivePrefix = {arXiv},
  primaryClass = {cs.CV},
  doi = {10.48550/arXiv.2406.10819}
}

@inproceedings{li2024on301,
  title = {On the Effects of Data Scale on {UI} Control Agents},
  author = {Li, Wei and Bishop, William and Li, Alice and Rawles, Chris and Campbell-Ajala, Folawiyo and Tyamagundlu, Divya and Riva, Oriana},
  booktitle = {Advances in Neural Information Processing Systems 37},
  year = {2024},
  pages = {92130--92154},
  doi = {10.52202/079017-2925},
  url = {https://doi.org/10.52202/079017-2925}
}

@inproceedings{bai2024digirl302,
  title = {{DigiRL}: Training {In-The-Wild} {Device-Control} Agents with Autonomous Reinforcement Learning},
  author = {Bai, Hao and Zhou, Yifei and Cemri, Mert and Pan, Jiayi and Suhr, Alane and Levine, Sergey and Kumar, Aviral},
  booktitle = {Advances in Neural Information Processing Systems 37},
  year = {2024},
  pages = {12461--12495},
  doi = {10.52202/079017-0397},
  url = {https://doi.org/10.52202/079017-0397}
}

@inproceedings{qian2024visual303,
  title = {Visual Grounding for User Interfaces},
  author = {Qian, Yijun and Lu, Yujie and Hauptmann, Alexander and Riva, Oriana},
  booktitle = {Proceedings of the 2024 Conference of the North American Chapter of the Association for Computational Linguistics: Human Language Technologies (Volume 6: Industry Track)},
  year = {2024},
  pages = {97--107},
  doi = {10.18653/v1/2024.naacl-industry.9},
  url = {https://doi.org/10.18653/v1/2024.naacl-industry.9}
}

@inproceedings{thil2024navigating304,
  title = {Navigating {WebAI}: Training Agents to Complete Web Tasks with Large Language Models and Reinforcement Learning},
  author = {Thil, Lucas-Andrei and Popa, Mirela and Spanakis, Gerasimos},
  booktitle = {Proceedings of the 39th ACM/SIGAPP Symposium on Applied Computing},
  year = {2024},
  pages = {866--874},
  doi = {10.1145/3605098.3635903},
  url = {https://doi.org/10.1145/3605098.3635903}
}

@misc{rahman2024v305,
  title = {{V-Zen}: Efficient {GUI} Understanding and Precise Grounding With A Novel Multimodal {LLM}},
  author = {Rahman, Abdur and Chawla, Rajat and Kumar, Muskaan and Datta, Arkajit and Jha, Adarsh and NS, Mukunda and Bhola, Ishaan},
  year = {2024},
  eprint = {2405.15341},
  archivePrefix = {arXiv},
  primaryClass = {cs.AI},
  doi = {10.48550/arXiv.2405.15341}
}

@article{you2024ferret306,
  title = {{Ferret-UI}: Grounded Mobile {UI} Understanding with Multimodal {LLMs}},
  author = {You, Keen and Zhang, Haotian and Schoop, Eldon and Weers, Floris and Swearngin, Amanda and Nichols, Jeffrey and Yang, Yinfei and Gan, Zhe},
  journal = {Lecture Notes in Computer Science},
  year = {2024},
  pages = {240--255},
  doi = {10.1007/978-3-031-73039-9\_14},
  url = {https://doi.org/10.1007/978-3-031-73039-9\_14}
}

@misc{chawla2024guide307,
  title = {{GUIDE}: Graphical User Interface Data for Execution},
  author = {Chawla, Rajat and Jha, Adarsh and Kumar, Muskaan and NS, Mukunda and Bhola, Ishaan},
  year = {2024},
  eprint = {2404.16048},
  archivePrefix = {arXiv},
  primaryClass = {cs.HC},
  doi = {10.48550/arXiv.2404.16048}
}

@misc{fereidouni2024grounded308,
  title = {Grounded Language Agent for Product Search via Intelligent Web Interactions},
  author = {Fereidouni, Moghis and Mosharrof, Adib and Siddique, A. B.},
  year = {2024},
  eprint = {2404.10887},
  archivePrefix = {arXiv},
  primaryClass = {cs.CL},
  doi = {10.48550/arXiv.2404.10887}
}

@misc{baechler2024screenai309,
  title = {{ScreenAI}: A {Vision-Language} Model for {UI} and Infographics Understanding},
  author = {Baechler, Gilles and Sunkara, Srinivas and Wang, Maria and Zubach, Fedir and Mansoor, Hassan and Etter, Vincent and Cărbune, Victor and Lin, Jason and Chen, Jindong and Sharma, Abhanshu},
  year = {2024},
  eprint = {2402.04615},
  archivePrefix = {arXiv},
  primaryClass = {cs.CV},
  doi = {10.48550/arXiv.2402.04615}
}

@inproceedings{li2024uinav310,
  title = {{UINav}: A Practical Approach to Train {On-Device} Automation Agents},
  author = {Li, Wei and Hsu, Fu-Lin and Bishop, William and Campbell-Ajala, Folawiyo and Lin, Max and Riva, Oriana},
  booktitle = {Proceedings of the 2024 Conference of the North American Chapter of the Association for Computational Linguistics: Human Language Technologies (Volume 6: Industry Track)},
  year = {2024},
  pages = {36--51},
  doi = {10.18653/v1/2024.naacl-industry.4},
  url = {https://doi.org/10.18653/v1/2024.naacl-industry.4}
}

@misc{zhang2023reinforced311,
  title = {Reinforced {UI} Instruction Grounding: Towards a Generic {UI} Task Automation {API}},
  author = {Zhang, Zhizheng and Xie, Wenxuan and Zhang, Xiaoyi and Lu, Yan},
  year = {2023},
  eprint = {2310.04716},
  archivePrefix = {arXiv},
  primaryClass = {cs.CV},
  doi = {10.48550/arXiv.2310.04716}
}

@misc{furuta2023multimodal312,
  title = {Multimodal Web Navigation with {Instruction-Finetuned} Foundation Models},
  author = {Furuta, Hiroki and Lee, Kuang-Huei and Nachum, Ofir and Matsuo, Yutaka and Faust, Aleksandra and Gu, Shixiang Shane and Gur, Izzeddin},
  year = {2023},
  eprint = {2305.11854},
  archivePrefix = {arXiv},
  primaryClass = {cs.LG},
  doi = {10.48550/arXiv.2305.11854}
}

@article{burns2022a313,
  title = {A Dataset for Interactive {Vision-Language} Navigation with Unknown Command Feasibility},
  author = {Burns, Andrea and Arsan, Deniz and Agrawal, Sanjna and Kumar, Ranjitha and Saenko, Kate and Plummer, Bryan A.},
  journal = {Lecture Notes in Computer Science},
  year = {2022},
  pages = {312--328},
  doi = {10.1007/978-3-031-20074-8\_18},
  url = {https://doi.org/10.1007/978-3-031-20074-8\_18}
}

\end{document}